\documentclass[12pt]{article}
\usepackage{mathrsfs}
\usepackage{fullpage}
\usepackage{color}
\usepackage{amsfonts}
\usepackage{graphicx}
\usepackage{amsmath}
\usepackage{amssymb}
\usepackage{bbm}
\usepackage{float}
\usepackage{rotating}
\allowdisplaybreaks
\usepackage[authoryear]{natbib}
\usepackage{tablefootnote}
\usepackage{bm}
\usepackage{xr}
\usepackage{booktabs}
\usepackage{color}
\usepackage{threeparttable}
\renewcommand\arraystretch{0.3}

\newtheorem{thm}{Theorem}
\newtheorem{lemma}{Lemma}

\def\proof {{\noindent\bf Proof.}\quad}
\def\var{\mathrm {var}}

\def\cov{\mathrm {cov}}

\def\diag{\mathrm {diag}}

\def\tr{\mathrm {tr}}
\def\U{{\bf U}}
\def\A{{\bf A}}
\def\V{{\bm V}}

\def\B{{\bf B}}
\def\z{{\bm z}}

\def\I{{\bf I}}

\def\bmu{{\bm \mu}}

\def\P{{\bf M}}

\def\F{{\bf F}}

\def\f{{\bm f}}

\def\X{{\bm X}}
\def\L{{\bf \Lambda}}
\def\O{{\bf \Omega}}
\def\D{{\bf D}}

\def\tr{\mathrm {tr}}

\def\bms{{\bm\Sigma}}

\def\cp{\mathop{\rightarrow}\limits^{p}}
\def\cd{\mathop{\rightarrow}\limits^{d}}
\def\mR{\mathcal{R}}
\def\bma{\bm \alpha}
\def\bmv{\bm \varepsilon}

\def\SUM{{\text{Sum}}}
\def\MAX{{\text{Max}}}

\def\dif{\mathop{}\hphantom{\mskip-\thinmuskip}\mathrm{d}}%
\let\daccent\d
\let\d\relax
\newcommand\d{\ifmmode\dif\else\expandafter\daccent\fi}
\date{\today}
\def\boxit#1{\vbox{\hrule\hbox{\vrule\kern6pt  \vbox{\kern6pt#1\kern6pt}\kern6pt\vrule}\hrule}}

\title{Testing Alpha in High Dimensional Linear Factor Pricing Models with Dependent Observations}
\author{Huifang Ma\\
			School of Statistics and Data Science, KLMDASR, LEBPS, and LPMC,\\ Nankai University
			\\~\\
			Long Feng\hspace{.2cm} \\
			School of Statistics and Data Science, KLMDASR, LEBPS, and LPMC,\\ Nankai University
		    \\~\\
	        Zhaojun Wang\hspace{.2cm} \\
	        School of Statistics and Data Science, KLMDASR, LEBPS, and LPMC,\\ Nankai University\\~\\
           Jigang Bao\hspace{0.2cm}\\
           Tsinghua Shenzhen International Graduate School, Tsinghua University\\
           Shenzhen Wukong Investment Management Co. Ltd.}
\begin{document}
\maketitle
\begin{abstract}
In this study, we introduce three distinct testing methods for testing alpha in high dimensional linear factor pricing model that deals with dependent data. The first method is a sum-type test procedure, which exhibits high performance when dealing with dense alternatives. The second method is a max-type test procedure, which is particularly effective for sparse alternatives. For a broader range of alternatives, we suggest a Cauchy combination test procedure. This is predicated on the asymptotic independence of the sum-type and max-type test statistics. Both simulation studies and practical data application demonstrate the effectiveness of our proposed methods when handling dependent observations.

\end{abstract}

\section{Introduction}

This study focuses on the detection of alpha in Linear Factor Pricing Models (LFPM), particularly when the quantity of securities significantly outweighs the temporal dimension of the return series. LFPM is a cornerstone in the field of finance. Inspired by the Arbitrage Pricing Theory \citep{ross1976arbitrage}, LFPM elucidates the relationship between security returns and market factors. LFPM manifests in various forms, such as the renowned single-factor model, namely the Sharpe-Lintner Capital Asset Pricing Model (CAPM) \citep{Sharpe1964CAPITALAP,Lintner1965THEVO}, the Fama-French three-factor model \citep{Fama1993Common}, and the Fama-French five-factor model \citep{Fama2014AFA}. Each factor in these models typically carries substantial economic significance and pricing capability.

When the number of securities $N$ is fixed, \citet{Gibbons1989ATO} introduced an exact multivariate F-test under the assumption of joint normality. This classic GRS test procedure was later expanded by \citet{Mackinlay1991UsingGM}, \citet{Zhou1993AssetpricingTU}, and \citet{Beaulieu2007}. In today’s financial markets, the number of securities often exceeds a thousand, which may surpass the length of the considered time period. Therefore, it’s not appropriate to assume a fixed number of securities. To address this, recent advancements have focused on devising tests that accommodate situations where the number of securities $N$ exceeds the time periods $T$. For instance, \citet{Pesaran2012TestingCW,Pesaran2017TestingFA,pesaran2023testing} introduced a sum-type test statistic that replaces the sample covariance matrix with an identity matrix in the traditional F-test, performing well under dense alternatives.  \citet{lan2018testing} proposed a random projection procedure which admits the covariance matrix of the idiosyncratic term to be nonsparse.  For sparse alternatives, \citet{Gungor2013TestingLF} and \cite{feng2022high} proposed  max-type test statistic. \cite{Fan2015} proposed a power enhancement procedure to strengthen the test power under the sparse alternatives. \citet{Yu2023PE} utilized the thresholding covariance estimator of \citet{Fan2011LargeCE} to propose a novel Wald-type test and proposed a new Cauchy combination test which combinates the Wald-type test and max-type test. \citet{xia2023adaptive} consider $L$-statistics for handling the sparse alternatives. For heavy-tailed distributions, \citet{liu2023high} proposed a robust spatial sign-based nonparametric test procedure. \cite{zhao2022high} extended \cite{liu2023high}'s results by considering a class of weighted spatial-sign tests and choosing an optimal test INST. 

In addition, Some literatures considered some more general LFPMs. For example, \citet{Gagliardini2016TimeVaryingRP} and \citet{Ma2018TestingAI} established two sum-type testing procedures which allow the existence of time-variation coefficients. \citet{ma2023adaptive} introduce a maximum-type test that performs well in scenarios where the alternative hypothesis is sparse and an adaptive test by combining the maximum-type test and \citet{Ma2018TestingAI}'s sum-type test. \cite{zhao2023robust} also proposed a robust spatial-sign-based nonparametric test procedure for the time-varying factor pricing model. In addressing observed non-traded factors, a substantial body of literature exists that estimates the risk premiums associated with these factors using a two-pass regression methodology. This approach, pioneered by \cite{1973Risk}, is also featured in recent works such as those by \cite{2021Factor} and \cite{0Arbitrage}. When it comes to unobserved factors, numerous methods have been developed to estimate these factors and their corresponding loadings, primarily through principal component analysis. For example, \cite{giglio2021thousands} devised a multiple testing procedure that operates on the residuals remaining after the influence of the estimated latent factors and loadings has been removed.

Most existing studies do not take into account the dependence structure among observations, instead requiring the assumption of identical and independent distribution for random errors. Only the works of \cite{Ma2018TestingAI} and \cite{ma2023adaptive} have extended their testing procedures under the assumption of martingale difference. In the financial market, observations often exhibit some form of dependence structure, particularly in daily and weekly data. As demonstrated in real data applications, the residuals of the traditional linear factor pricing model are time-dependent, making the assumption of identical and independent distribution unsuitable for random error. In high-dimensional settings, \cite{ayyala2017mean} proposed a sum-type test procedure for the $M-$dependent stationary Gaussian process, with corrections to the proofs of \cite{ayyala2017mean}'s main theorems provided by \cite{CLAPR19}. For the max-type test procedure, significant efforts have been made towards Gaussian approximation for high-dimensional time series, as seen in the works of \cite{ZW17}, \cite{ZC18}, and \cite{chang2024central}. An overview of these efforts can be found in \cite{chernozhukov2023high}. To the best of our knowledge, no literature exists that considers the high-dimensional alpha test of linear factor pricing models with dependent observations.

In this study, we initially introduce a sum-type test statistic, which is derived by simply summing the squares of the estimated alphas for each security. We compute the expectation and asymptotic variance of this sum-type test statistic and affirm its asymptotic normality under the assumption of an $MA(\infty)$ process. Notably, we relax the assumptions of normal distribution and $M$-dependence as presented in \cite{ayyala2017mean}. Subsequently, we provide a consistent estimator of the expectation and asymptotic variance and propose a sum-type test procedure that performs exceptionally well under dense alternatives, i.e., when many alphas of the securities are non-zero. In cases where only a few alphas of the securities are non-zero, we propose a max-type test statistic, which is the maximum of the square of estimated alphas divided by its corresponding long-run variances. For dependent observations, we demonstrate that the limit null distribution of the proposed max-type test statistic also follows a Gumbel distribution and establish its consistency under sparse alternatives.

It's important to note that the underlying truth in real applications is typically unknown, as the density or sparsity is contingent on the properties of the securities involved.  Recent literature has demonstrated that the maximum-type test statistic is asymptotically independent of the sum-type test statistic, leading to the proposal of a corresponding combination test in numerous high-dimensional problems. These include high-dimensional mean testing problems by \cite{xu2016adaptive}, \citet{he2021} and \citet{Feng2022AsymptoticIO}, cross-sectional independence test in high-dimensional panel data models by \citet{feng2022a}, testing of high-dimensional covariance matrix by \citet{yu2022jasa}, high-dimensional change point inference by \citet{wang2023} and testing alpha of high dimensional time-varying factor model by \cite{ma2023adaptive}. All the aforementioned papers affirm the asymptotic independence between the sum-type test statistic and the max-type test statistic, based on the assumption of identical and independent distribution. However, this assumption may not be directly applicable in our context. In this study, we are the first to establish the asymptotic independence between the sum-type test statistic and the max-type test statistic under the assumption of time-dependence. We then introduce a Cauchy combination test procedure, which is capable of handling both dense and sparse alternatives simultaneously. Furthermore, we provide a power analysis of the proposed Cauchy combination test procedure. The advantages of our proposed tests over existing methods are demonstrated through extensive Monte Carlo experiments and an empirical application.

In summary, there are three main contributions of this paper:
    \begin{itemize}
    \item[1.] We have relaxed the assumption of Gaussian distribution and $M$-dependence as presented in \cite{ayyala2017mean}, and have established the asymptotic normality of the sum-type test statistic in the presence of dependent observations.
    \item[2.] For the max-type test, we have also established the limit null distribution and its consistency under sparse alternatives, assuming time-dependent observations.
    \item[3.] We are the first to establish the asymptotic independence between the sum-type test statistic and the max-type test statistic under the assumption of time-dependence. We have proposed a Cauchy combination test procedure, which performs very well under neither dense nor sparse alternatives.
    \end{itemize}

The remainder of this paper is structured as follows. In Sections 2, we present the sum-type test, the max-type test procedures and the Cauchy combination test procedure. Section 3 showcases some simulation studies. An empirical application is provided in Section 4. Section 5 contains some discussions. All technical details are included in the Appendix. 

\section{Test Procedures}
\subsection{Sum-type test}

Assume that there are a total of $N$ securities and each security has $T$ observations of its excess return that are regularly collected from different time periods.
Let $y_{it}$ be the excess return of security $i\in \{1,\cdots,N\}$ at time $t\in \{1,\cdots,T\}$.  Let $\f_t=(f_{t1},\cdots,f_{tp})^\top\in\mathcal{R}^p$ denote the $p$-vector of observed factors at time $t$,
which represents the excess return over the excess return on the $p$ observed factors. In this paper, we assume that the number of factors $p$ is fixed, which depends on the specific multi-factor model studied.

Assume that security returns are governed by the following multi-factor model:
\begin{align}\label{mod}
y_{it}=\alpha_i+\bm\beta_i^\top \f_t+\varepsilon_{it},~i\in \{1,\cdots,N\},~t\in \{1,\cdots,T\},
\end{align}
where $\alpha_i$ is a scalar representing the security specific intercept, $\bm\beta_i=(\beta_{i1},\cdots,\beta_{ip})^\top\in\mathcal{R}^p$
is a vector of multiple regression betas of security $i$ with respect to the $p$ factors, and the $\varepsilon_{it}$'s are the corresponding idiosyncratic error terms. Note that for the classic CAPM \citep{Sharpe1964CAPITALAP,Lintner1965THEVO}, the single common factor is the market risk, while for the Fama and French three-factor model \citep{Fama1993Common}, the common factors are the market risk, SMB and HML, where SMB and HML measure the historic excess returns of small-cap stocks over big-cap stocks and of value stocks over growth stocks, respectively.

Stacking by cross-sectional observations, we can write the above multi-factor model as
\begin{align}\label{mod_mat}
{\bf y}_{t}=\bm\alpha+{\bf B}\f_t+\bm\varepsilon_{t},~t\in \{1,\cdots,T\},
\end{align}
where ${\bf y}_{t}=(y_{1t}, \cdots, y_{Nt})^\top\in\mR^N$, $\bm\alpha=(\alpha_1, \cdots, \alpha_N)^\top\in\mR^N$,
${\bf B}=(\bm \beta_1,\cdots,\bm \beta_N)^\top\in\mR^{N\times p}$ and $\bmv_{t}=(\varepsilon_{1t}, \cdots, \varepsilon_{Nt})^\top\in\mR^N$. In addition, there are some notation to be used in later sections: ${\bf y}_{i\cdot}=(y_{i1}, \cdots, y_{iT})^\top\in\mR^T$,
$\bmv_{i\cdot}=(\varepsilon_{i1}, \cdots, \varepsilon_{iT})^\top\in\mR^T$, for each $i\in \{1,\cdots,N\}$;
${\bf Y}=({\bf y}_1, \cdots, {\bf y}_N)\in\mR^{T\times N}$, $\bmv=(\bmv_1, \cdots, \bmv_N)\in\mR^{T\times N}$,
$\F=(\f_1, \cdots, \f_T)^\top\in\mR^{T\times p}$.

To evaluate the marginal return associated with an additional strategy that is not explained by existing factors, many researchers employ the specification test for a factor model
by testing
\begin{align} \label{h1}
H_0: \bm \alpha=\mathbf{0}~~ \text{versus}~~ H_1: \bm \alpha\not=\mathbf{0}.
\end{align}
This test has a special explanation in the seminal single-factor model CAPM. Specifically, if a single factor or a particular portfolio is mean-variance efficient,
i.e. it minimizes variance for a given level of expected return, then $E(y_{it})=\bm\beta_i^\top E(\f_t)$ must be satisfied
with $\bm\beta_i=\cov(y_{it}, \f_t)\var^{-1}(\f_t)\in\mR^p$ standing for the systematic risk on security $i$, which
is equivalent to $\bm\alpha=\mathbf{0}$ under the distributional assumption given by \eqref{mod}.
Thus, in CAPM the above test in \eqref{h1} is actually a test of the mean-variance efficiency.

Let $\bm 1_T=(1,\cdots,1)^\top\in\mathcal{R}^T$, $\I_T$ be the $T\times T$ identity matrix and $\P_{\F}=\I_T-\F(\F^\top\F)^{-1}\F^\top$;
$\hat{\bm \alpha}=(\hat{\alpha}_1,\cdots,\hat{\alpha}_N)^\top$ be the Ordinary Least Squares (OLS) estimator of, where
$$\hat{\alpha}_i=\mathbf{y}_{i\cdot}^\top \left(\frac{\P_{\F}\bm 1_T}{\bm 1_T^\top \P_{\F}\bm 1_T}\right).$$
And $\hat{\bms}=\frac{1}{T}\sum_{t=1}^T\hat{\bmv}_{t}\hat{\bmv}_{t}^\top$ with $\hat{\bmv}_{t}=(\hat{\bmv}_{1t},\cdots,\hat{\bmv}_{Nt})^\top$,
where $\hat{\bmv}_{it}$ is the OLS residual from the regression of $y_{it}$ on an intercept and $\f_t$,
i.e. $$\hat{\bmv}_{i\cdot}=(\hat{\varepsilon}_{i1}, \cdots, \hat{\varepsilon}_{iT})^\top=\P_{\F}(\mathbf{y}_i-\hat{\alpha}_i).$$

To handle dense alternatives, we propose a sum-type test, named the SUM test, with the test statistic constructed as
\begin{align}\label{sum_test}
T_{\SUM}=\hat{\bm \alpha}^\top \hat{\bm \alpha}.
\end{align}
Note that, if the error term is identical and independent distributed, \cite{pesaran2023testing} also construct a similar sum-type test statistic
\begin{align*}
T_{{\rm PY}}&=\frac{N^{-1/2}\sum_{i=1}^N
  \{t_i^2-v/(v-2)\}}{v/(v-2)\sqrt{2(v-1)/(v-4)\{1+(N-1)\tilde{\rho}_{MT}^2\}}}, \\
t_i^2&=\frac{\hat{\alpha}_i^2(\bm 1_T^\top \P_{\F}\bm
  1_T)}{v^{-1}\hat{\bmv}_{i\cdot}^\top\hat{\bmv}_{i\cdot}},
\end{align*}
where $v=T-p-1$ and
$\tilde{\rho}_{MT}=2/\{N(N-1)\}\sum_{i=2}^N\sum_{j=1}^{i-1}\tilde{\rho}^2_{ij}$
is the corresponding correlation estimator of
$\rho^2=2/\{N(N-1)\}\sum_{i=2}^N\sum_{j=1}^{i-1}\rho^2_{ij}$
with $\tilde{\rho}_{ij}$ denoting the multiple testing estimator of
the correlation
$\rho_{ij}=\sigma_{ij}/(\sigma_{ii}^{1/2}\sigma_{jj}^{1/2})$
\citep{bailey2019a}.
However, when the errors $\bm \varepsilon_t$ are time-dependent, a bias term that cannot be ignored appears in $T_{{\rm PY}}$. That is, under the null hypothesis, the expectation of $T_{{\rm PY}}$ is not zero, i.e., $E(T_{{\rm PY}})\not=0$. Furthermore, the asymptotic variance of $T_{{\rm PY}}$ will no longer be one. Therefore, $T_{{\rm PY}}$ may not perform well with dependent observations.

In our proposed test, we allow both $N$ and $T$ to be large, and allow $\bmv_t$'s be dependent linear processes. We need the conditions for the error terms and factors given below.
\begin{itemize}
\item[(C1)] (\romannumeral1) $\{\f_1,\cdots,\f_T\}$ is an strictly stationary. (\romannumeral2)
$\bms_{\f}\equiv\cov(\f_t)$ is strictly positive definite. (\romannumeral3) $\f_t^\top \f_t$ is bounded for all $t$.
\end{itemize}
Under Condition (C1), we have $T^{-1}\bm 1_T^\top \P_{\F}\bm 1_T\cp \omega$, $T^{-1}\F^\top \bm 1_T\cp E(\f_t)=: \bmu_{\f}$, $T^{-1}\F^\top \F\cp \mu_{\f}\mu_{\f}^\top +\bms_{\f}=: {\bf \Lambda}_{\f}$. Define $e_t:=\frac{1}{\omega }(1-\f_t^\top{\bf \Lambda}_{\f}^{-1}\bmu_{\f})$. Because $\f_t$ is strictly stationary, so $e_t$ is also strictly stationary. 
Define $\X_t:=\bmv_te_t$. We consider the following dependent model for $\X_t$:
\begin{itemize}
\item[(C2)] $\X_{1},\cdots,\X_T\in \mathbb{R}^N$ follow an $MA(\infty)$ process of the form \citep{ZPG18}
\begin{align}\label{modelx}
    \X_t=\bm\Sigma^{1/2} \sum_{k=0}^{\infty} b_k \z_{t-k}.
\end{align}  
(\romannumeral1) $\{\z_t\}$ is a sequence of $N$-dimensional random vectors such that, if the coordinates of $\z_{t}$ are $\left\{z_{i t}\right\},$ then the two dimensional array $\left\{z_{i t}: 1 \leq i \leq N, t \geq 1\right\}$ of variables are i.i.d. satisfying the moment conditions $\mathbb{E} z_{i t}=0, \mathbb{E}\left|z_{i t}\right|^{2}=1$ and $\mathbb{E}\left|z_{i t}\right|^{4}=\nu_{4}<\infty$. 

(\romannumeral2) $\sum_{k=0}^\infty|b_k|<\infty$, $\sum_{k=0}^\infty b_k=s\ne 0$ and $b_k=o(T^{-5})$, i.e. $\lim_{T\to\infty}T^5b_T=0$. 
\end{itemize}
Unlike the $M-$dependent Gaussian stationary assumption in \cite{ayyala2017mean}, we allow the error term can follows a $MA(\infty)$ process and $z_{it}$ could be not normal distributed.
Under Condition (C2), $\X_t$ has mean zero and auto-covariance structure given by $\Gamma_h,h\in\mathbb{Z}$, i.e. $\mathbb{E}(\X_t\X_{t+h}^\top)=:\Gamma_h=a_h \bm\Sigma$ for $h=0,1,2,\cdots$ and $\Gamma_h=\Gamma_{-h}^\top$ with $a_h:=\sum_{k=0}^\infty b_k b_{k+h}$. Next, we give some conditions of structure of the auto-covariance matrix.
\begin{itemize}
    \item[(C3)] There exist two positive constants $M_0$ and $M_1$ such that $\Vert\bm\Sigma\Vert_2\le M_0$ and $\tr(\bm\Sigma)/N\ge M_1$, and $\tr^2(\bm\Sigma)/(T\tr(\bm\Sigma^2))=o(1)$. Here $\|\cdot\|_2$ stands for either the spectral norm of a matrix or the Euclidean norm of a vector.
    
    \item[(C4)] $\O_T:= T\cov(\bar{\X}_T)=\Gamma_0+2\sum_{h=1}^T(1-\frac{h}{T})\Gamma_h\to \Gamma_0$, where $\Gamma_0$ is the spectral matrix evaluated at the zero frequency.

    \item[(C5)] $N=O(T)$, and $N\to\infty$ as $T\to \infty$.
\end{itemize}
Condition (C3) reflects Assumption (A3) as stated in \cite{ZPG18}, suggesting that the correlation among these variables should not be overly high. Condition (C4) is a standard assumption made for multivariate time series. In this study, we permit the number of securities $N$ to be comparable to the length of the time period $T$, which is generally sufficient for most empirical studies.

We propose to use a standardized version of $T_{\SUM}$ as the test statistic. From simple mathematical derivation, we have
\begin{align*}\label{tilde_sum_test}
    \hat{\bm \alpha}=\bm\alpha+\frac{1}{T}\sum_{t=1}^T\bmv_t \eta_t
\end{align*}
with $\bm\eta=(\eta_t,\cdots,\eta_T)^\top=\frac{\P_{\F}\bm 1_T}{T^{-1}\bm 1_T^\top \P_{\F}\bm 1_T}$. Define 
\begin{align}
    \tilde{T}_{\SUM}=\tilde{\bm \alpha}^\top \tilde{\bm \alpha} ~~\text{with}~~ \tilde{\bm \alpha}=\bm\alpha+\frac{1}{T}\sum_{t=1}^T\bmv_t e_t.
\end{align}
In the appendix, we will show the difference between $T_{\SUM}$ and $\tilde{T}_{\SUM}$ are negligible.
So, we only need to consider the mean and variance of $\tilde{T}_{\SUM}$ under $H_0$, given below:
\begin{align*}
    &\mathbb{E}(\tilde{T}_{\SUM})=\mathbb{E}(\bar{\X}_T^\top \bar{\X}_T)=\frac{1}{T}\left\{\tr(\Gamma_0)+2\sum_{h=1}^T\left(1-\frac{h}{T}\right)\tr(\Gamma_h)\right\}=: \mu_T,\\
    &\var(\tilde{T}_{\SUM})\approx \frac{2}{T^2} \tr(\O_T^2)=:\sigma_T^2,
\end{align*}
We use $\mu_T'=\frac{1}{T}\{\tr(\Gamma_0)+2\sum_{h=1}^M(1-\frac{h}{T})\tr(\Gamma_h)\}$ replace $\mu_T$, where $M=\lceil\min(N,T)^{1/8}\rceil$. The reasons for such substitution will be explained later.

First, we give the asymptotic normality of the test statistic $T_{\SUM}$.
\begin{thm}\label{th1}
Under Conditions (C1)-(C5) hold and assuming $M=\lceil\min(N,T)^{1/8}\rceil$. Under $H_0$, we have $(T_{\SUM}-\mu_T')/\sigma_T \cd N(0,1)$ as $\min(N,T)\to\infty$.
\end{thm}

Next, we consider the following alternative hypothesis:
\begin{itemize}
\item[(C6)] $\bma^\top (\Gamma_h\Gamma_{-h})^{1/2}\bma=o\{(M+1)^{-1}T^{-1}\tr(\O_T^2)\}$, $h=0,\cdots,M$.
\end{itemize}

\begin{thm}\label{th2}
Under Conditions (C1)-(C6) hold and assuming $M=\lceil\min(N,T)^{1/8}\rceil$, we have $(T_{\SUM}-\mu_T')/\sigma_T\cd N\left(\frac{\bma^\top\bma}{\sqrt{2T^{-2}\tr(\O_T^2)}},1\right)$ as $\min(N,T)\to\infty$.
\end{thm}

It is worth noting that $(T_{\SUM}-\mu_T')/\sigma_T$ is usually unknown since it involves population parameters. Thus it cannot be used as a test statistic in practice. We therefore need to find consistent estimators of $\mu_T'$ and $\sigma_T^2$. Since $\hat{\varepsilon}_{it}\cp \varepsilon_{it}$ and $\eta_t-e_t=o_p(1)$, the following formula gives an estimator of the auto-covariance matrix $\Gamma_h$:
\begin{align*}
    \hat{\Gamma}_h=\frac{1}{T-h}\sum_{t=1}^{T-h}\hat{\bmv}_{t+h}\hat{\bmv}_t^\top\eta_{t+h}\eta_t.
\end{align*}
Then, we consider the plug-in estimator of $\mu_T'$:
\begin{align*}
    \hat{\mu}_T=\frac{1}{T-p-1}\left\{\tr(\hat{\Gamma}_0)+2\sum_{h=1}^M\left(1-\frac{h}{T-p-1}\right)\tr(\hat{\Gamma}_h)\right\}.
\end{align*}

We next find a ratio consistent estimator of $\sigma_T^2$. For testing unit root, \citet{ZPG18} proposed an estimator for $a_0^2\tr(\bm\Sigma^2)$ and prove that the estimator is suitable for some special correlation structure. Their main idea was to construct a special set of subscripts such that only pairwise correlations exist between the four samples in each component of the estimator. Similarly, we consider the following estimator for $a_{h_1}a_{h_2}\tr(\bm \Sigma^2)=\tr(\Gamma_{h_1}\Gamma_{h_2})$:
\begin{align*}
    \widehat{\tr(\Gamma_{h_1}\Gamma_{h_2})}\equiv S_{h_1,h_2}=\frac{\sum_{t=1}^{[T/2]-h_2}\sum_{s=t+[T/2]}^{T-h_2}\hat{\bmv}_t^\top \hat{\bmv}_s \hat{\bmv}_{t+h_1}^\top \hat{\bmv}_{s+h_2}\eta_t\eta_s\eta_{t+h_1}\eta_{s+h_2} }{(T-h_2/2-\frac{3}{2}[T/2]+1/2)([T/2]-h_2)}.
\end{align*}
Accordingly, 
\begin{align*}
    \hat{\sigma}_T^2=\frac{2}{T^2}\left(S_{0,0}+2\sum_{r=1}^MS_{0,r}+2\sum_{r=1}^MS_{r,0}+4\sum_{r=1}^M\sum_{s=1}^MS_{r,s} \right).
\end{align*}
Next, we show that the above two estimators are consistent.
\begin{thm}\label{th3}
Assume the conditions in Theorems 1 and 2 hold. Then as $\min(N,T)\to\infty$, $\hat{\mu}_T-\mu_T'=o_p(\sigma_T)$ and $\hat{\sigma}_T^2/\sigma_T^2\cp 1$. 
\end{thm}

According to Theorems \ref{th1}-\ref{th3}, the $p$-value associated with $T_{\SUM}$ is 
\begin{equation*}
    p_{\SUM}=1-\Phi\left( \frac{T_{\SUM}-\hat{\mu}_T}{\hat{\sigma}_T}\right),
\end{equation*}
where $\Phi(\cdot)$ is the cumulative distribution function (CDF) of ${N}(0,1)$. If the $p$-value is below some pre-specified significant level, say $\gamma\in(0,1)$, then we rejected the null hypothesis.

According to Theorem \ref{th1} and \ref{th2}, the power function of $T_{\SUM}$ is
\begin{align}\label{sumpower}
\beta_{\SUM}(\bm \alpha)=\Phi\left(-z_{\gamma}+\frac{T\bma^\top\bma}{\sqrt{2\tr(\O_T^2)}}\right)
\end{align}
where $z_{\gamma}$ is the upper $\gamma$ quantile of $N(0,1)$. Note that the power function $\beta_{\SUM}(\bm \alpha)$ has the similar form of the power function of \cite{chen2010}'s test for the high dimensional two sample mean problem. If the error term $\bm \varepsilon_t$ are identical and independent distributed and the variances of each $\varepsilon_{it}, i=1,\cdots,N$ are equal, $\beta_{\SUM}$ are the same as the power function of $T_{{\rm PY}}$.

\subsection{Max-type test}

The power function (\ref{sumpower}) indicates that the sum-type test $T_{\SUM}$ can achieve strong performance when a significant number of alphas are nonzero. Conversely, if only a limited number of alphas are nonzero, $T_{\SUM}$ may not exhibit sufficient power. In scenarios involving sparse alternatives, numerous studies have shown that the max-type test procedure is highly effective. For instance, \cite{TonyCai2014TwosampleTO} demonstrated its utility in high-dimensional two-sample problems, \cite{feng2022high} applied it to high-dimensional alpha testing in linear factor pricing models, and \cite{ma2023adaptive} used it for high-dimensional alpha testing in time-varying factor pricing models.
  
Consequently, we proposed the following max-type test
\begin{align*}
T_{\MAX}=\max_{1\le i\le N}\frac{T\hat{\alpha}_i^2}{\hat{\sigma}_i},
\end{align*}
where $\hat{\sigma}_i=\sum_{h\in\mathcal{M}}\omega(h/M)\hat{\phi}_{i,h}$, $\hat{\phi}_{i,h}=\frac{1}{T-h}\sum_{t=h+1}^T\hat{\varepsilon}_{it}\hat{\varepsilon}_{i,t-h}\eta_t\eta_{t-h}$ and $\mathcal{M}=\{0, \pm 1,\dots,\pm M\}$. Here $\omega(\cdot)$ is a weighted function. For simplicity, we use the plain estimate in the article, i.e. $\omega(\cdot)=1$. Here $\hat{\sigma}_i$ is the long-run variance estimator.

Next, we will demonstrate that when $(N,T)\to\infty$, $T_{\MAX}-2\log(N) +\log\{\log(N)\}$ has a type \uppercase\expandafter{\romannumeral1} extreme value distribution. To proceed, we first introduce some conditions. 

Let $\D_0\equiv\diag(\sigma_1^0,\dots,\sigma_N^0)$ be the diagonal matrix of the long-run covariance of $\X_t$, $\O:=\Gamma_0+2\sum_{h=0}^\infty\Gamma_h$, and $\bm R_0:=\D_0^{-1/2}\O\D_0^{-1/2}$. Let $[\A]_{ij}$ denote the $(i,j)$-th element of matrix $\A$. 
\begin{itemize}
\item[(C7)] $\max_{1\le i\le N}\sum_{j=1}^N[\bm\Sigma^{1/2}]_{ij}^4<C$ for some positive constant $C$.

\item[(C8)] $\max_{1\le i<j\le N}[\bm R_0]_{ij}\le r<1$ and $\max_{1\le j\le N}\sum_{i=1}^N[\bm R_0]_{ij}^2\le c$ for some constants $r,c>0$.

\item[(C9)] $M^{1+a}T^{1/q-1/2}\log^2(N)\log^{3/2}(NT)=o(1)$ and $M^{qa^2+(q+7)a+2}T^{-qa/2}\log^{2+5qa/2}(N)\log^{7a}(NT)=o(1)$ for some $q\ge 4$ and $a>1/2-1/q$.
\end{itemize}
Condition (C7) and (C8) are similar to the conditions in Lemma 6 in \cite{TonyCai2014TwosampleTO}, which assume the correlations between those variables are not very large. In the special case with $M\asymp T^{1/8}$,  Condition (C9) becomes $\log(N)=o(T^\tau)$ where $\tau=\frac{2}{5}(\frac{1}{2}-\frac{1}{q}-\frac{1+a}{8})\wedge\frac{-qa^2+(3q-7)a-2}{16+(56+20q)a}$ with $q\ge 4$ and $a>1/2-1/q$.

\begin{thm}\label{th4}
Under Conditions (C1)-(C2) and (C7)-(C9), and assuming $M=\lceil\min(N,T)^{1/8}\rceil$, we have as $\min(N,T)\to\infty$,
\begin{align*}
P_{H_0}\left(T_{\MAX}-2\log(N)+\log\{\log(N)\}\le x\right)\to \exp\left\{-\frac{1}{\sqrt{\pi}}\exp\left(-\frac{x}{2}\right)\right\}\equiv F(x).
\end{align*}
\end{thm}
 
Here, $\mathbb{P}_{H_0}$ denotes the probability measure under the null hypothesis $H_0$. According to the limiting null distribution derived in Theorem \ref{th4}, we can easily obtain the $p$-value associated with $T_{\MAX}$, namely,
\begin{equation*}
    p_{\MAX}=1-F(T_{\MAX}-2\log(N) +\log\{\log(N)\}).
\end{equation*}
Again, small values of $p_{\MAX}$ indicate rejecting the null hypothesis. Next, we turn to the power analysis of the proposed max-type test.

\begin{thm}\label{th5}
Under Conditions (C1)-(C2) and (C7)-(C9), and assuming $M=\lceil\min(N,T)^{1/8}\rceil$, we have as $\min(T,N)\to
\infty$,
\begin{align*}
\inf_{\bm \alpha\in \mathcal{A}(c)}P(\Phi_{\gamma}=1) \to 1,
\end{align*}
where $\Phi_{\gamma}\equiv I[T_{\MAX}-2\log(N)+\log\{\log(N)\}\ge
q_{\gamma}]$, $q_\gamma \equiv-\log (\pi)-2 \log \left\{\log (1-\gamma)^{-1}\right\}$, $c=4$ and
\begin{align}\label{bound}
\mathcal{A}(c)\equiv\left\{\bm\alpha =(\alpha'_1,\cdots,\alpha'_N)\in
  \mathcal{R}^N: \max_{1\le i\le N} \alpha_i/\sqrt{\sigma_i^0}\ge
  c\sqrt{\log(N)/{T}}\right\}.
\end{align}
\end{thm}

Theorem \ref{th5} demonstrates that if at least one alpha is significantly large, the suggested max-type test maintains consistency. As showed in Theorem 3 in \cite{feng2022high}, the rate $\sqrt{\log (N)/T}$ is minimax rate optimal, which means any $\gamma$-level test is unable to detect the alternative hypothesis in probability uniformly over 
$$\left\{\boldsymbol{\alpha} \in \mathcal{R}^N: \sum_{i=1}^N I\left(\alpha_i \neq 0\right)=k_N\right\}\bigcap\{\boldsymbol{\alpha} \in \left.\mathcal{R}^N: \max _{1 \leq i \leq N}\left|\alpha_i\right|>c_4 \sqrt{\log (N) / T}\right\}$$
 where $k_N=O(N^r),r<1/4$. Hence the order of the lower bound $\max _{1 \leq i \leq N} \alpha_i / \sqrt{\sigma_i^0} \geq c \sqrt{\log (N) / {T}}$ cannot be improved.

\subsection{Cauchy combination test}
In real-world scenarios, it's often unclear whether the intercept vector is sparse or dense. To accommodate different alternative behaviors, we integrate the Sum-test and Max-test. The crucial point is that these two test statistics are asymptotically independent under certain mild conditions and the null hypothesis.

\begin{thm}\label{th6}
    Under Conditions (C1)-(C9) and assuming $M=\lceil\min(N,T)^{1/8}\rceil$, under $H_0$, we have as $\min(N,T)\to\infty$, 
    \begin{equation*}
    	\mathbb{P}_{H_0}\left( \frac{T_{\SUM}-\hat{\mu}_T}{\hat{\sigma}_T} \leq x, T_{\MAX}-2\log(N)+\log\{\log (N)\} \leq y\right) \to \Phi(x)F(y).
    \end{equation*}
\end{thm}

According to Theorem \ref{th6}, we suggest combining the corresponding $p$-values by using Cauchy Combination Method \citep{liu2020}, to wit,
\begin{equation*}
    p_{CC}=1-G[0.5\tan\{(0.5-p_{\MAX})\pi\}+0.5\tan\{(0.5-p_{\SUM})\pi\}],
\end{equation*}
where $G(\cdot)$ is the CDF of the the standard Cauchy distribution. If the final $p$-value is less than some pre-specified significant level $\gamma\in(0,1)$, then we reject $H_0$.

Next, we analyze the power of the adaptive testing procedure. We consider the following sequence of alternative hypotheses, to wit,
\begin{equation}\label{alternative}
     H_{1,NT}: ||\bma||_0=o[N/\log^2\{\log(N)\}]~~\text{and}~~||\bma||_2=O\{T^{-1/2}\tr^{1/4}(\O_T^2)\}.
\end{equation}
In fact, the asymptotic independence between the maximum-type and sum-type statistics still hold under the hypotheses given in \eqref{alternative}.
\begin{thm}\label{th7}
    Under the same condition as Theorem \ref{th6}, we have as $(N,T)\to\infty$, under $H_{1,NT}$,
    \begin{align*}
    	&\mathbb{P}\left( \frac{T_{\SUM}-\hat{\mu}_T}{\hat{\sigma}_T} \leq x, T_{\MAX}-2\log(N)+\log\{\log (N)\} \leq y\right)\\
    	&\qquad \to\mathbb{P}\left( \frac{T_{\SUM}-\hat{\mu}_T}{\hat{\sigma}_T} \leq x\right)\mathbb{P}\left(T_{\MAX}-2\log(N)+\log\{\log (N)\} \leq y\right).
    \end{align*}
\end{thm}

\cite{li2023} show that the power of Cauchy combination-based test would be powerful than that of the test based on $\min\{p_{\MAX},p_{\SUM}\}$ (referred to as the minimal $p$-value combination), say $\beta_{M\wedge S,  \gamma}=P(\min\{{  p}_{\MAX},{  p}_{\SUM}\}\leq 1-\sqrt{1- \gamma})$. Obviously,
    \begin{align}\label{power_H1}
   \beta_{CC, \gamma}\ge \beta_{M\wedge S,  \gamma} &\ge P(\min\{{  p}_{\MAX},{  p}_{\SUM}\}\leq  \gamma/2)\nonumber\\
    &= \beta_{{\MAX}, \gamma/2}+\beta_{{\SUM}, \gamma/2}-P({  p}_{\MAX}\leq  \gamma/2, {  p}_{\SUM}\leq  \gamma/2)\nonumber\\
    &\ge \max\{\beta_{{\MAX}, \gamma/2},\beta_{{\SUM}, \gamma/2}\}.
    \end{align}
On the other hand, under $H_{1}$ in \eqref{alternative}, we have
    \begin{align}\label{power_H1np}
   \beta_{CC, \gamma}\ge  \beta_{{\MAX}\wedge S,  \gamma} \ge \beta_{{\MAX}, \gamma/2}+\beta_{{\SUM}, \gamma/2}-\beta_{{\MAX}, \gamma/2}\beta_{{\SUM}, \gamma/2}+o(1),
    \end{align}
due to the asymptotic independence entailed by Theorem \ref{th7}. For a small $ \gamma$, the difference between $\beta_{{\MAX}, \gamma}$ and $\beta_{{\MAX}, \gamma/2}$ should be small, and the same fact applies to $\beta_{S, \gamma}$. Consequently, by \eqref{power_H1}--\eqref{power_H1np}, the power of the Cauchy combination test would be no smaller than or even significantly larger than that of either max-type or sum-type test.

Next, we present some specific cases to demonstrate the power performance of each test. For simplicity, we assume that $\sigma_i^0=1,i=1,\cdots,N$ and $\tr(\O_T^2)=c_ON$. Additionally, $\alpha_i=a,i=1,\cdots,s$ and $\alpha_i=0$ for $i=s+1,\cdots,N$. Let $\xi_ \gamma:=\lim\beta_{\MAX, \gamma}$, $\eta_ \gamma:=\lim\beta_{\SUM, \gamma}$. Table \ref{tabpower3} summarizes the asymptotic powers of these tests under different configurations of the change magnitude $a$ and the sparsity level $s$. Here we provide a detailed analysis as follows:
\begin{itemize}
\item[(I)] $Ta^2/\log N\to\infty$. Then, $\xi_ \gamma=1$ by Theorem \ref{th5} and  $\beta_{CC,\gamma}=1$ by (\ref{power_H1}).
\item[(II)] $a=C\sqrt{(\log N)/T}$. If $C\ge 4$, $\xi_ \gamma=1$ by Theorem \ref{th5}.  Next, we consider three case for $s$. 
\begin{itemize}
\item[(i)] $s=o(N^{1/2}/\log N)$. Now $T\bm \alpha^T\bm \alpha/\sqrt{2\tr(\O_T^2)}\to 0$, so $\eta_ \gamma=\gamma$ by (\ref{sumpower}). And $\beta_{CC,\gamma}\ge\xi_{ \gamma/2}+\eta_{ \gamma/2}-\xi_{ \gamma/2}\eta_{ \gamma/2}$ by (\ref{power_H1np});
\item[(ii)] $s\sim N^{1/2}/\log N$; Both $T_{\SUM}$ and $T_{\MAX}$ possess a certain level of power. And $\beta_{CC,\gamma} \ge\xi_{ \gamma/2}+\eta_{ \gamma/2}-\xi_{ \gamma/2}\eta_{ \gamma/2}$ could yield higher power than either sum-type or max-type procedure.
\item[(iii)] $N^{-1/2}s\log N\to \infty$. Now $T\bm \alpha^T\bm \alpha/\sqrt{2\tr(\O_T^2)}\to \infty$, so $\eta_ \gamma=1$ by (\ref{sumpower}) and  $\beta_{CC,\gamma}=1$ by (\ref{power_H1}).
\end{itemize}
\item[(III)] $a=o\{\sqrt{(\log N)/T}\}$. The max-type test would not have any power, i.e. $\xi_{\gamma}=\gamma$. Specially, if $N^{-1/2}Tsa^2\to\infty$, we have $T\bm \alpha^T\bm \alpha/\sqrt{2\tr(\O_T^2)}\to \infty$, so, similar to II-(iii), $\eta_ \gamma=1$  and  $\beta_{CC,\gamma}=1$.
\end{itemize}

\begin{table}
\begin{threeparttable}[ht]
\caption{Asymptotic powers of the max-type and sum-type tests, and the Cauchy combination test, respectively.}
\centering
     \tabcolsep 1pt
\begin{tabular}{cccccc}
\toprule
Case & $a$ & $s$ & $\lim\beta_{\MAX, \gamma}$ & $\lim\beta_{\SUM, \gamma}$ & $\lim\beta_{CC, \gamma}$\\
\midrule
I & $Ta^2/\log N\to\infty$ & any & $1$ & $\eta_ \gamma$ & $1$\\
\smallskip\\
II--(i) & $a\sim\sqrt{(\log N)/T}$ & $s=o(N^{1/2}/\log N)$ & $\xi_ \gamma$ & $ \gamma$ & $\ge\xi_{ \gamma/2}+ \gamma/2(1-\xi_{ \gamma/2})$\\
II--(ii) & &$s\sim N^{1/2}/\log N$ & $\xi_ \gamma$ & $\eta_ \gamma$ & $\ge\xi_{ \gamma/2}+\eta_{ \gamma/2}-\xi_{ \gamma/2}\eta_{ \gamma/2}$\\
II--(iii) & &$N^{-1/2}s\log N\to \infty$ & $\xi_ \gamma$ & $1$ & $1$\\
II' & $a=C\sqrt{(\log N)/T}$\tnote{*} & any & $1$ & $\eta_ \gamma$ & $1$\\
\smallskip\\
III & $a=o\{\sqrt{(\log N)/T}\}$ & any & $ \gamma$ & $\eta_ \gamma$ & $\ge\eta_{ \gamma/2}+ \gamma/2(1-\eta_{ \gamma/2})$\\
III' & $a=o\{\sqrt{(\log N)/T}\}$ & $N^{-1/2}Tsa^2\to\infty$ & $ \gamma$ & $1$ & $1$\\
\bottomrule
\end{tabular}
\label{tabpower3}
\begin{tablenotes}
    \item $^*$ $C>0$ is a large enough constant
\end{tablenotes}
\end{threeparttable}
\end{table}

\section{Simulation}
In this section, we compare our proposed methods with three test procedures for the independent case: the sum-type test procedure proposed by \citet{Pesaran2017TestingFA} (abbreviated as PY), the max-type test procedure proposed by \cite{feng2022high} (abbreviated as FLM) and the combination test procedure proposed by \cite{feng2022high} (abbreviated as COM).

This simulation is designed to mimic the commonly used Fama-French three-factor model,
where the factors $\f_t$ have strong serial correlation and heterogeneous variance.
Specifically, we consider the example studied
in \citet{feng2022high}. The response $Y_{it}$ are generated according to the LFPM in (\ref{mod}) with $p=3$:
\begin{align*}
Y_{it}=\alpha_i+\sum_{j=1}^p \beta_{ij} f_{tj} +\varepsilon_{it},
\end{align*}
where the three factors, $f_{t1}$, $f_{t2}$ and $f_{t3}$, are the
Fama-French three factors (Market factor, SMB, HML).
We generate each factor from an autoregressive conditional
heteroskedasticity process and the GARCH(1,1) model respectively.
Specifically
\begin{align*}
f_{t1}=&0.53+0.06 f_{t-1, 1}+ h_{t1}^{1/2}\zeta_{t1}, ~\mbox{Market factor},\\
f_{t3}=&0.19+0.19 f_{t-1, 2}+ h_{t2}^{1/2}\zeta_{t2}, ~\mbox{SMB factor},\\
f_{t3}=&0.19+0.05 f_{t-1, 3}+ h_{t3}^{1/2}\zeta_{t3}, ~\mbox{HML factor},
\end{align*}
where for  $j=1,2,3$, $\zeta_{tj}$ are generated independently from a standard normal distribution,
and the variance term $h_{tj}$ is generated as follows:
\begin{align*}
h_{t1}&=0.89+0.85 h_{t-1, 1}+ 0.11 \zeta_{t-1,1}^2, ~\mbox{Market factor},\\
h_{t2}&=0.62+0.74 h_{t-1, 2}+ 0.19\zeta_{t-1,2}^2, ~\mbox{SMB},\\
h_{t3}&=0.80+0.76 h_{t-1, 3}+ 0.15 \zeta_{t-1,3}^2, ~\mbox{HML}.
\end{align*}
The above process is simulated over the periods $t\in\{-49, \cdots, 0,
1, \cdots, T\}$ with the initial values
$f_{-50,j}=0$ and $h_{-50,j}=1$ for any $j\in\{1,2,3\}$, and the
generated data that belong to the periods $\{1, \cdots, T\}$ is
extracted as the simulation data.

The three groups of coefficients corresponding to the
three factors, $\beta_{i1}$, $\beta_{i2}$ and $\beta_{i3}$,
are generated independently from $U(0.2,2)$, $U(-1,1.5)$ and $U(-1.5,1.5)$, respectively.
Then, we set $\bm\alpha=\bm 0$ under the null hypothesis.

The error $\bm \varepsilon_t$ are generated from $\bm \varepsilon_t=\sum_{h=0}^M A_h \z_{t-h}$, where $A_0,\cdots, A_M$ are $N\times N$ matrices which determine the autocovariance structure. And $\z_t=\bms^{1/2}\bm \zeta_t$ where $\bm \zeta_t=(\zeta_{t1},\cdots,\zeta_{tN})$ are independently and identically distributed from $N(0,1)$ and $t(3)$. Here we consider $A_0=\I_N$,
\begin{align*}
A_h(i,j)&=\left\{
\begin{array}{cl}
h^{-1}\phi_1& \text{if}~ i=i,h\ge 1\\
\frac{\phi_1}{h(i-j)^2} & \text{if}~ 1\le |i-j|\le \omega N,\\
0& \text{if}~|i-j|>\omega N.
\end{array}
\right. \text{if}~~ h=1,2.\\
\bms(i,j)&=\left\{
\begin{array}{cl}
1& \text{if}~ i=j,\\
\phi_2& \text{if}~ i=i,h\ge 1\\
\frac{\phi_2}{(i-j)^2} & \text{if}~ 1\le |i-j|\le \omega N,\\
0& \text{if}~|i-j|>\omega N.
\end{array}
\right.
\end{align*}
For $h>2$, we simply set $A_h=e^{-2h}$. We set $T=400,800$, $N=250,500$ and $(\omega,\phi_1,\phi_2)=(0.9,0.6,0.4)$. And we consider two case for $M$: (i) independent case, $M=0$; (ii) 2-dependent case, $M=2$; (iii) $\infty$-dependent case, $M=T-1$. Table \ref{tab1} presents the empirical dimensions of each test. In the case of independence, all tests effectively manage the empirical dimensions in most scenarios. However, in the dependent case, the empirical dimensions of the PY, FLM, and COM tests exceed the nominal level, which is expected as these tests do not account for the error terms’ dependent structure. On the other hand, our proposed test procedures - SUM, MAX, CC - demonstrate strong performance in terms of empirical dimensions, particularly when dealing with large sample sizes.

\begin{figure}
	\centering
	\includegraphics[width=0.8\textwidth]{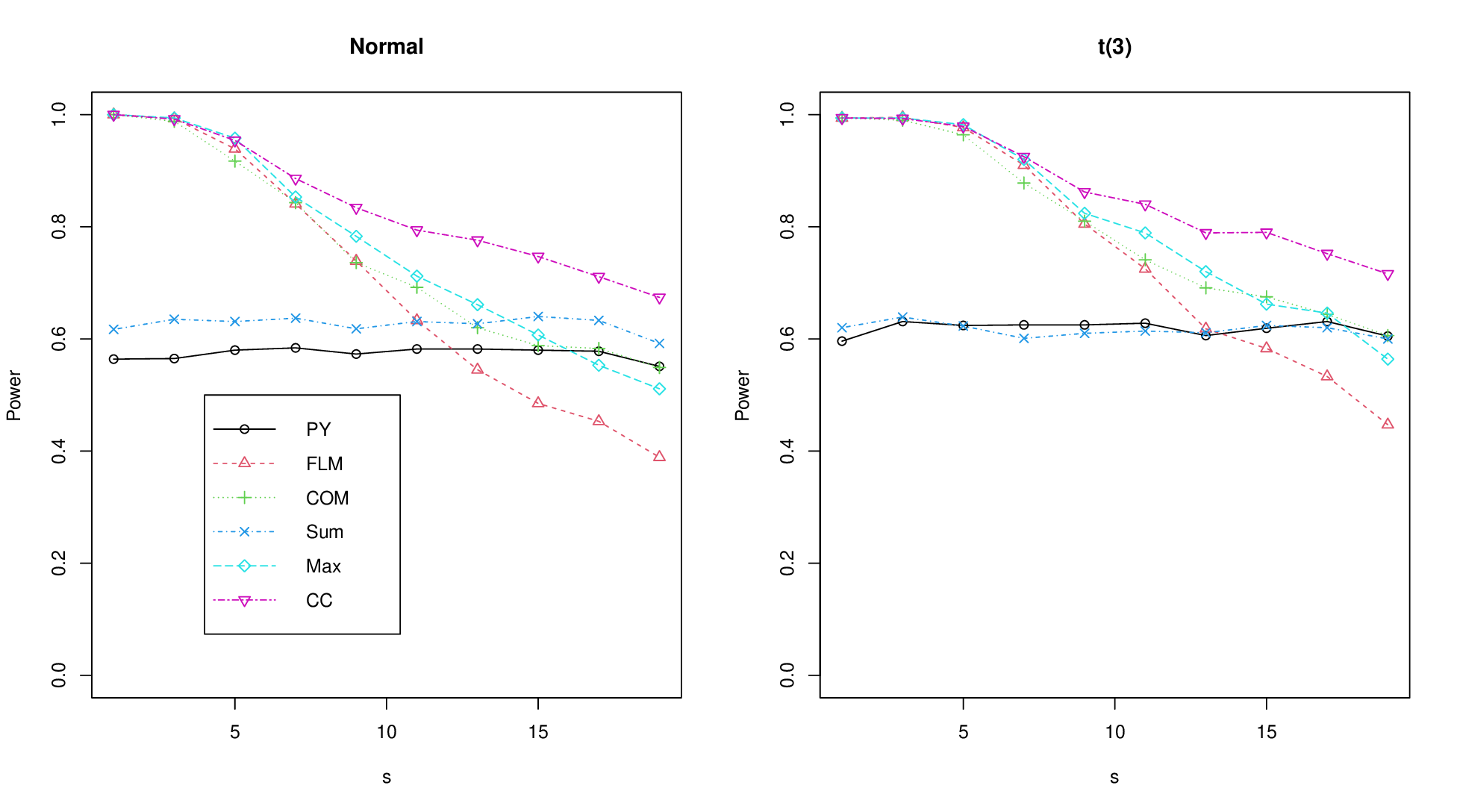}
	\caption{Power comparison of different methods with different sparsity under independent case $M=0$ with $(T,N)=(400,500)$.}
	\label{fig1}
\end{figure}

\begin{figure}
	\centering
	\includegraphics[width=0.8\textwidth]{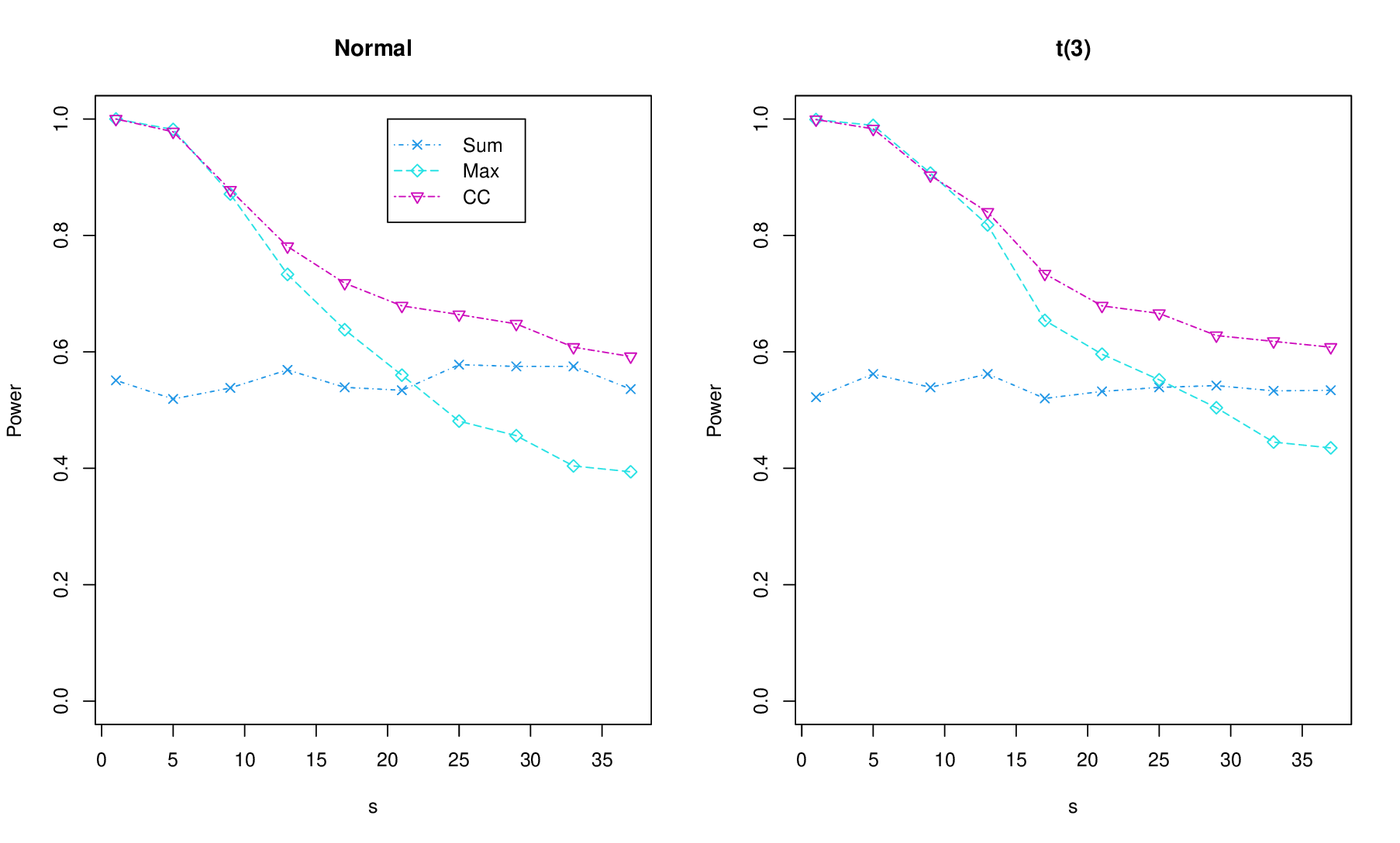}
	\caption{Power comparison of different methods with different sparsity under 2-dependent case $M=2$ with $(T,N)=(400,500)$.}
	\label{fig2}
\end{figure}

\begin{figure}
	\centering
	\includegraphics[width=0.8\textwidth]{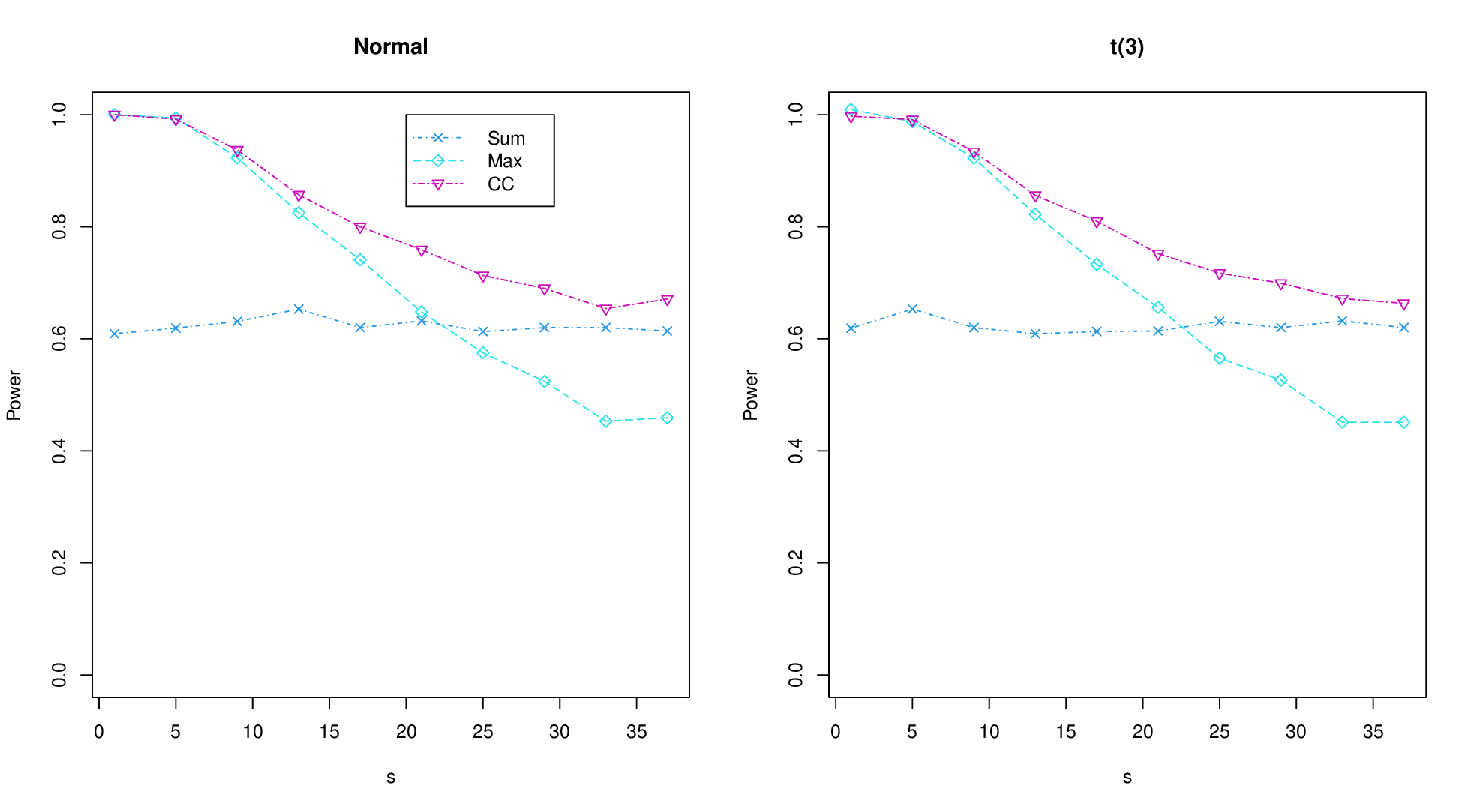}
	\caption{Power comparison of different methods with different sparsity under $\infty$-dependent case $M=\infty$ with $(T,N)=(400,500)$.}
	\label{figm}
\end{figure}

For power comparison, we generate $\alpha_i,i\in S$ independently from $U(0,\sqrt{s^{-1}c_M\log(N)/T})$ where each element in $S$ is uniformly and randomly drawn from $\{1,\cdots,N\}$ and $|S|=s$. And $c_0=12,c_2=80,c_{\infty}=90$ for a fair power comparison. Figure \ref{fig1} illustrates the power curves of each test under varying levels of sparsity in the independent case where $M=0$. It was observed that the SUM and MAX tests perform comparably to the PY and FLM tests, respectively. The CC test surpasses the COM test in all scenarios due to the superior efficiency of the Cauchy combination test procedure over the minimal value procedure \citep{li2023}. As the PY, FLM, and COM tests are unable to control the empirical sizes in the dependent case, we only consider our proposed three test procedures for $M=2,\infty$. Figure \ref{fig2}-\ref{figm} depicts the power curves of the SUM, MAX, and CC tests under varying sparsity. The SUM test outperforms the MAX test when signals are very dense, while the MAX test excels when signals are very sparse. These observations align with intuitive expectations and are corroborated by several other studies, such as \cite{Feng2022AsymptoticIO}. Furthermore, our proposed CC test performs similarly to the SUM test under very dense alternatives and the MAX test under very sparse alternatives. When the signals are neither very dense nor sparse, the CC test exhibits the best performance, underscoring the superiority of the Cauchy combination test procedure.

\begin{table}[ht]
	\centering
   \renewcommand{\arraystretch}{1}
   	\caption{Empirical sizes of different tests.}
	\begin{tabular}{c|cccc|cccc|cccc}\hline \hline
 &\multicolumn{4}{c}{$M=0$}&\multicolumn{4}{c}{$M=2$}&\multicolumn{4}{c}{$M=\infty$}\\ \hline
 &\multicolumn{2}{c}{$T=400$} &\multicolumn{2}{c}{$T=800$}&\multicolumn{2}{c}{$T=400$} &\multicolumn{2}{c}{$T=800$}&\multicolumn{2}{c}{$T=400$}&\multicolumn{2}{c}{$T=800$}\\ \hline
 $N$&250&500&250&500&250&500&250&500&250&500&250&500\\ \hline
& \multicolumn{12}{c}{$\zeta_{ti}\sim N(0,1)$}\\ \hline
PY&6.1&4.6&5.8&4.6&100&100&100&100&100&100&100&100\\
FLM&4.8&5.8&3.6&5.4&97.3&99.6&95.8&99.5&100&100&100&100\\
COM&6.1&5.2&4.9&5.2&100&100&100&100&100&100&100&100\\
SUM&6.2&5.1&5.1&5.7&5.7&5.5&5.7&5.8&5.6&6.1&4.9&5.8\\
MAX&7.3&6.7&5.2&6.7&6.7&6.6&4.1&6.2&6.3&6.5&5.4&5.9\\
CC&6.7&6.0&5.6&6.2&6.6&5.8&6.0&5.7&5.4&6.2&5.7&5.0\\ \hline
& \multicolumn{12}{c}{$\zeta_{ti}\sim t(3)$}\\ \hline
PY&6.7&5.3&5.8&5.2&100&100&100&100&100&100&100&100\\
FLM&4.2&4.5&3.4&4.8&96.9&99.8&95.5&99.7&100&100&100&100\\
COM&5&4.4&4.8&5.3&100&100&100&100&100&100&100&100\\
SUM&6.4&5.7&5.8&6.3&5.6&6.3&5.7&5.4&6.0&5.8&5.4&5.3\\
MAX&5.2&5.8&2.9&6.1&5.9&6.4&3.1&5.4&6.5&5.9&4.7&5.1\\
CC&6.7&6.5&4.8&6.1&6.1&5.5&5.2&4.6&5.7&5.5&5.1&4.8\\ \hline \hline
	\end{tabular}
	\label{tab1}
\end{table}

\begin{figure}[ht]
	\centering
	\includegraphics[width=0.8\textwidth]{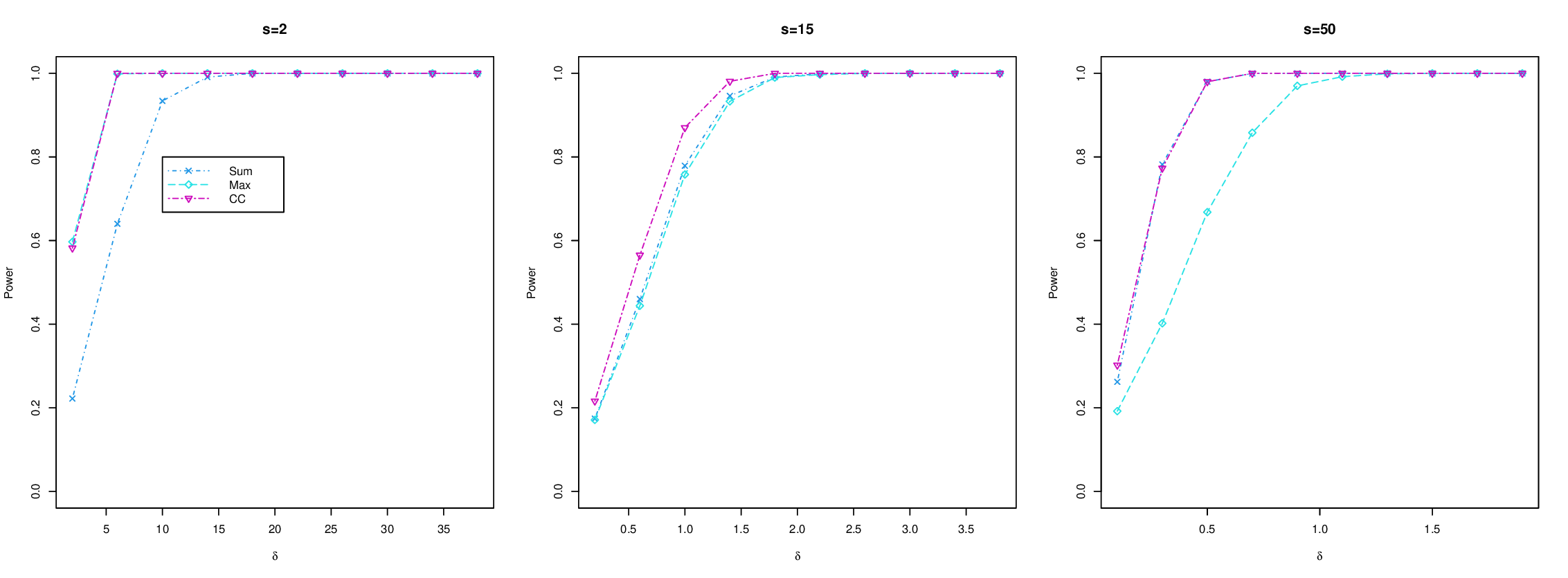}
	\caption{Power comparison of different methods with different signal strength under independent case $M=0$ with $(T,N)=(400,500)$.}
	\label{fig3}
\end{figure}

\begin{figure}[ht]
	\centering
	\includegraphics[width=0.8\textwidth]{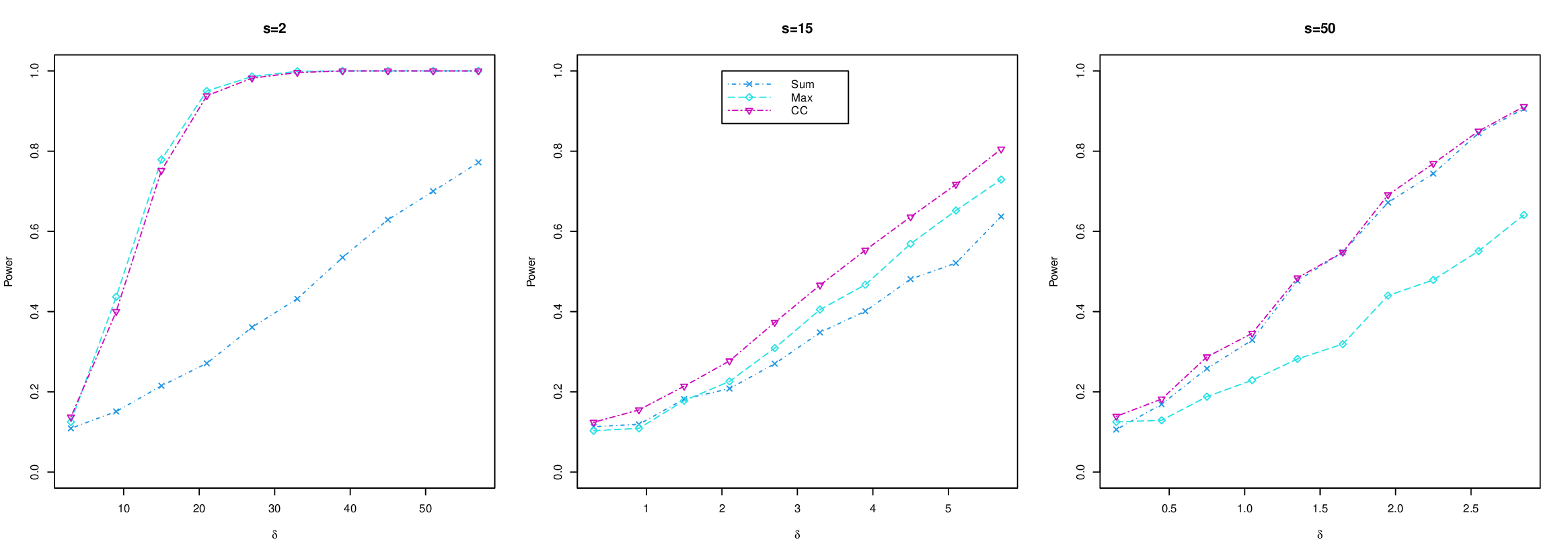}
	\caption{Power comparison of different methods with different signal strength under 2-dependent case $M=2$ with $(T,N)=(400,500)$.}
	\label{fig4}
\end{figure}

\begin{figure}[ht]
	\centering
	\includegraphics[width=0.8\textwidth]{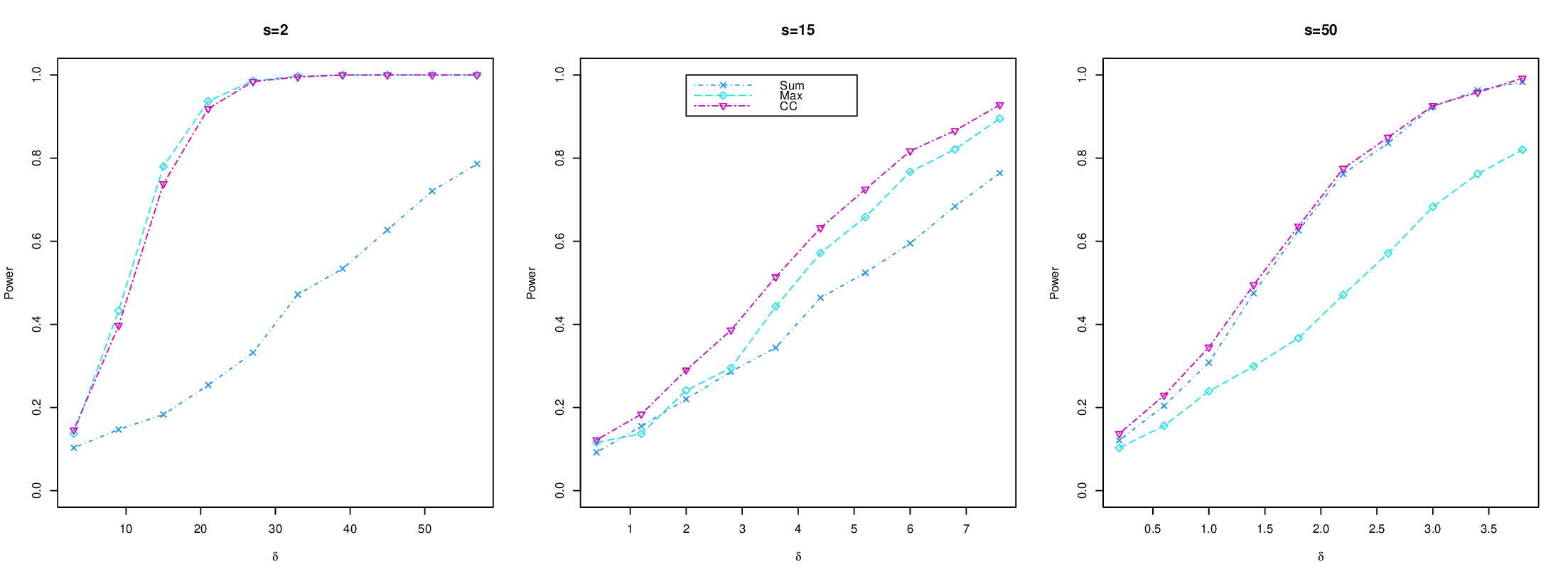}
	\caption{Power comparison of different methods with different signal strength under $\infty$-dependent case $M=\infty$ with $(T,N)=(400,500)$.}
	\label{fig5}
\end{figure}

Continuing with the same settings, we examine the power performance of the tests involved across varying signal strengths. For the alternative hypothesis, we contemplate three scenarios for the sparsity level $s=2,15,50$. The nonzero $\alpha_i$s are drawn from $U(0,\sqrt{\delta \log N/T})$. Here we only consider $\zeta_{ti}\sim N(0,1)$. The outcomes would be analogous if we consider $\zeta_{ti}\sim t(3)$. Figures \ref{fig3}-\ref{fig5} depict the power curves of our proposed three methods for $M=0,2,\infty$, respectively. We observe that the power of each test increases as the signal strength escalates, demonstrating the consistency of each test. Moreover, we find that the Cauchy combination tests outperform the other methods, particularly when the sparsity level is moderate.

\section{Real Data Application}

In this study, we focus on the stocks listed in the S\&P 500 index, a globally recognized benchmark for U.S. equities. We have compiled weekly returns for all the securities in the S\&P 500 index from January 14, 2005, to November 24, 2023, excluding any missing values. This gives us a total of $T=888$ observations over this extensive period. As the composition of the index changes over time, we have limited our analysis to $N=393$ securities that were consistently part of the S\&P 500 index throughout the entire period. The time series data for the risk-free rate of return and other factors were sourced from Ken French's data library web page.
We have chosen the one-month U.S. treasury bill rate as the risk-free rate, denoted as $r_{ft}$. As a proxy for the market return $r_{mt}$, we used the value-weighted return on all stocks listed on the NYSE, AMEX, and NASDAQ, as provided by CRSP. We calculated the Small Minus Big (SMB) factor, $\text{SMB}_t$, as the average return on three small portfolios minus the average return on three big portfolios. Similarly, the High Minus Low (HML) factor, $\text{HML}_t$, is calculated as the average return on two value portfolios minus the average return on two growth portfolios. Both these factors are based on the stocks listed on the NYSE, AMEX, and NASDAQ.

We use the Fama-French three-factor model to describe the above panel data.
Recall that the Fama-French three-factor model is
\begin{align}\label{ff3}Y_{it}=r_{it}-r_{ft}=\alpha_i+\beta_{i1}
  (r_{mt}-r_{ft})+\beta_{i2}SMB_t+\beta_{i3} HML_t+\epsilon_{it}, \end{align}
for each $i\in\{1,\cdots,N\}$ and
$t\in\{\tau,\cdots,\tau+T'-1\}$, where $r_{mt}-r_{ft}$ is also referred
to as the market factor. We are interested in testing
\begin{align}\label{fs}
H_0: \alpha_1=\dots\alpha_N=0, \text{versus}~H_1:
  \exists~i\in\{1,\cdots,N\}\text{ s.t. } \alpha_i\neq 0.
\end{align}

\begin{figure}
	\centering
	\includegraphics[width=0.6\textwidth]{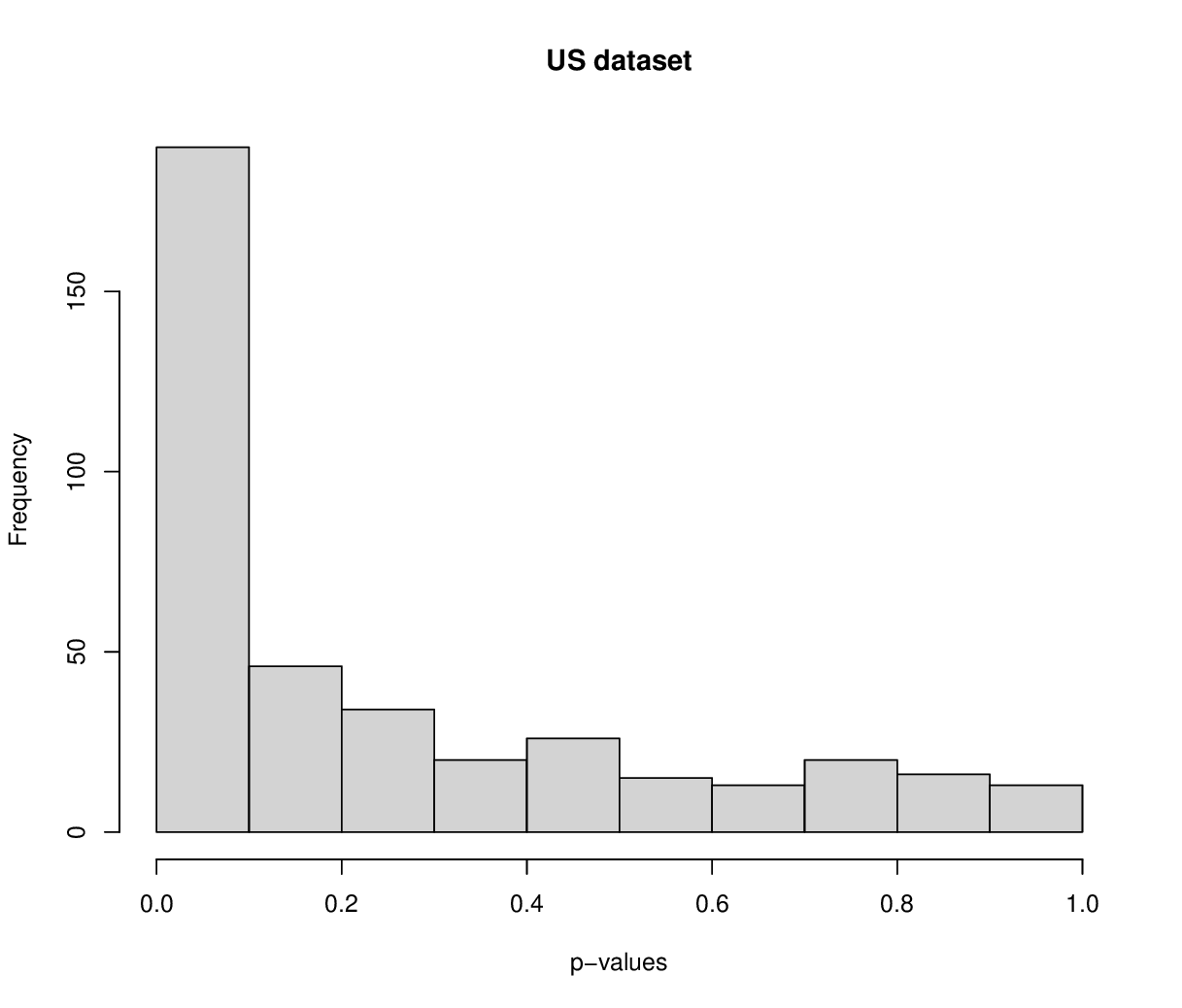}
	\caption{Histogram of $p$-values of the Box-Pierce of residuals of Fama-French three factor model for U.S.'s datasets.}
	\label{figbox}
\end{figure}

Initially, we examine the presence of time series dependence in the residual sequences of the dataset. We apply the Box-Pierce test, a conventional autocorrelation test, to the residuals of each security under the Fama-French three-factor model with the total sample sizes. The histogram of p-values from the Box-Pierce test for the U.S. datasets is shown in Figure \ref{figbox}. We notice a significant number of $p$-values below 0.1, leading us to believe that some residual sequences may exhibit autocorrelation. Furthermore, we employ the max-type high-dimensional white noise test procedure, as described in \cite{feng2022testingwhite}, to verify if the residuals are white noise. The $p$-value of this test is 2e-11, indicating some time-dependence among $\epsilon_{it}$. Consequently, test procedures based on the i.i.d assumption may not perform well and could yield inaccurate results. To analyze this dataset, we apply three test procedures proposed in this paper - SUM, MAX, and CC, which are designed to handle such time-dependence.

\begin{figure}
	\centering
	\includegraphics[width=\textwidth]{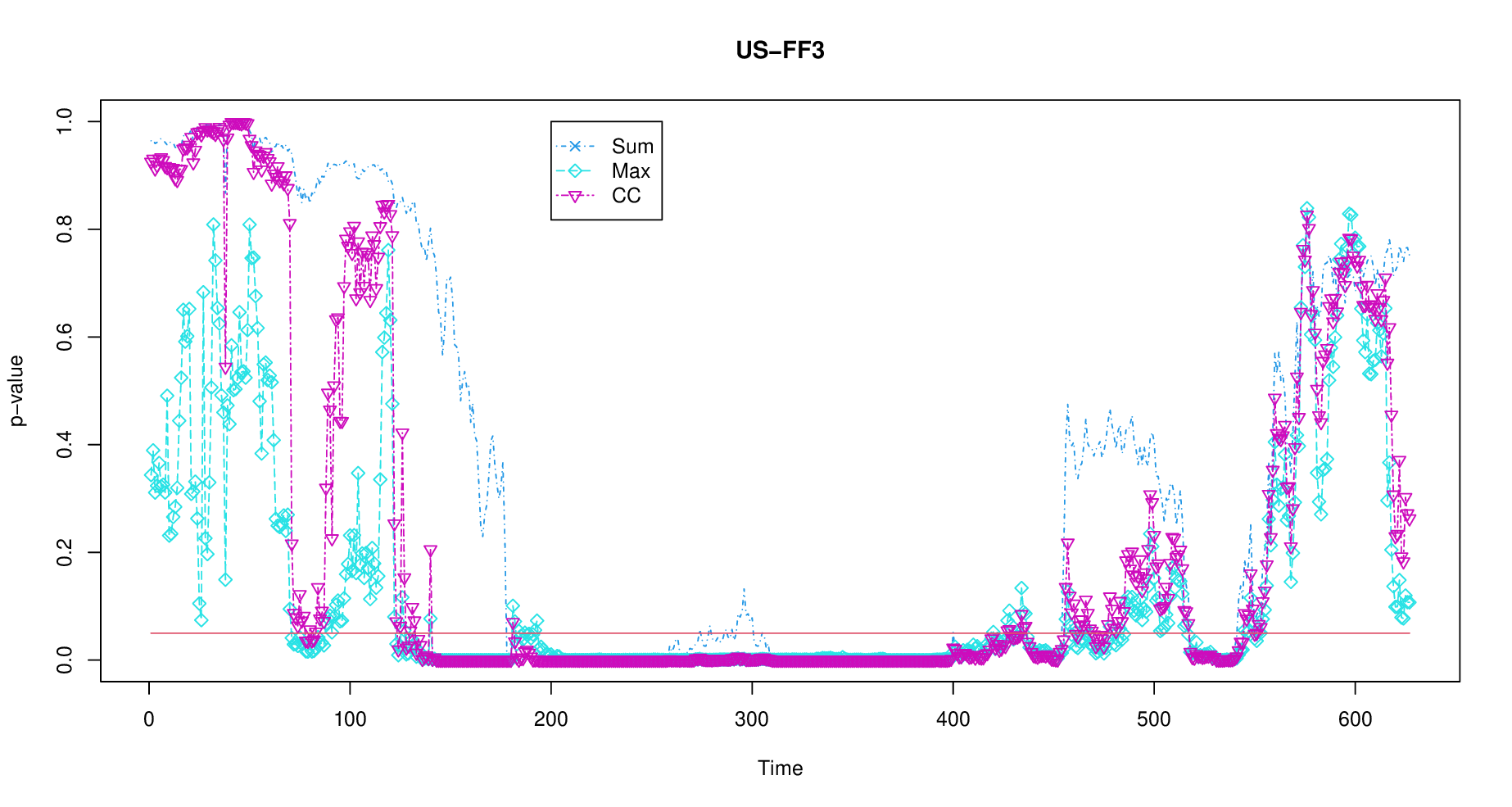}
	\caption{$p$-values of alpha tests for U.S.'s datasets with Fama-French three factor model.}
	\label{figus}
\end{figure}

We next examine the hypothesis (\ref{fs}) using a rolling window procedure with a window length of $T=260$ (equivalent to 5 years). The p-values for each alpha test procedure over the time period are depicted in Figure \ref{figus}. Our observations reveal that for the initial seven years and the final six years, nearly all tests do not reject the null hypothesis. This suggests that the three observed factors can effectively account for the stock returns. However, during the five-year period in the middle, the CC test frequently rejects the null hypothesis, leading us to infer that the markets were inefficient during these periods. Interestingly, there are instances over the long time period where the MAX or SUM test does not reject the null hypothesis. This demonstrates that our CC test procedure is more robust and efficient under various sparsity alternatives.

\section{Conclusion}
Initially, we introduced the sum-type and max-type tests for a high-dimensional linear pricing model with dependent observations. These tests perform well against both dense and sparse alternatives. To broaden the application, we introduced a Cauchy combination test that combines the sum-type and max-type tests. Both simulation studies and real data applications have shown the necessity of considering the time-dependence of the error term. Ignoring this could lead to significant size distortion in methods based on the i.i.d assumption.

Furthermore, given the importance of allowing for time-variation in the risk-exposure coefficients under LFPMs, our future work will focus on developing sum-type, max-type, and combination test procedures that allow for such time-variation. We plan to use a more complex and powerful technical framework to establish the asymptotic theory, as opposed to the current one.

\section{Appendix}
\subsection{Proof of Theorems 1-3}
Let $M=\lceil\min(N,T)^{1/8}\rceil$. We can define a M-dependent approximation sequence for  $\{\bm \X_t\}$:
\begin{align*}
    \bm \gamma_t:= \mathbb{E}(\bm \X_t|\z_{t-M},\cdots,\z_t)=\bm \Sigma^{1/2}\sum_{k=0}^M b_k\z_{t-k}.
\end{align*}
By replacing $\X_t$ in $\tilde{T}_{\SUM}$ with $\bm \gamma_t$, we define $T_{\SUM}^{NG}:=\bar{\bm \gamma}_T^\top \bar{\bm \gamma}_T$. Let $a_{h,M}:=\sum_{k=0}^{M-h}b_kb_{k+h}$, then $\Gamma_{h,M}:= \mathbb{E}(\bm\gamma_t \bm \gamma_{t+h}^\top)=a_{h,M}\bm \Sigma$, and $\O_{T,M}:= T\mathbb{E}(\bar{\bm \gamma}_T\bar{\bm \gamma}_T^\top)=\sum_{h\in\mathcal{M}}(1-\frac{|h|}{T})\Gamma_{h,M}$, where $\mathcal{M}=\{0,\pm1,\dots,\pm M\}$. 

Let $\{\bm \delta_t\}$ be a Gaussian sequence which is independent of $\{\bm \gamma_t\}$ and preserves the same auto-covariance structure as $\{\bm \gamma_t\}$, i.e. $\mathbb{E}(\bm \delta_t \bm \delta_{t+h}^\top)=\Gamma_{h,M}$. Similarly, define $T_{\SUM}^{G}:=\bar{\bm \delta}_T^\top \bar{\bm \delta}_T$.

Firstly, we establish some useful results in the following lemmas.
\begin{lemma}\label{key_inequality.1}
Suppose $z_{it}$ are independent and identically distributed random variables satisfying $\mathbb{E}(z_{it})=0$, $\mathbb{E}(z_{it}^2)=1$ and $\mathbb{E}(z_{it}^4)=\mu_4<\infty$. Let $W=\sum_{t=0}^\infty a_t\z_t$, where $\z_t=(z_{1t},\dots,z_{Nt})^\top$ and $\sum_{i=0}^\infty |a_i|<\infty$. Then, there exists a positive constant $\tau_1\ge 3$ such that for all $N\times N$ positive semi-definite matrix $B$ and absolute convergent $\{a_i\}_{i\ge 0}$,
\begin{align*}
    \mathbb{E}[(W^\top B W)^2]\le\tau_1[\mathbb{E}(W^\top B W)]^2.
\end{align*}
\end{lemma}
\proof Let $[B]_{ij}$ denote the $(i,j)$-th element of $B$. Since $B$ is positive semi-definite, we can decompose $B$ as $B=D\Lambda D^\top$, where $D$ is an orthogonal matrix and $\Lambda=\diag(\lambda_1,\dots,\lambda_N)$ with $\lambda_1\ge\dots\ge\lambda_N\ge 0$ being the eigenvalues of $B$. Define $\tilde{\z}_t=D^\top \z_t$, then
\begin{align*}
    &\mathbb{E}(\tilde{z}_{it}^2)=\mathbb{E}\left\{\left(\sum_{j=1}^N[D]_{ji}z_{jt}\right)^2\right\}=1,~~\text{and}\\
    &\mathbb{E}(\tilde{z}_{it}^4)=\mathbb{E}\left\{\left(\sum_{j=1}^N[D]_{ji}z_{jt}\right)^4\right\}=\sum_{j=1}^N[D]_{ij}^4(\mu_4-3)+3\le \mu_4.
\end{align*}
Thus,
\begin{align*}
    \mathbb{E}(W^\top B W)=\sum_{r=1}^N[B]_{rr}\mathbb{E}\left\{\left(\sum_{t=0}^\infty a_tz_{rt}\right)^2\right\}=\tr(B)\sum_{t=0}^\infty a_t^2,
\end{align*}
and
\begin{align*}
    \mathbb{E}[(W^\top B W)^2]=&\sum_{1\le r_1,r_2,r_3,r_4\le N}[B]_{r_1r_2}[B]_{r_3r_4}\mathbb{E}\left\{\prod_{l=1}^4\left(\sum_{t=0}^\infty a_tz_{r_l,t}\right)\right\}\\
    =&\sum_{r=1}^N[B]_{rr}^2\mathbb{E}\left\{\left(\sum_{t=0}^\infty a_tz_{it}\right)^4\right\}\\
    &+\sum_{1\le r\ne s\le N}(2[B]_{rs}^2+[B]_{rr}[B]_{ss})\mathbb{E}\left\{\left(\sum_{t=0}^\infty a_tz_{rt}\right)^2 \left(\sum_{t=0}^\infty a_tz_{st}\right)^2\right\}\\
    =&\sum_{r=1}^N[B]_{rr}^2\left(\sum_{t=0}^\infty a_t^4\mu_4+3\sum_{t\ne s\ge 0}a_t^2a_s^2 \right)+\sum_{1\le r\ne s\le N}(2[B]_{rs}^2+[B]_{rr}[B]_{ss})\left(\sum_{t=0}^\infty a_t^2\right)^2\\
    \le &\tr^2(B)\left(\sum_{t=0}^\infty a_t^4\mu_4+3\sum_{t\ne s\ge 0}a_t^2a_s^2 \right)+(2\tr(B^2)+\tr^2(B))\left(\sum_{t=0}^\infty a_t^2\right)^2\\
    \le &(\mu_4+6)\left(\sum_{t=0}^\infty a_t^2\right)^2\tr^2(B).
\end{align*}
Then, taking $\tau_1=\mu_4+6$ completes the proof.

\begin{lemma}\label{key_inequality.2}
    Under Conditions (C1)-(C5), for all $0\le k_1,k_2,k_3,k_4\le T$, there exists a positive constant $\tau_2$ such that 
    \begin{align*}
        &|\mathbb{E}\{(\bm \X_{k_1}^\top\bm \X_{k_2}-\tr(\Gamma_{|k_1-k_2|})) (\bm \X_{k_3}^\top\bm \X_{k_4}-\tr(\Gamma_{|k_3-k_4|}))\} |\le \tau_2\tr(\Omega_T^2),~~\text{and}\\
        &|\mathbb{E}\{(\bm \gamma_{k_1}^\top\bm \gamma_{k_2}-\tr(\Gamma_{|k_1-k_2|,M})) (\bm \gamma_{k_3}^\top\bm \gamma_{k_4}-\tr(\Gamma_{|k_3-k_4|,M}))\}|\le \tau_2\tr(\Omega_{T,M}^2).
    \end{align*}
\end{lemma}
\proof To prove the first result, since
\begin{align*}
    &|\mathbb{E}\{(\bm \X_{k_1}^\top\bm \X_{k_2}-\tr(\Gamma_{|k_1-k_2|})) (\bm \X_{k_3}^\top\bm \X_{k_4}-\tr(\Gamma_{|k_3-k_4|}))\} |\\
    \le& [\mathbb{E}\{(\bm \X_{k_1}^\top\bm \X_{k_2}-\tr(\Gamma_{|k_1-k_2|}))^2\} \mathbb{E}\{(\bm \X_{k_3}^\top\bm \X_{k_4}-\tr(\Gamma_{|k_3-k_4|}))^2\}]^{1/2},
\end{align*}
it suffices to show that for all $h\ge 0$,
\begin{align*}
    \mathbb{E}\{(\bm \X_0^\top\bm \X_h-\tr(\Gamma_h))^2\}\le \tau_2\tr(\Omega_T^2).
\end{align*}
We split $\bm\X_h$ into two independent parts as 
\begin{align*}
    \bm\X_h=\bm\Sigma^{1/2}\sum_{k=0}^\infty b_{k+h}\z_{-k}+\bm\Sigma^{1/2}\sum_{k=0}^{h-1}b_k\z_{h-k}=:\bm\X_{h,(1)}+\bm\X_{h,(2)},
\end{align*}
then 
\begin{align*}
    \mathbb{E}\{(\bm \X_0^\top\bm \X_h-\tr(\Gamma_h))^2\}=\mathbb{E}\{(\bm \X_0^\top\bm \X_{h,(1)}-\tr(\Gamma_h))^2\}+\mathbb{E}\{(\bm \X_0^\top\bm \X_{h,(2)})^2\}.
\end{align*}
By the independence of $\z_{it}$'s, we have
\begin{align*}
    \mathbb{E}\{(\z_r\bm\Sigma\z_s)^2\}&=\mathbb{E}\{\tr(\bm\Sigma \z_s\z_s^\top \bm\Sigma \z_r\z_r^\top)\}=\tr(\bm\Sigma^2),~~\text{and}\\
    \mathbb{E}\{(\z_k^\top\bm\Sigma\z_k-\tr(\bm\Sigma))^2\}&=\mathbb{E}\left[\left\{\sum_{1\le i,j\le N}(z_{ik}z_{jk}-\mathbb{I}_{(i=j)})(\bm\Sigma)_{ij}\right\}^2\right]\\
    &=\sum_{i=1}^N[\bm\Sigma]_{ii}^2(\mu_4-1)+2\sum_{i\ne j}[\bm\Sigma]_{ij}^2\le (\mu_4+1)\tr(\bm\Sigma^2).
\end{align*}
Hence, we have
\begin{align*}
    \mathbb{E}\{(\bm \X_0^\top\bm \X_{h,(2)})^2\}=\left(\sum_{k=0}^{h-1}b_k^2\right)\left(\sum_{k=0}^\infty b_k^2\right)\tr(\bm\Sigma^2),
\end{align*}
and
\begin{align*}
    &\mathbb{E}\{(\bm \X_0^\top\bm \X_{h,(1)}-\tr(\Gamma_h))^2\}\\
    =&\sum_{k=0}^\infty b_k^2b_{k+h}^2\mathbb{E}\{(\z_r\bm\Sigma\z_s)^2\}+\sum_{r\ne s\ge 0}b_r b_{s+h} b_s b_{r+h}\mathbb{E}(\z_r^\top \bm\Sigma \z_s \z_s^\top \bm\Sigma \z_r )+\sum_{r\ne s\ge 0}b_r^2b_{s+h}^2\mathbb{E}(\z_r^\top \bm\Sigma \z_s \z_r^\top \bm\Sigma \z_s )\\
    \le &\sum_{k=0}^\infty b_k^2b_{k+h}^2(\mu_4+1)\tr(\bm\Sigma^2)+ \sum_{r\ne s\ge 0} (b_r b_{s+h} b_s b_{r+h} +b_r^2 b_{s+h}^2) \tr(\bm\Sigma^2)\\
    \le & \left\{\sum_{k=0}^\infty b_k^2b_{k+h}^2(\mu_4+1)+ \left(\sum_{k=0}^\infty b_k^2\right)\left(\sum_{k=0}^\infty b_{k+h}^2\right)+\left(\sum_{k=0}^\infty b_k b_{k+h}\right)^2\right\}\tr(\bm\Sigma^2)\\
    \le &(\mu_4+3)\left(\sum_{k=0}^\infty b_k^2\right)^2\tr(\bm\Sigma^2).
\end{align*}
These results above, together with the fact that $\tr(\O_T^2)=\sum_{h_1,h_2\in\mathcal{T}}(1-\frac{|h_1|}{T})(1-\frac{|h_2|}{T})a_{h_1}a_{h_2}\tr(\bm\Sigma^2)$, imply the desired result. 

Let $b_{k,M}=b_k\mathbb{I}_{(k\le M)}$ and repeat the above proof process, then we get the second result.

\begin{lemma}\label{key_equality.3}
    Let $\{b_n\}_{n\ge 0}$ be a sequence of numbers. If there exists a constant $k>1$ such that $b_n=o(n^{-k})$, then 
    \begin{align*}
        \sum_{m=n}^\infty b_m=o(n^{-k+1}).
    \end{align*}
\end{lemma}
\proof For all $\epsilon>0$, there exists $N_0\in\mathbb{N}$ such that for all $n>N_0$, $n^kb_n<\epsilon$. Then for all $n>N_0$,
\begin{align*}
    n^{k-1}\sum_{m=n}^\infty b_m=\sum_{l=1}^\infty \sum_{m=ln}^{(l+1)n-1}\frac{(nl)^k}{nl^k} b_m\le\sum_{l=1}^\infty \sum_{m=ln}^{(l+1)n-1}\frac{1}{nl^k} m^kb_m\le \epsilon \sum_{l=1}^\infty \frac{1}{l^k}.
\end{align*}

Next, we prove Theorem 1 in the following three parts: first, we prove the asymptotic normality of $T_{\SUM}^{G}$, then we use some Gaussian approximation approach to prove the asymptotic normality of $T_{\SUM}^{NG}$. Finally, we use $T_{\SUM}^{NG}$ to approximate $T_{\SUM}$ and complete the proof of Theorem 1.

\begin{lemma}\label{clt}
    Under Conditions (C1)-(C5), we have
    \begin{align*}
        \frac{T_{\SUM}^{G}-T^{-1}\tr(\O_{T,M})}{\sqrt{2T^{-2}\tr(\O_{T,M}^2)}}\cd N(0,1).
    \end{align*}
\end{lemma}
\proof Let $\lambda_1 \ge\dots\ge \lambda_N$ be the eigenvalues of $\bm\Sigma$. By Condition (C3), we have $\lambda_1^2\le M_0$ and $\tr(\bm\Sigma)\ge NM_1$. Furthermore, using the fact that $\tr^2(\bm\Sigma)\le N\tr(\bm\Sigma^2)$, we have
\begin{align*}
    \frac{M^4\tr(\bm\Sigma^4)}{\tr^2(\bm\Sigma^2)} \le\frac{N^{1/2}\sum_{i=1}^N\lambda_i^4}{(N^{-1}\tr^2(\bm\Sigma))^2} \le\frac{N^{1/2}NM_0^2}{(NM_1^2)^2}=N^{-1/2}M_0^2M_1^{-4}=o(1),
\end{align*}
which implies that
\begin{align*}
    \tr(\Gamma_{a,M}\Gamma_{b,M}\Gamma_{c,M}\Gamma_{d,M})=o\{M^{-4}\tr^2(\O_{T,M}^2)\},\forall a,b,c,d\in\mathcal{M}.
\end{align*}
Hence, recalling that $T_{\SUM}^{G}=\bar{\bm\delta}_T^\top \bar{\bm\delta}_T$ and $\mathbb{E}(T_{\SUM}^{G})=T^{-1}\tr(\O_{T,M})$, and according to equation (10) in the proof of theorem 2.1 in \citet{CLAPR19}, we have
\begin{align*}
    \frac{T_{\SUM}^{G}-T^{-1}\tr(\O_{T,M})}{\sqrt{\var(T_{\SUM}^{G})}}\cd N(0,1).
\end{align*}

Next, we analyze the variance of $T_{\SUM}^{G}$. Let $\tilde{\bm\delta}:= (\bm\delta_1^\top,\dots,\bm\delta_T^\top)^\top\in\mathcal{R}^{TN}$ and $\tilde{\bm\Sigma}:= \sum_{h\in\mathcal{M}}D_h\otimes \Gamma_{h,M}$, where $D_h$ is an $T\times T$ matrix with $[D_h]_{ts}=1$ if $t-s=h$ and all remaining elements equaling zero. Let $\mathbbm{1}\in\mathcal{R}^{T\times T}$ be the all $1$'s matrix and $\I_N$ be the $N\times N$ identity matrix. Here we use a property about the fourth moment of multivariate Gaussian distribution: if $\X\in\mathcal{R}^k$ is Gaussian with $\mathbb{E}(\X)=\bm 0$ and $\var(\X)=\bm\Sigma$, then 
\begin{align*}
    \cov(\X^\top \A\X,\X^\top \B\X)=2\tr(\A\bm\Sigma \B\bm\Sigma).
\end{align*}
Then, we have
\begin{align*}
    \var(T_{\SUM}^{G})=&\var(T^{-2}\tilde{\bm\delta}^\top (\mathbbm{1}\otimes \I_N)\tilde{\bm\delta})\\
    =&2T^{-4}\tr((\mathbbm{1}\otimes \I_N)\tilde{\bm\Sigma}(\mathbbm{1}\otimes \I_N)\tilde{\bm\Sigma})\\
    =&2T^{-4}\sum_{a\in\mathcal{M}}\sum_{b\in\mathcal{M}}\tr(\mathbbm{1}D_a\mathbbm{1}D_b)\tr(\Gamma_{a,M}\Gamma_{b,M})\\
    =&2T^{-4}\sum_{a\in\mathcal{M}}\sum_{b\in\mathcal{M}}(T-|a|)(T-|b|)\tr(\Gamma_{a,M}\Gamma_{b,M})\\
    =&2T^{-2}\tr(\O_{T,M}^2).
\end{align*}
In summary, the proof is finished.

\begin{lemma}\label{Gaussian_approximation}
    Under Conditions (C1)-(C5), we have $T_{\SUM}^{G}-T_{\SUM}^{NG}=o_p\{\sqrt{T^{-2}\tr(\O_{T,M}^2)}\}$.
\end{lemma}
\proof Let $\mathcal{F}_{ts}^{G}:=T^{-2}(\bm\delta_t^\top \bm\delta_s-\tr(\Gamma_{|t-s|,M}))$, then $\sum_{1\le t,s\le T}\mathcal{F}_{ts}^{G}=\bar{\bm\delta}_T^\top \bar{\bm\delta}_T-T^{-1}\tr(\O_{T,M})$. For any $T$, choose $\zeta\in(0,1)$ and $c>0$ such that $w_T=cT^\zeta>M$ and $T=w_Tq_T+r_T$, where $0\le r_T<w_T$. For $1\le t,s\le q_T$, define
\begin{align*}\label{A.1}
    &B_{ts}^{G}:=\sum_{k=(t-1)w_T+1}^{tw_T-M}\sum_{l=(s-1)w_T+1}^{sw_T-M}\mathcal{F}_{kl}^{G}, \\
    &D_{ts}^{G}:=\sum_{k=(t-1)w_T+1}^{tw_T}\sum_{l=(s-1)w_T+1}^{sw_T}\mathcal{F}_{kl}^{G}-B_{ts}^{G},\\
    &F^{G}:=\sum_{(k,l)\in\{1,\dots,T\}^2-\{1,\dots,q_Tw_T\}^2 }\mathcal{F}_{kl}^{G}.\tag{A.1}
\end{align*}
Then, $\sum_{1\le t,s\le T}\mathcal{F}_{ts}^{G}=\sum_{1\le t,s\le q_T}(B_{ts}^{G}+D_{ts}^{G})+F^{G}$.
For non-Gaussian sequence $\{\bm\delta_t\}_{t=1}^T$, we can define $B_{ts}^{NG}$, $D_{ts}^{NG}$ and $F^{NG}$ similarly. We use the unmarked symbol $B_{ts}$, $D_{ts}$ and $F$ when we do not emphasize the difference between Gaussian situation and non-Gaussian situation. Define
\begin{align*}
    S_1:=\sum_{1\le t<s\le q_T}B_{ts},~~ S_2:=\sum_{t=1}^{q_T}B_{tt},~~S_3:=\sum_{1\le t<s\le q_T}D_{ts},~~S_4:=F,
\end{align*}
and $\Delta S_i=S_i^{G}-S_i^{NG}$. Then, $T_{\SUM}^{G}-T_{\SUM}^{NG}=\Delta S_1+\Delta S_2/2+\Delta S_3/2+\Delta S_4/2$. Therefore, it suffices to show that $\Delta S_i=o_p\{\sqrt{T^{-2}\tr(\O_{T,M}^2)}\}$ for $i=1,2,3,4$.

\noindent\textbf{Step 1.} Show that 
\begin{align*}\label{A.2}
    \Delta S_1=o_p\left\{\sqrt{T^{-2}\tr(\O_{T,M}^2)}\right\}.\tag{A.2}
\end{align*}
Let $\mathcal{C}_b^3(\mathcal{R})$ be the class of bounded functions with continuous derivatives up to order $3$. For $f\in\mathcal{C}_b^3(\mathcal{R})$, let $f^{(i)}$ denote the $i$-th derivative of $f$, $i=1,2,3$. Define 
\begin{align*}
    \bm\zeta_t:=\frac{1}{w_T-M}\sum_{k=(t-1)w_T+1}^{tw_T-M}\bm\delta_k~~\text{and}~~\bm\xi_t:=\frac{1}{w_T-M}\sum_{k=(t-1)w_T+1}^{tw_T-M}\bm\gamma_k.
\end{align*}
Then, $\{\bm\zeta_t\}_{t=1}^{q_T}$ and $\{\bm\xi_t\}_{t=1}^{q_T}$ are both independent series with $\mathbb{E}(\bm\zeta_t)=\bm 0=\mathbb{E}(\bm\xi_t)$ and $\var(\bm\zeta_t)=\var(\bm\xi_t)=\frac{1}{w_T-M}\O_{w_T-M}$, where $\O_{w_T-M}=\sum_{h\in\mathcal{M}}(1-\frac{|h|}{w_T-M})\Gamma_{h,M}$. Then for $t<s$, we can rewrite $B_{ts}^{G}$ and $B_{ts}^{NG}$ as 
\begin{align*}
    &B_{ts}^{G}=\sum_{k=(t-1)w_T+1}^{tw_T-M}\sum_{l=(s-1)w_T+1}^{sw_T-M}T^{-2}\bm\delta_k^\top \bm\delta_l=\frac{(w_T-M)^2}{T^2}\bm\zeta_t^\top \bm\zeta_s,~~\text{and}\\
    &B_{ts}^{NG}=\sum_{k=(t-1)w_T+1}^{tw_T-M}\sum_{l=(s-1)w_T+1}^{sw_T-M}T^{-2}\bm\gamma_k^\top \bm\gamma_l=\frac{(w_T-M)^2}{T^2}\bm\xi_t^\top \bm\xi_s.
\end{align*}
Define 
\begin{align*}\label{A.3}
    W(\bm\zeta_1,\dots,\bm\zeta_{q_T}):=\sum_{1\le t<s\le q_T}B_{ts}^{G}/\sqrt{2T^{-2}\tr(\O_{T,M}^2)}=\sum_{1\le t<s\le q_T}u(\bm\zeta_t,\bm\zeta_s),\tag{A.3}
\end{align*}
where $u(\bm\zeta_t,\bm\zeta_s)=\frac{(w_T-M)^2}{T\sqrt{2\tr(\O_{T,M}^2)}}\bm\zeta_t^\top \bm\zeta_s$. Since $\zeta_t$ and $\bm\xi_t$ are linear combinations of $\z$ and $\mathbb{E}(z_{it}^4)<\infty$, using the results of Lemma \ref{key_inequality.1}, it is easy to check that they have the following properties: for $1\le t,s\le q_T$, 
\begin{itemize}
    \item $\forall \bm a\in\mathcal{R}^N,~~ \mathbb{E}(u(\bm\zeta_t,\bm a))=\mathbb{E}(u(\bm a,\bm\zeta_t))=0,~~\mathbb{E}(u(\bm\xi_t,\bm a))=\mathbb{E}(u(\bm a,\bm\xi_t))=0$.

    \item $\forall \bm a,\bm b\in\mathcal{R}^N$,
    \begin{align*}
        &\mathbb{E}(u(\bm a,\bm\xi_t)u(\bm b,\bm\xi_t))=\mathbb{E}(u(\bm a,\bm\zeta_t)u(\bm b,\bm\zeta_t)),\\
        &\mathbb{E}(u(\bm a,\bm\xi_t)u(\bm\xi_t,\bm b))=\mathbb{E}(u(\bm a,\bm\zeta_t)u(\bm\zeta_t,\bm b)),\\
        &\mathbb{E}(u(\bm\xi_t,\bm a)u(\bm b,\bm\xi_t))=\mathbb{E}(u(\bm\zeta_t,\bm a)u(\bm b,\bm\zeta_t)).
    \end{align*}

    \item Let $c_{ts}^2=\mathbb{E}(u(\bm \xi_t,\bm\xi_s)^2)$ and $\rho_0=\max(\tau_1,3)$, then
    \begin{align*}
        \max(\mathbb{E}(u(\bm \xi_t,\bm\xi_s)^4),\mathbb{E}(u(\bm \xi_t,\bm\zeta_s)^4),\mathbb{E}(u(\bm \zeta_t,\bm\zeta_s)^4))\le \rho_0c_{ts}^4<\infty.
    \end{align*}
\end{itemize}

It is known that a sequence of random variables $\{Z_n\}_{n=1}^\infty$ converges weakly to a random variable $Z$ if and only if for every $f\in\mathcal{C}_b^3(\mathbb{R})$, $\mathbb{E}(f(Z_n))\to\mathbb{E}(f(Z))$, see, e.g. \citet{Pollard1984ConvergenceOS}. Hence, it suffices to show that
\begin{align*}
    |\mathbb{E}f(W(\bm\zeta_1,\dots,\bm\zeta_{q_T}))-\mathbb{E}f(W(\bm\xi_1,\dots,\bm\xi_{q_T}))|\to 0,
\end{align*}
for every $f\in \mathcal{C}_b^3(\mathbb{R})$ as $(N,T)\to \infty$. Define 
\begin{align*}\label{A.4}
    &W_k:=W(\bm\zeta_1,\dots,\bm\zeta_{k-1},\bm\xi_k,\dots,\bm\xi_{q_T})~~\text{for}~~ k=1,\dots,q_T+1,~~\text{and}\\
    &W_{k,0}:=\sum_{1\le t<s\le k-1}u_T(\bm\zeta_t,\bm\zeta_s)+\sum_{k+1\le t<s\le q_T}u_T(\bm\xi_t,\bm\xi_s)+\sum_{1\le t\le k-1,k+1\le s\le q_T}u_T(\bm\zeta_t,\bm\xi_s).\tag{A.4}
\end{align*}
Then, 
\begin{align*}
    |\mathbb{E}f(W(\bm\zeta_1,\dots,\bm\zeta_{q_T}))-\mathbb{E}f(W(\bm\xi_1,\dots,\bm\xi_{q_T}))|\le \sum_{k=1}^{q_T}|\mathbb{E}f(W_k)-\mathbb{E}f(W_{k+1})|.
\end{align*}
By Taylor's expansion, we have
\begin{align*}
    &f(W_k)-f(W_{k,0})=\sum_{i=1}^2\frac{1}{i!}f^{(i)}(W_{k,0})(W_k-W_{k,0})^i+O(|W_k-W_{k,0}|^3),~~\text{and}\\
    &f(W_{k+1})-f(W_{k,0})=\sum_{i=1}^2\frac{1}{i!}f^{(i)}(W_{k,0})(W_{k+1}-W_{k,0})^i+O(|W_{k+1}-W_{k,0}|^3).
\end{align*}
According to the properties of $u(\bm\xi_t,\bm\zeta_s)$, we have
\begin{align*}\label{A.5}
    &\mathbb{E}(W_k-W_{k,0}|\bm\zeta_1,\dots,\bm\zeta_{k-1},\bm\xi_{k+1},\dots,\bm\xi_{q_T})=\sum_{t=1}^{k-1}\mathbb{E}(u(\bm\zeta_t,\bm\xi_k)|\bm\zeta_t)+\sum_{s=k+1}^{q_T}\mathbb{E}(u(\bm\xi_k,\bm\xi_s)|\bm\xi_s)=0,\\
    &\mathbb{E}(W_{k+1}-W_{k,0}|\bm\zeta_1,\dots,\bm\zeta_{k-1},\bm\xi_{k+1},\dots,\bm\xi_{q_T})=\sum_{t=1}^{k-1}\mathbb{E}(u(\bm\zeta_t,\bm\zeta_k)|\bm\zeta_t)+\sum_{s=k+1}^{q_T}\mathbb{E}(u(\bm\zeta_k,\bm\xi_s)|\bm\xi_s)=0,\\
    &\mathbb{E}[(W_k-W_{k,0})^2|\bm\zeta_1,\dots,\bm\zeta_{k-1},\bm\xi_{k+1},\dots,\bm\xi_{q_T}]=\mathbb{E}[(W_{k+1}-W_{k,0})^2|\bm\zeta_1,\dots,\bm\zeta_{k-1},\bm\xi_{k+1},\dots,\bm\xi_{q_T}].\tag{A.5}
\end{align*}
Furthermore,
\begin{align*}
    \mathbb{E}\left[\sum_{i=1}^2\frac{1}{i!}f^{(i)}(W_{k,0})(W_k-W_{k,0})^i\right]=\mathbb{E}\left[\sum_{i=1}^2\frac{1}{i!}f^{(i)}(W_{k,0})(W_{k+1}-W_{k,0})^i\right].
\end{align*}
Thus, there exists a positive constant $C$ such that 
\begin{align*}
    |\mathbb{E}f(W(\bm\zeta_1,\dots,\bm\zeta_{q_T}))-\mathbb{E}f(W(\bm\xi_1,\dots,\bm\xi_{q_T}))|\le &C\sum_{k=1}^{q_T}(\mathbb{E}|W_k-W_{k,0}|^3+\mathbb{E}|W_{k+1}-W_{k,0}|^3)\\
    \le &C\sum_{k=1}^{q_T}[(\mathbb{E}|W_k-W_{k,0}|^4)^{3/4}+(\mathbb{E}|W_{k+1}-W_{k,0}|^4)^{3/4}].
\end{align*}

Next, we derive the upper bounds for $\mathbb{E}|W_k-W_{k,0}|^4$ and $\mathbb{E}|W_{k+1}-W_{k,0}|^4$. According to the properties of $u(\bm\xi_t,\bm\zeta_s)$, we have
\begin{align*}\label{A.6}
    &\mathbb{E}|W_k-W_{k,0}|^4\\=&\sum_{t=1}^{k-1}\mathbb{E}(u(\bm\zeta_t,\bm\xi_k)^4)+\sum_{s=k+1}^{q_T}\mathbb{E}(u(\bm\xi_k,\bm\xi_s)^4)+6\sum_{t_1=1}^{k-1}\sum_{t_2=1}^{k-1}\mathbb{E}(u(\bm\zeta_{t_1},\bm\xi_k)^2u(\bm\zeta_{t_2},\bm\xi_k)^2)\\
    &+6\sum_{s_1=k+1}^{q_T}\sum_{s_2=k+1}^{q_T}\mathbb{E}(u(\bm\xi_k,\bm\xi_{s_1})^2u(\bm\xi_k,\bm\xi_{s_2})^2)+6\sum_{t=1}^{k-1}\sum_{s=k+1}^{q_T}\mathbb{E}(u(\bm\zeta_t,\bm\xi_k)^2u(\bm\xi_k,\bm\xi_s)^2)\\
    \le &\rho_0\left(\sum_{t=1}^{k-1}c_{tk}^4+\sum_{s=k+1}^{q_T}c_{ks}^4+6\sum_{t_1=1}^{k-1}\sum_{t_2=1}^{k-1}c_{t_1,k}^2c_{t_2,k}^2+6\sum_{s_1=k+1}^{q_T}\sum_{s_2=k+1}^{q_T}c_{k,s_1}^2c_{k,s_2}^2+6\sum_{t=1}^{k-1}\sum_{s=k+1}^{q_T}c_{tk}^2c_{ks}^2 \right)\\
    \le &3\rho_0\left(\sum_{t=1}^{k-1}c_{tk}^2+\sum_{s=k+1}^{q_T}c_{ks}^2\right)^2.\tag{A.6}
\end{align*}
Similarly, we get $\mathbb{E}|W_{k+1}-W_{k,0}|^4\le3\rho_0(\sum_{t=1}^{k-1}c_{tk}^2+\sum_{s=k+1}^{q_T}c_{ks}^2)^2$. Thus, 
\begin{align*}
    |\mathbb{E}f(W(\bm\zeta_1,\dots,\bm\zeta_{q_T}))-\mathbb{E}f(W(\bm\xi_1,\dots,\bm\xi_{q_T}))|=O\left\{\sum_{k=1}^{q_T}\left(\sum_{t=1}^{k-1}c_{tk}^2+\sum_{s=k+1}^{q_T}c_{ks}^2\right)^{3/2}\right\}.
\end{align*}
Since $\{\bm\xi_t\}_{t=1}^{q_T}$ has the same first two moment as $\{\bm\zeta_t\}_{t=1}^{q_T}$, we have
\begin{align*}
    c_{ts}^2=&\frac{(w_T-M)^2}{2T^2\tr(\O_{T,M}^2)}\mathbb{E}[(\bm\xi_t^\top \bm\xi_s)^2]\\
    =&\frac{1}{2T^2\tr(\O_{T,M}^2)}\sum_{k_1=(t-1)w_T+1}^{tw_T-M}\sum_{k_2=(s-1)w_T+1}^{sw_T-M}\sum_{k_3=(t-1)w_T+1}^{tw_T-M}\sum_{k_4=(s-1)w_T+1}^{sw_T-M}\mathbb{E}(\bm\gamma_{k_1}^\top \bm\gamma_{k_2} \bm\gamma_{k_3}^\top \bm\gamma_{k_4})\\
    =&\frac{1}{2T^2\tr(\O_{T,M}^2)}\tr\left\{\left(\sum_{k_1=(t-1)w_T+1}^{tw_T-M}\sum_{k_2=(t-1)w_T+1}^{tw_T-M}\Gamma_{k_1-k_2,M}\right)^2\right\}\\
    =&\frac{(w_T-M)^2\tr(\O_{w_T-M}^2)}{2T^2\tr(\O_{T,M}^2)}\{1+o(1)\}.
\end{align*}
Furthermore, 
\begin{align*}
    |\mathbb{E}f(W(\bm\zeta_1,\dots,\bm\zeta_{q_T}))-\mathbb{E}f(W(\bm\xi_1,\dots,\bm\xi_{q_T}))|=O\left[q_T\left\{q_T\frac{(w_T-M)^2\tr(\O_{w_T-M}^2)}{2T^2\tr(\O_{T,M}^2)}\right\}^{3/2}\right]=o(1).
\end{align*}
In summary, \eqref{A.4} is proved.

\noindent\textbf{Step 2.} Show that 
\begin{align*}\label{A.7}
    \Delta S_2=\sum_{t=1}^{q_T}B_{tt}^{G}-\sum_{t=1}^{q_T}B_{tt}^{NG}=o_p\left\{\sqrt{T^{-2}\tr(\O_{T,M}^2)}\right\}.\tag{A.7}
\end{align*}
By the classical CLT, it holds that $\sum_{t=1}^{q_T}B_{tt}/\sqrt{q_T\var(B_{11})}\cd N(0,1)$. Since $\var(B_{11})=(w_T-M)^2T^{-4}\tr(\O_{w_T-M}^2)$, we have
\begin{align*}
    \frac{\sum_{t=1}^{q_T}B_{tt}}{\sqrt{T^{-2}\tr(\O_{T,M}^2)}}=\frac{1}{\sqrt{q_T}}\sqrt{\frac{q_T^2(w_T-M)^2T^{-4}\tr(\O_{w_T-M}^2)}{T^{-2}\tr(\O_{T,M}^2)}}\frac{\sum_{t=1}^{q_T}B_{tt}}{\sqrt{q_T\var(B_{11})}}=o_p(1).
\end{align*}

\noindent\textbf{Step 3.} Show that 
\begin{align*}\label{A.8}
    \Delta S_3=\sum_{1\le t,s\le q_T}D_{ts}^{G}-\sum_{1\le t,s\le q_T}D_{ts}^{NG}=o_p\left\{\sqrt{T^{-2}\tr(\O_{T,M}^2)}\right\}.\tag{A.8}
\end{align*}
Define
\begin{align*}
    &L_{ts}:=\sum_{k=tw_T-M+1}^{tw_T}\sum_{l=(s-1)w_T+1}^{sw_T-M}\mathcal{F}_{kl},\\
    &R_{ts}:=\sum_{k=(t-1)w_T+1}^{tw_T-M}\sum_{l=sw_T-M+1}^{sw_T}\mathcal{F}_{kl},\\
    &C_{ts}:=\sum_{k=tw_T-M+1}^{tw_T}\sum_{l=sw_T-M+1}^{sw_T}\mathcal{F}_{kl}.
\end{align*}
Then, $D_{ts}=L_{ts}+R_{ts}+C_{ts}=2L_{ts}+C_{ts}$. Define
\begin{align*}
    \mathcal{L}_1=\sum_{t=1}^{q_T-1}\sum_{s=t+2}^{q_T}L_{ts},~~\mathcal{L}_2=\sum_{t=2}^{q_T}\sum_{s=1}^{t-1}L_{ts},&~~\mathcal{L}_3=\sum_{t=1}^{q_T}L_{tt},~~\mathcal{L}_3=\sum_{t=1}^{q_T}L_{t,t+1}\\
    \mathcal{C}_1=\sum_{t=1}^{q_T}\sum_{s=t+1}^{q_T}C_{ts},~~&\mathcal{C}_2=\sum_{t=1}^{q_T}C_{tt}.
\end{align*}
Then, $S_3=2\mathcal{L}_1+2\mathcal{L}_2+2\mathcal{L}_3+2\mathcal{L}_4+2\mathcal{C}_1+\mathcal{C}_2$. 

Next, we analyze the second moment of $\mathcal{L}_i$'s and $\mathcal{C}_i$'s. For $k_1,k_2,k_3,k_4\in\{1,\dots,T\}^4$, by the M-dependence of $\bm\gamma_t$'s, $\mathbb{E}(\bm\gamma_{k_1}^\top \bm\gamma_{k_2}\bm\gamma_{k_3}^\top \bm\gamma_{k_4})\ne 0$ only if $k_{(2)}-k_{(1)}\le M$ and $k_{(4)}-k_{(3)}\le M$, where $k_{(1)},\dots,k_{(4)}$ are the sorted $k_1,\dots,k_4$. Accordingly, 
\begin{align*}
    &\mathbb{E}\{(\mathcal{L}_1^{NG})^2\}\\=&\sum_{t_1=1}^{q_T-1}\sum_{s_1=t_1+2}^{q_T}\sum_{t_2=1}^{q_T-1}\sum_{s_2=t_2+2}^{q_T}\mathbb{E}(L_{t_1s_1}^{NG}L_{t_2s_2}^{NG})\\
    =&\sum_{t_1=1}^{q_T-1}\sum_{s_1=t_1+2}^{q_T}\sum_{t_2=1}^{q_T-1}\sum_{s_2=t_2+2}^{q_T}\sum_{k_1=t_1w_T-M+1}^{t_1w_T}\sum_{k_2=(s_1-1)w_T+1}^{s_1w_T-M}\sum_{k_3=t_2w_T-M+1}^{t_2w_T}\sum_{k_4=(s_2-1)w_T+1}^{s_2w_T-M}\frac{\mathbb{E}(\bm\gamma_{k_1}^\top \bm\gamma_{k_2}\bm\gamma_{k_3}^\top \bm\gamma_{k_4})}{T^4}\\
    =&\frac{1}{T^4}\sum_{t=1}^{q_T-1}\sum_{s=t+2}^{q_T}\mathbb{E}\left\{\left(\sum_{k=(t-1)w_T-M+1}^{tw_T}\sum_{l=(s-1)W_T+1}^{sw_T-M}\bm\gamma_k\top \bm\gamma_l\right)^2 \right\}\\
    =&\frac{1}{T^4}\frac{(q_T-1)(q_T-2)}{2}M(w_T-M)\tr(\O_M\O_{w_T-M})\\
    \le &\frac{1}{T^4}\frac{(q_T-1)(q_T-2)}{2}M(w_T-M)\{\tr(\O_M^2)\tr(\O_{w_T-M}^2)\}^{1/2}\\
    =&O\{q_TMT^{-3}\tr(\O_{T,M}^2)\}=o\{T^{-2}\tr(\O_{T,M}^2)\}.
\end{align*}
Here, we explain the third equation as follows. Since $s_1\ge t_1+2$, which implies $k_2-k_1\ge M$, $\mathbb{E}(\bm\gamma_{k_1}^\top \bm\gamma_{k_2}\bm\gamma_{k_3}^\top \bm\gamma_{k_4})\ne 0$ only if $t_1=t_2$ or $t_1=s_2$ or $t_1=s_2-1$. If $t_1=t_2$, then $\mathbb{E}(\bm\gamma_{k_1}^\top \bm\gamma_{k_2}\bm\gamma_{k_3}^\top \bm\gamma_{k_4})\ne 0$ only if $s_1=s_2$. If $t_1=s_2$ or $t_1=s_2-1$, then $t_2\le s_2-2\le t_1-1$, which implies $k_{(2)}-k_{(1)}=\min(k_1-k_3,k_4-k_3)>M$. Therefore, $\mathbb{E}(\bm\gamma_{k_1}^\top \bm\gamma_{k_2}\bm\gamma_{k_3}^\top \bm\gamma_{k_4})\ne 0$ only if $t_1=t_2$ and $s_1=s_2$. Then, we get $\mathcal{L}_1^{NG}=o_p\{\sqrt{T^{-2}\tr(\O_{T,M}^2)}\}$. Similarly, we get $\mathcal{L}_2^{NG}=o_p\{\sqrt{T^{-2}\tr(\O_{T,M}^2)}\}$,  $\mathcal{C}_1^{NG}=o_p\{\sqrt{T^{-2}\tr(\O_{T,M}^2)}\}$, $\mathcal{L}_1^{G}=o_p\{\sqrt{T^{-2}\tr(\O_{T,M}^2)}\}$, $\mathcal{L}_2^{G}=o_p\{\sqrt{T^{-2}\tr(\O_{T,M}^2)}\}$ and $\mathcal{C}_1^{G}=o_p\{\sqrt{T^{-2}\tr(\O_{T,M}^2)}\}$.
\begin{align*}
    \mathbb{E}\{(\mathcal{L}_3^{NG})^2\}=&\sum_{t_1=1}^{q_T}\sum_{t_2=1}^{q_T}\mathbb{E}(L_{t_1t_1}^{NG}L_{t_2t_2}^{NG})\\
    =&\sum_{t_1=1}^{q_T}\sum_{t_2=1}^{q_T}\sum_{k_1=t_1w_T-M+1}^{t_1w-T}\sum_{k_2=(t_1-1)w_T+1}^{t_1w_T-M}\sum_{k_3=t_2w_T-M+1}^{t_2w-T}\sum_{k_4=(t_2-1)w_T+1}^{t_2w_T-M}\mathbb{E}(\mathcal{F}_{k_1k_2}^{NG}\mathcal{F}_{k_3k_4}^{NG})\\
    =&\sum_{t=1}^{q_T}\mathbb{E}\left\{\left(\sum_{k=tw_T-M+1}^{tw_T}\sum_{l=(t-1)w_T+1}^{tw_T-M}\mathcal{F}_{kl}^{NG}\right)^2\right\}\\
    \le &T^{-4}q_TM^2(w_T-M)^2\tau_2\tr(\O_{T,M}^2)=o(T^{-2}\tr(\O_{T,M}^2)).
\end{align*}
Here, the third equation follows from the independence of $\mathcal{F}_{k_1k_2}$ and $\mathcal{F}_{k_3k_4}$, and the fourth inequality follows from Lemma \ref{key_inequality.2}. Then, we get $\mathcal{L}_3^{NG}=o_p\{\sqrt{T^{-2}\tr(\O_{T,M}^2)}\}$. Similarly, we get $\mathcal{L}_4^{NG}=o_p\{\sqrt{T^{-2}\tr(\O_{T,M}^2)}\}$, $\mathcal{C}_2^{NG}=o_p\{\sqrt{T^{-2}\tr(\O_{T,M}^2)}\}$, $\mathcal{L}_3^{G}=o_p\{\sqrt{T^{-2}\tr(\O_{T,M}^2)}\}$, $\mathcal{L}_4^{G}=o_p\{\sqrt{T^{-2}\tr(\O_{T,M}^2)}\}$ and $\mathcal{C}_2^{G}=o_p\{\sqrt{T^{-2}\tr(\O_{T,M}^2)}\}$. In summary, we get \eqref{A.8}.

\noindent\textbf{Step 4.} Show that 
\begin{align*}\label{A.9}
    \Delta S_4=F^{G}-F^{NG}=o_p\left\{\sqrt{T^{-2}\tr(\O_{T,M}^2)}\right\}.\tag{A.9}
\end{align*}
Define $\mathcal{A}=\{q_Tw_T+1,\dots,T\}\times\{1,\dots,T\}\times\{q_Tw_T+1,\dots,T\}\times\{1,\dots,T\}$, $\mathcal{A}_1=\{(k_1.k_2,k_3,k_4)\in\{1,\dots,T\}^4:|k_1-k_3|\ge 3M,\min(k_1,k_3)>q_Tw_T\}$ and $\mathcal{A}_2=\{(k_1.k_2,k_3,k_4)\in\{1,\dots,T\}^4:|k_1-k_3|\ge 3M,\min(k_1,k_3)>q_Tw_T,\min(|k_1-k_2|,|k_1-k_4|,|k_3-k_2|,|k_3-k_4|)\le M\}$. Then, for all $(k_1.k_2,k_3,k_4)\in\mathcal{A}_1-\mathcal{A}_2$, we have $\mathbb{E}(\mathcal{F}_{k_1k_2}\mathcal{F}_{k_3k_4})=0$. Since $\mathcal{A}_2\subset\{q_Tw_T+1,\dots,T\}^4$, $|\mathcal{A}_2|\le (w_T+M)^4$. Similarly, we have $|\mathcal{A}-\mathcal{A}_1|\le 6MTw_T^2$. Accordingly, we have 
\begin{align*}
    \mathbb{E}\{(F^{NG})^2\}=&\mathbb{E}\left\{\left(2\sum_{k=1}^{q_Tw_T}\sum_{l=q_Tw_T+1}^T\mathcal{F}^{NG} +\sum_{k=q_Tw_T+1}^T\sum_{l=q_Tw_T+1}^T\mathcal{F}^{NG} \right)^2\right\}\\
    \le &4\sum_{k_1=1}^T\sum_{k_2=q_Tw_T+1}^T\sum_{k_3=1}^T\sum_{k_4=q_Tw_T+1}^T|\mathbb{E}(\mathcal{F}_{k_1k_2}\mathcal{F}_{k_3k_4})|\\
    =&4\left(\sum_{(k_1,k_2,k_3,k_4)\in\mathcal{A}_1}+\sum_{(k_1,k_2,k_3,k_4)\in\mathcal{A}_2}\right)|\mathbb{E}(\mathcal{F}_{k_1k_2}\mathcal{F}_{k_3k_4})|\\
    \le &4\{6MTw_T^2+(w_T+M)^4\}T^{-4}\tr(\Omega_{T,M}^2)=o(T^{-2}\tr(\Omega_{T,M}^2)).
\end{align*}
Thus, $F^{NG}=o_p\{\sqrt{T^{-2}\tr(\O_{T,M}^2)}\}$. Similarly, we get $F^{G}=o_p\{\sqrt{T^{-2}\tr(\O_{T,M}^2)}\}$. Then, $\Delta S_4=o_p\{\sqrt{T^{-2}\tr(\O_{T,M}^2)}\}$.

In summary, \eqref{A.2} and \eqref{A.7}-\eqref{A.9} yield Lemma 5.

\begin{lemma}\label{error of mean}
    Under Conditions (C1)-(C5), we have $\mathbb{E}(T_{\SUM}^{NG})-\mu_T'=o_p\{\sqrt{T^{-2}\tr(\O_{T,M}^2)}\}$.
\end{lemma}
\proof By the definitions and the fact that $\tr^2(\bm\Sigma)\le N\tr(\bm\Sigma^2)$, we have
\begin{align*}
    |\mu_T'-\mathbb{E}(T_{\SUM}^{NG})|=&\left|\frac{1}{T}\sum_{h\in\mathcal{M}}\left(1-\frac{|h|}{T}\right)\tr(\Gamma_h-\Gamma_{h,M})\right|\\
    =&\left|\frac{1}{T}\tr(\bm\Sigma)\sum_{h\in\mathcal{M}}\left(1-\frac{|h|}{T}\right)\sum_{k=M-h+1}^\infty b_kb_{k+h}\right|\\
    \le &\frac{2}{T}\{N\tr(\bm\Sigma^2)\}^{1/2}\left(\sum_{k=0}^\infty |b_k|\right)\left(\sum_{k=M+1}^\infty|b_k|\right)\\
    =&o\{N^{1/2}T^{-1}\tr^{1/2}(\bm\Sigma^2)M^{-4}\}=o_p\{\sqrt{T^{-2}\tr(\O_{T,M}^2)}\},
\end{align*}
where the last equality is because 
\begin{align*}\label{A.10}
    \sum_{k=M}^\infty |b_k|=o(M^{-4})\tag{A.10}
\end{align*}
due to Condition (C2) (\romannumeral2) and Lemma \ref{key_equality.3}.

\begin{lemma}\label{M-dependence_approximation}
    Under Conditions (C1)-(C5), we have $\tilde{T}_{\SUM}-T_{\SUM}^{NG}=o_p\{\sqrt{T^{-2}\tr(\O_{T,M}^2)}\}$.
\end{lemma}
\proof By the definitions, we have
\begin{align*}
    \tilde{T}_{\SUM}-T_{\SUM}^{NG}=&\frac{1}{T^2}\sum_{t=1}^T\sum_{s=1}^T(\X_t^\top \X_s-\bm\gamma_t^\top \bm\gamma_s)\\
    =&\frac{1}{T^2}\sum_{t=1}^T\sum_{s=1}^T(\X_t-\bm\gamma_t)^\top (\X_s-\bm\gamma_s)+\frac{2}{T^2}\sum_{t=1}^T\sum_{s=1}^T\bm\gamma_t^\top (\X_s-\bm\gamma_s).
\end{align*}
Define $\mathcal{S}_{t_1s_1t_2s_2}=\{(k_1,k_2,k_3,k_4):t_1-M\le k_1\le t_1,k_2\le s_1-M-1,t_2-M\le k_3\le t_2,k_4\le s_2-M-1\}$, $\mathcal{S}_1=\{(k_1,k_2,k_3,k_4):k_1=k_2\ne k_3=k_4\}$, $\mathcal{S}_2=\{(k_1,k_2,k_3,k_4):k_1=k_3\ne k_2=k_4\}$, $\mathcal{S}_3=\{(k_1,k_2,k_3,k_4):k_1=k_4\ne k_2=k_3\}$ and $\mathcal{S}_4=\{(k_1,k_2,k_3,k_4):k_1=k_2= k_3=k_4\}$. Then, 
\begin{align*}
    &\var\left\{\sum_{t=1}^T\sum_{s=1}^T\bm\gamma_t^\top (\X_s-\bm\gamma_s)\right\}\\=&\sum_{\substack{1\le t_1,s_1,t_2,s_2\le T\\(k_1,k_2,k_3,k_4)\in\mathcal{S}_{t_1s_1t_2s_2}}}b_{t_1-k_1}b_{s_1-k_2}b_{t_2-k_3}b_{s_2-k_4}\mathbb{E}\left\{(\z_{k_1}^\top \bm\Sigma\z_{k_2}-\mathbb{I}_{(k_1=k_2)}\tr(\bm\Sigma))(\z_{k_3}^\top \bm\Sigma\z_{k_4}-\mathbb{I}_{(k_3=k_4)}\tr(\bm\Sigma))\right\}\\
    =&\sum_{\substack{1\le t_1,s_1,t_2,s_2\le T\\(k_1,k_2,k_3,k_4)\in\mathcal{S}_{t_1s_1t_2s_2}\cap\mathcal{S}_1}}b_{t_1-k_1}b_{s_1-k_1}b_{t_2-k_3}b_{s_2-k_3}\mathbb{E}\left\{(\z_{k_1}^\top \bm\Sigma\z_{k_1}-\tr(\bm\Sigma))(\z_{k_3}^\top \bm\Sigma\z_{k_3}-\tr(\bm\Sigma))\right\}\\
    &+\sum_{\substack{1\le t_1,s_1,t_2,s_2\le T\\(k_1,k_2,k_3,k_4)\in\mathcal{S}_{t_1s_1t_2s_2}\cap\mathcal{S}_2}}b_{t_1-k_1}b_{s_1-k_2}b_{t_2-k_1}b_{s_2-k_2}\mathbb{E}\{(\z_{k_1}^\top \bm\Sigma\z_{k_2})^2\}\\
    &+\sum_{\substack{1\le t_1,s_1,t_2,s_2\le T\\(k_1,k_2,k_3,k_4)\in\mathcal{S}_{t_1s_1t_2s_2}\cap\mathcal{S}_3}}b_{t_1-k_1}b_{s_1-k_2}b_{t_2-k_2}b_{s_2-k_1}\mathbb{E}\{(\z_{k_1}^\top \bm\Sigma\z_{k_2})(\z_{k_2}^\top \bm\Sigma\z_{k_1})\}\\
    &+\sum_{\substack{1\le t_1,s_1,t_2,s_2\le T\\(k_1,k_2,k_3,k_4)\in\mathcal{S}_{t_1s_1t_2s_2}\cap\mathcal{S}_4}}b_{t_1-k_1}b_{s_1-k_1}b_{t_2-k_1}b_{s_2-k_1}\mathbb{E}\{(\z_{k_1}^\top \bm\Sigma\z_{k_1}-\tr(\bm\Sigma))^2\}\\
    \le &\sum_{\substack{1\le t_1,s_1,t_2,s_2\le T\\(k_1,k_2,k_3,k_4)\in\mathcal{S}_{t_1s_1t_2s_2}\cap\mathcal{S}_2}}|b_{t_1-k_1}b_{s_1-k_2}b_{t_2-k_1}b_{s_2-k_2}|\tr(\bm\Sigma^2)\\
    &+\sum_{\substack{1\le t_1,s_1,t_2,s_2\le T\\(k_1,k_2,k_3,k_4)\in\mathcal{S}_{t_1s_1t_2s_2}\cap\mathcal{S}_3}}|b_{t_1-k_1}b_{s_1-k_2}b_{t_2-k_2}b_{s_2-k_1}|\tr(\bm\Sigma^2)\\
    &+\sum_{\substack{1\le t_1,s_1,t_2,s_2\le T\\(k_1,k_2,k_3,k_4)\in\mathcal{S}_{t_1s_1t_2s_2}\cap\mathcal{S}_4}}|b_{t_1-k_1}b_{s_1-k_1}b_{t_2-k_1}b_{s_2-k_1}|\tau_1\tr^2(\bm\Sigma)\\
    \le &T^2\left\{\left(\sum_{t_1,t_2\ge 0}|b_{t_1}b_{t_2}|\right)\left(\sum_{s_1,s_2\ge M}|b_{s_1}b_{s_2}|\right)+\left(\sum_{t_1\ge 0,s_2\ge M}|b_{t_1}b_{s_2}|\right)\left(\sum_{t_2\ge 0,s_1\ge M}|b_{t_2}b_{s_1}|\right)\right\}\tr(\bm\Sigma^2)\\
    &+\tau_1NT\sum_{t_1,t_2\ge 0,s_11,s_2\ge M}|b_{t_1}b_{t_2}b_{s_1}b_{s_2}|\tr(\bm\Sigma^2)=o\{T^{-2}\tr(\O_{T,M}^2)\},
\end{align*}
where the first inequality comes from Lemma \ref{key_inequality.1} and the last equation comes from \eqref{A.10}. Hence, we get $T^{-2}\sum_{t=1}^T\sum_{s=1}^T\bm\gamma_t^\top (\X_s-\bm\gamma_s)=o_p\{\sqrt{T^{-2}\tr(\O_{T,M}^2)}\}$. Similarly, we can prove $T^{-2}\sum_{t=1}^T\sum_{s=1}^T(\X_t-\bm\gamma_t)^\top (\X_s-\bm\gamma_s)=o_p\{\sqrt{T^{-2}\tr(\O_{T,M}^2)}\}$. This completes the proof.

\begin{lemma}\label{error of variance}
    Under Conditions (C1)-(C5), we have $T^{-2}\tr(\O_{T,M}^2)=\sigma_T^2\{1+o(1)\}$.
\end{lemma}
\proof Recalling the definitions, we have 
\begin{align*}
    \O_T-\O_{T,M}=\left[\left\{a_0+2\sum_{h=1}^T\left(1-\frac{h}{T}\right)a_h\right\}-\left\{a_{0,M}+2\sum_{h=1}^M\left(1-\frac{h}{T}\right)a_{h,M}\right\}\right]\bm\Sigma=:B_0\bm\Sigma,
\end{align*}
and 
\begin{align*}
    |B_0|\le& 2\left|\sum_{h=0}^T\left(1-\frac{h}{T}\right)a_h-\sum_{h=0}^M\left(1-\frac{h}{T}\right)a_{h,M}\right|\\
    =&2\left|\sum_{h=M+1}^T\left(1-\frac{h}{T}\right)\sum_{k=0}^\infty b_kb_{k+h}+\sum_{h=0}^M\left(1-\frac{h}{T}\right)\sum_{k=M-h+1}^\infty b_kb_{k+h}\right|\\
    \le &4\left(\sum_{k=0}^\infty |b_k|\right)\left(\sum_{l=M+1}^\infty |b_l|\right)=o(M^{-4})
\end{align*}
due to \eqref{A.10}. Then, using the fact that
\begin{align*}
    \sigma_T^2-T^{-2}\tr(\O_{T,M}^2)=&T^{-2}\{\tr(\O_T^2)-\tr(\O_{T,M}^2)\}\\
    =&T^{-2}\tr\{(\O_T-\O_{T,M})^2\}+2T^{-2}\tr\{\O_{T,M}(\O_T-\O_{T,M})\},
\end{align*}
we have $T^{-2}\tr(\O_{T,M}^2)=\sigma_T^2\{1+o(1)\}$.

\begin{lemma}\label{error of e_t}
    Under Conditions (C1)-(C6), we have $T_{\SUM}-\tilde{T}_{\SUM}=o_p\{\sqrt{T^{-2}\tr(\O_{T,M}^2)}\}$.
\end{lemma}
\proof Since $|1-\f_t^\top (\F^\top \F)^{-1}\F^\top \bm 1_T|\le 1+\sum_{k=1}^p|f_{tk}|\cdot \Vert(\F^\top \F/T)^{-1}\Vert_\infty \cdot \Vert\F^\top \bm 1_T/T\Vert_\infty\le Cp(1+\Vert\f_t\Vert)\le C'$ for some positive constants $C,C'$, we have $\bm 1_T^\top \P_\F\bm 1_T/T=O(1)$ and $\eta_t=O(1)$. Define $\zeta_t:=\eta_t-e_t$. By the definition,
\begin{align*}
  \zeta_t=&\frac{T}{\bm 1_T^\top \P_\F\bm 1_T}\{1-\f_t^\top(\F^\top \F)^{-1}\F^\top \bm 1_T\}-\frac{1}{\omega}(1-\f_t^\top\L_{\f}^{-1}\bmu_{\f})\\
  =&\frac{T}{\bm 1_T^\top \P_\F\bm 1_T}\{\f_t^\top\L_{\f}^{-1}\bmu_{\f}-\f_t^\top(\F^\top \F)^{-1}\F^\top \bm 1_T\}+\left(\frac{T}{\bm 1_T^\top \P_\F\bm 1_T}-\frac{1}{\omega}\right)(1-\f_t^\top\L_{\f}^{-1}\bmu_{\f})
\end{align*}
By Condition (C1), we have $T^{-1}\bm 1_T^\top \P_\F\bm 1_T=\omega+O_p(T^{-1/2})$, $T^{-1}\F^\top \bm 1——T=\bmu_{\f}+O_p(T^{-1/2})$, and $T^{-1}\F^\top \F=\L_{\f}+O_p(T^{-1/2})$. Additionally, $\Vert\f_t\Vert$ is bounded, so 
\begin{align*}\label{A.11}
    \zeta_t=O_p(T^{-1/2})~~\text{and}~~e_t=O_p(1).\tag{A.11}
\end{align*}
By the definitions, we have
\begin{align*}
T_{\SUM}-\tilde{T}_{\SUM}=&(\tilde{\bma}+\frac{1}{T}\sum_{t=1}^T\bmv_t\zeta_T)^\top (\tilde{\bma}+\frac{1}{T}\sum_{t=1}^T\bmv_t\zeta_T)-\tilde{\bma}^\top \tilde{\bma}\\
=&\frac{2}{T}\sum_{t=1}^T\bmv_t^\top\tilde{\bma}\zeta_t+\frac{1}{T^2}\sum_{t=1}^T\sum_{s=1}^T\bmv_t^\top\bmv_s\zeta_t\zeta_s\\
=&\frac{2}{T}\sum_{t=1}^T\bmv_t^\top{\bma}\zeta_t+\frac{2}{T^2}\sum_{t=1}^T\sum_{s=1}^T\bmv_t^\top\bmv_s\zeta_te_s+\frac{1}{T^2}\sum_{t=1}^T\sum_{s=1}^T\bmv_t^\top\bmv_s\zeta_t\zeta_s\\
=: &A_1+A_2+A_3.
\end{align*}

We consider the first term $A_1$. $\mathbb{E}(A_1)=0$ and
\begin{align*}
E(A_1^2)=&E\left(\frac{4}{T^2}\sum_{t=1}^T\sum_{s=1}^T\bma^\top\bmv_t\bmv_s^\top\bma\zeta_t\zeta_s\right)\\
=&E\left(\frac{4}{T^2}\sum_{t=1}^T\sum_{s=1}^T\bma^\top\X_t\X_s^\top\bma\zeta_t\zeta_se_t^{-1}e_s^{-1}\right)\\
=&O(T^{-3})\sum_{t=1}^T\sum_{s=1}^T\bma^\top\Gamma_{t-s}\bma\\
=&O(T^{-3}\bma^\top \Omega_T\bma)=o_p\{\sqrt{T^{-2}\tr(\O_{T,M}^2)}\},
\end{align*}
by Condition (C6) (see the detail proof of this term in the proof of Theorem 2). Hence, $A_1=o_p\{\sqrt{T^{-2}\tr(\O_{T,M}^2)}\}$. Similarly, we have
\begin{align*}
E(A_2)=\frac{2}{T^2}\sum_{t=1}^T\sum_{s=1}^TE(\X_t^\top\X_s\zeta_te_t^{-1})=O(T^{-3/2})\tr(\O_T)=o\{\sqrt{T^{-2}\tr(\O_{T,M}^2)}\},
\end{align*}
by condition (C2) (\romannumeral3). And $\var(A_2)=O\{T^{-3}\tr(\O_T^2)\}=o\{T^{-2}\tr(\O_{T,M}^2)\}$. Taking the same procedure, we have $E(A_3)=O\{T^{-2}\tr(\O_T)\}=o\{\sqrt{T^{-2}\tr(\Omega_{T,M}^2)}\}$ and $\var(A_3)=O\{T^{-4}\tr(\O_T^2)\}=o\{T^{-2}\tr(\Omega_{T,M}^2)\}$. Here we complete the proof.

\subsubsection{Proof of Theorem 1}
\proof Note that
\begin{align*}\label{A.12}
    \frac{T_{\SUM}-\mu_T'}{\sigma_T}=&\sqrt{\frac{2T^{-2}\tr(\Omega_{T,M}^2)}{\sigma_T^2}}\frac{1}{\sqrt{2T^{-2}\tr(\Omega_{T,M}^2)}}\{(T_{\SUM}-\tilde{T}_{\SUM})+(\tilde{T}_{\SUM}-T_{\SUM}^{NG})\\
    &+(T_{\SUM}^{NG}-T_{\SUM}^{G})+(T_{\SUM}^{G}-\mathbb{E}(T_{\SUM}^{G}))+(\mathbb{E}(T_{\SUM}^{G})-\mu_T')\}.\tag{A.12}
\end{align*}
According to Lemmas \ref{clt}-\ref{error of e_t}, we can easily obtain the result by Slutsky's Theorem.

\subsubsection{Proof of Theorem 2}
\proof According to the proof of Theorem 1, we have
\begin{align*}
  T_{\SUM}= \bar{\bm\gamma}_T^\top \bar{\bm\gamma}_T+\bm \alpha^\top \bm\alpha+\bar{\bm\gamma}_T^\top \bm \alpha+o_p(\sigma_T).
\end{align*}
Thus, to proof the result, we only need to show that $\mathbb{E}\{(\bar{\bm\gamma}_T^\top \bm \alpha)^2\}=\frac{1}{T}\bma^\top \O_{T,M}\bma=o(\sigma_T^2)$. Note that $\O_{T,M}=\sum_{h\in\mathcal{M}}(1-\frac{|h|}{T})\Gamma_{h,M}$. For each $h$, we consider the singular value decomposition of $\Gamma_{h,M}$ as $\Gamma_{h,M}=\U\Lambda\V^\top$, where $\Lambda$ is the diagonal matrix with positive square root of eigenvalues of $\Gamma_{h,M}\Gamma_{h,M}^\top$. It yields that $\Gamma_{h,M}\Gamma_{-h,M}=\U\Lambda^2\U^\top$ which gives $\U\Lambda\U^\top=(\Gamma_{h,M}\Gamma_{-h,M})^{1/2}$. Similarly, we have $\V\Lambda\V^\top=(\Gamma_{-h,M}\Gamma_{h,M})^{1/2}$. Thus,
\begin{align*}
    \bma^\top \Gamma_{h,M}\bma=&\bma^\top \U\Lambda\V^\top\bma=(\Lambda^{1/2}\U^\top\bma)^\top (\Lambda^{1/2}\V^\top\bma)\\
    \le &\sqrt{\bma^\top(\Gamma_{h,M}\Gamma_{-h,M})^{1/2}\bma}\sqrt{\bma^\top(\Gamma_{-h,M}\Gamma_{h,M})^{1/2}\bma},
\end{align*}
where the inequality comes from the Cauchy-Schwarz inequality and 
\begin{align*}
    &(\Lambda^{1/2}\U^\top\bma)^\top(\Lambda^{1/2}\U^\top\bma)=\bma^\top\U\Lambda\U^\top\bma=\bma^\top(\Gamma_{h,M}\Gamma_{-h,M})^{1/2}\bma,\\
    &(\Lambda^{1/2}\V^\top\bma)^\top(\Lambda^{1/2}\V^\top\bma)=\bma^\top\V\Lambda\V^\top\bma=\bma^\top(\Gamma_{-h,M}\Gamma_{h,M})^{1/2}\bma.
\end{align*}
Then, the upper bound on $\bma^\top\O_{T,M}\bma$ is given by 
\begin{align*}
    \bma^\top\O_{T,M}\bma=&\sum_{h\in\mathcal{M}}\left(1-\frac{|h|}{T}\right)\bma^\top \Gamma_{h,M}\bma\\
    \le &\sum_{h\in\mathcal{M}}\left(1-\frac{|h|}{T}\right)\sqrt{\bma^\top(\Gamma_{h,M}\Gamma_{-h,M})^{1/2}\bma}\sqrt{\bma^\top(\Gamma_{-h,M}\Gamma_{h,M})^{1/2}\bma}.
\end{align*}
Thus, 
\begin{align*}
    &\frac{T^{-1}\bma^\top \O_{T,M}\bma}{T^{-2}\tr(\O_{T,M}^2)}\\
    \le &\frac{1}{{T^{-1}\tr(\O_{T,M}^2)}}\sum_{h\in\mathcal{M}}\left(1-\frac{|h|}{T}\right)\sqrt{\bma^\top(\Gamma_{h,M}\Gamma_{-h,M})^{1/2}\bma}\sqrt{\bma^\top(\Gamma_{-h,M}\Gamma_{h,M})^{1/2}\bma}\\
    \lesssim& \sum_{h\in\mathcal{M}}\left(1-\frac{|h|}{T}\right)\sqrt{\frac{\bma^\top(\Gamma_h\Gamma_{-h})^{1/2}\bma}{{T^{-1}\tr(\O_T^2)}}}\sqrt{\frac{\bma^\top(\Gamma_{-h}\Gamma_h)^{1/2}\bma}{{T^{-1}\tr(\O_T^2)}}}\to 0,
\end{align*}
by Condition (C6). This, together with the result of Lemma \ref{error of variance}, implies the desired result.

\begin{lemma}\label{consistence of mean estimator}
    Under Conditions (C1)-(C5), we have $\hat{\mu}_T-\mu_t'=o_p\{\sqrt{T^{-2}\tr(\O_{T,M}^2)}\}$.
\end{lemma}
\proof Define
\begin{align*}
    \hat{\mu}_T':=\frac{1}{T}\left\{\tr(\hat{\Gamma}_0)+2\sum_{h=1}^M\left(1-\frac{h}{T}\right)\tr(\hat{\Gamma}_h)\right\}.
\end{align*}
Note that 
\begin{align*}\label{A.13}
    \hat{\bmv}_t=\bm{y}_t-\hat{\bma}-\hat{\B}\f_t=\bmv_t+(\bma-\hat{\bma})+(\B-\hat{\B})\f_t,\tag{A.13}
\end{align*}
where $\hat{\B}=(\hat{\bm \beta}_1,\dots,\hat{\bm\beta}_N)^\top$ with $\hat{\bm\beta}_i=(\F^\top\F)^{-1}\F^\top(\bm y_{i.}-\alpha_i)$. Then,
\begin{align*}
    \hat{\mu}_T'=&\frac{1}{T^2}\sum_{t=1}^T\sum_{s=1}^T\hat{\bmv}_t^\top\hat{\bmv}_s\eta_t\eta_s\mathbb{I}_{(|t-s|\le M)}\\
    =&\frac{1}{T^2}\sum_{t=1}^T\sum_{s=1}^T\{\bmv_t^\top\bmv_s+(\bmv_t+\bmv_s)^\top(\bma-\hat{\bma})+\bmv_t^\top(\B-\hat{\B})\f_s+\bmv_s^\top(\B-\hat{\B})\f_t\\
    &+(\bma-\hat{\bma})^\top(\bma-\hat{\bma})+(\bma-\hat{\bma})^\top(\B-\hat{\B})(\f_t+\f_s)\\
    &+\f_t^\top(\B-\hat{\B})^\top(\B-\hat{\B})\f_s\}\eta_t\eta_s\mathbb{I}_{(|t-s|\le M)}.
\end{align*}
Define $\tilde{\mu}_T=T^{-2}\sum_{t=1}^T\sum_{s=1}^T\bmv_t^\top\bmv_s\eta_t\eta_s\mathbb{I}_{(|t-s|\le M)}$ and $\check{\mu}_T=T^{-2}\sum_{t=1}^T\sum_{s=1}^T\X_t^\top\X_s\mathbb{I}_{(|t-s|\le M)}$. 

\noindent\textbf{Step 1.} According to the proof of Theorem 3 in \citet{FLM11}, we have
\begin{align*}\label{A.14}
    &\max_{1\le i\le N}|\hat{\alpha}_i-\alpha_i|=O_p\{\sqrt{\log (N)/T}\}=\max_{1\le i\le N}\Vert\hat{\bm\beta}_i-\bm\beta_i\Vert,~~\text{and}\\
    &\max_{1\le i\le N}\left|T^{-1}\sum_{t=1}^T\varepsilon_{it}\right|=O_p\{\sqrt{\log (N)/T}\}=\max_{1\le i\le N}\left\| T^{-1}\sum_{t=1}^T\varepsilon_{it}\f_t\right\|,\tag{A.14}
\end{align*}
which, together with the fact that $\tr(\O_{T,M}^2)\lesssim \tr(\bm\Sigma^2)=O(N)$ by Condition (C3), imply that
\begin{align*}
    \frac{\hat{\mu}_T'-\tilde{\mu}_T}{\sqrt{T^{-2}\tr(\O_{T,M}^2)}}\le C\frac{T^{-2}\cdot TM\cdot N\log N/T}{\sqrt{T^{-2}N}}=CMN^{1/2}T^{-1}\log N\to 0
\end{align*}
with probability tending to one, i.e. $\hat{\mu}_T'-\tilde{\mu}_T=o_p\{\sqrt{T^{-2}\tr(\O_{T,M}^2)}\}$.

\noindent\textbf{Step 2.} Note that $\mathbb{E}(\check{\mu}_T)=\mu_T'$ and 
\begin{align*}
    \check{\mu}_T=\frac{1}{T}\sum_{h\in\mathcal{M}}\left(1-\frac{|h|}{T}\right)\tr(\hat{\Gamma}_{h,M}),
\end{align*}
where $\hat{\Gamma}_{h,M}=\frac{1}{T-h}\sum_{t=1}^{T-h}\X_{t+h}^\top\X_t$ and $\hat{\Gamma}_{-h,M}=\hat{\Gamma}_{h,M}^\top$ for $h\ge 0$. Then, 
\begin{align*}
    \var(\check{\mu}_T)=&\frac{1}{T^4}\sum_{h_1\in\mathcal{M}}\sum_{h_2\in\mathcal{M}}\cov((T-|h_1|)\tr(\hat{\Gamma}_{h_1,M}),(T-|h_2|)\tr(\hat{\Gamma}_{h_2,M}))\\
    \lesssim &\frac{4}{T^4}\sum_{h_1=0}^M\sum_{h_2=0}^M\cov((T-h_1)\tr(\hat{\Gamma}_{h_1,M}),(T-h_2)\tr(\hat{\Gamma}_{h_2,M})).
\end{align*}
\begin{align*}
    &\cov((T-h_1)\tr(\hat{\Gamma}_{h_1,M}),(T-h_2)\tr(\hat{\Gamma}_{h_2,M}))\\=&\mathbb{E}\left[\sum_{t_1=1}^{T-h_1}\sum_{t_2=1}^{T-h_2}\{\X_{t_1}^\top \X_{t_1+h_1}-\mathbb{E}(\X_{t_1}^\top \X_{t_1+h_1})\}\{\X_{t_2}^\top \X_{t_2+h_2}-\mathbb{E}(\X_{t_2}^\top \X_{t_2+h_2})\}\right]\\
    =&\sum_{\substack{1\le t_1\le T-h_1,1\le t_2\le T-h_2\\k_1\le t_1,k_2\le t_1+h_1,k_3\le t_2,k_4\le t_2+h_2}}b_{t_1-k_1}b_{t_1+h_1-k_2}b_{t_2-k_3}b_{t_2+h_2-k_4}\\
    &\qquad\times \mathbb{E}\{(\z_{k_1}^\top\bm\Sigma\z_{k_2}-\mathbb{I}_{(k_1=k_2)}\tr(\bm\Sigma))(\z_{k_3}^\top\bm\Sigma\z_{k_4}-\mathbb{I}_{(k_3=k_4)}\tr(\bm\Sigma))\}.
\end{align*}
Consider the value of $\rho_{k_1k_2k_3k_4}:=\mathbb{E}\{(\z_{k_1}^\top\bm\Sigma\z_{k_2}-\mathbb{I}_{(k_1=k_2)}\tr(\bm\Sigma))(\z_{k_3}^\top\bm\Sigma\z_{k_4}-\mathbb{I}_{(k_3=k_4)}\tr(\bm\Sigma))\}$.
\begin{equation*}
    \rho_{k_1k_2k_3k_4}
    \begin{cases}
        =\tr(\bm\Sigma^2), & k_1=k_3=r,k_2=k_4=s,r\ne s,\\
        =\tr(\bm\Sigma^2), & k_1=k_4=r,k_2=k_3=s,r\ne s,\\
        \le \tau_1\tr^2(\bm\Sigma),& k_1=k_2=k_3=k_4=r,\\
        =0,& \text{otherwise}.
    \end{cases}
\end{equation*}
So we just have to look at the coefficient of $\rho_{k_1k_2k_3k_4}$. Denote it as $C_{k_1k_2k_3k_4}$. Then,
\begin{align*}
    \sum_{t_1=1}^{T-h_1}\sum_{t_2=1}^{T-h_2}|C_{rsrs}|\le \sum_{\substack{1\le t_1\le T-h_1,1\le t_2\le T-h_2\\r\le t_1\wedge t_2,s\le (t_1+h_1)\wedge(t_2+h_2)}}|b_{t_1-r}b_{t_1+h_1-s}b_{t_2-r}b_{t_2+h_2-s}|.
\end{align*}
Consider the number of $|b_{k_1}b_{k_2}b_{k_3}b_{k_4}|$ appears in the right hand side. This can be equivalently written as 
\begin{align*}
    t_1-r=k_1,~~t_1+h_1-s=k_1,~~t_2-r=k_3,~~t_2+h_2-s=k_4.
\end{align*}
Fix $t_1,k_1,k_2,k_3,k_4$, $(r,s,t_2)$ can be uniquely represented by $t_1,k_1,k_2,k_3,k_4$ or not exist. Therefore, $|b_{k_1}b_{k_2}b_{k_3}b_{k_4}|$ appears at most $T-h_1$ times. Hence, 
\begin{align*}
    \sum_{t_1=1}^{T-h_1}\sum_{t_2=1}^{T-h_2}|C_{rsrs}|\le T\left(\sum_{k=0}^\infty |b_k|\right)^4.
\end{align*}
Similarly, we get
\begin{align*}
    \sum_{t_1=1}^{T-h_1}\sum_{t_2=1}^{T-h_2}|C_{rssr}|\le T\left(\sum_{k=0}^\infty |b_k|\right)^4,~~\sum_{t_1=1}^{T-h_1}\sum_{t_2=1}^{T-h_2}|C_{rrrr}|\le \left(\sum_{k=0}^\infty |b_k|\right)^4.
\end{align*}
These above results, together with the fact that $\tr^2(\bm\Sigma)\le N\tr(\bm\Sigma^2)$, yiled that
\begin{align*}
    \var(\check{\mu}_T)=O(M^2(T^{-3}+NT^{-4})\tr(\bm\Sigma^2))=o\{T^{-2}\tr(\O_{T,M}^2)\},
\end{align*}
which implies that $\check{\mu}_T-\mu_T'=o_p\{\sqrt{T^{-2}\tr(\O_{T,M}^2)}\}$.

\noindent\textbf{Step 3.} By \eqref{A.11}, we have $1-\frac{\eta_t\eta_s}{e_te_s}=\frac{e_t(e_s-\eta_s)+\eta_s(e_t-\eta_t)}{e_te_s}=O_p(T^{-1/2})$. Furthermore, we have
\begin{align*}
    \check{\mu}_T-\tilde{\mu}_T=\frac{1}{T^2}\sum_{t=1}^T\sum_{s=1}^T\X_t^\top \X_s\left(1-\frac{\eta_t\eta_s}{e_te_s}\right)\mathbb{I}_{(|t-s|\le M)}=o_p\{T^{-1/2}\sqrt{T^{-2}\tr(\O_{T,M}^2)}\}.
\end{align*}

\noindent\textbf{Step 4.} Recalling the definitions of $\hat{\mu}_T$ and $\hat{\mu}_T'$, we have
\begin{align*}
    \hat{\mu}_T-\hat{\mu}_T'=\frac{p+1}{T-p-1}\hat{\mu}_T'-\sum_{h=1}^M\frac{2(p+1)h}{T(T-p-1)^2}\tr(\hat{\Gamma}_h).
\end{align*}
According to the results of Step 1-3 and Lemma \ref{error of mean}, we have 
\begin{align*}
    \frac{p+1}{T-p-1}\hat{\mu}_T'/\sqrt{T^{-2}\tr(\O_{T,M}^2)}=O\left\{\frac{T^{-2}\tr(\O_{T,M})}{\sqrt{T^{-2}\tr(\O_{T,M}^2)}}\right\}=O(T^{-1}N^{1/2})=o(1),
\end{align*}
and
\begin{align*}
    \sum_{h=1}^M\frac{2(p+1)h}{T(T-p-1)^2}\tr(\hat{\Gamma}_h)/\sqrt{T^{-2}\tr(\O_{T,M}^2)}=O\left\{\frac{MT^{-3}\tr(\O_{T,M})}{\sqrt{T^{-2}\tr(\O_{T,M}^2)}}\right\}=O(MT^{-2}N^{1/2})=o(1),
\end{align*}
which yield that $\hat{\mu}_T-\hat{\mu}_T'=o_p\{\sqrt{T^{-2}\tr(\O_{T,M}^2)}\}$.

In summary, we have $\hat{\mu}_T-\mu_t'=o_p\{\sqrt{T^{-2}\tr(\O_{T,M}^2)}\}$.

\begin{lemma}\label{consistence of variance estimator}
    Under Conditions (C1)-(C5), we have $\hat{\sigma}_T^2/\sigma_T^2\cp 1$.
\end{lemma}
\proof Define
\begin{align*}
    \tilde{\sigma}_T^2:=\frac{2}{T^2}\left(\tilde{S}_{0,0}+2\sum_{r=1}^M\tilde{S}_{0,r}+2\sum_{r=1}^M\tilde{S}_{r,0}+4\sum_{r=1}^M\sum_{s=1}^M\tilde{S}_{r,s} \right),
\end{align*}
where
\begin{align*}
    \tilde{S}_{h_1,h_2}:=\frac{\sum_{t=1}^{[T/2]-h_2}\sum_{s=t+[T/2]}^{T-h_2}\X_t^\top \X_s\X_{t+h_1}^\top \X_{s+h_2}}{(T-h_2/2-\frac{3}{2}[T/2]+1/2)([T/2]-h_2)},
\end{align*}
which can be expanded in terms of $z_{it}$'s, because
\begin{align*}
    \X_t^\top \X_s\X_{t+h_1}^\top \X_{s+h_2}=\sum_{k_1\le t,k_2\le s,k_3\le t+h_1,k_4\le s+h_2}b_{t-k_1}b_{s-k_2}b_{t+h_1-k_3}b_{s+h_2-k_4}\z_{k_1}^\top\bm\Sigma\z_{k_2}\z_{k_3}^\top\bm\Sigma\z_{k_4}.
\end{align*}
Next, write $\tilde{S}_{h_1,h_2}$ a sum of the terms involving the high order of $z_{it}$ and the terms involving the low order of $z_{it}$. Specifically, write $\tilde{S}_{h_1,h_2}=\tilde{S}_{h_1,h_2,H}+\tilde{S}_{h_1,h_2,L}$. 

\noindent\textbf{Step 1.} Here 
\begin{align*}
    &(T-h_2/2-\frac{3}{2}[T/2]+1/2)([T/2]-h_2)\tilde{S}_{h_1,h_2,H}\\
    =&\sum_{\substack{1\le t\le [T/2]-h_1\\t+[T/2]\le s\le T-h_2}}\sum_{1\le i,j\le N}[\bm\Sigma]_{ii}[\bm\Sigma]_{ij}\{(\sum_{\substack{r\le t\wedge s\wedge t+h_1\\k\le s+h_2}}b_{t-r}b_{s-r}b_{t+h_1-r}b_{s+h_1-k}\\
    &+\sum_{\substack{r\le t\wedge s\wedge s+h_2\\k\le t+h_1}}b_{t-r}b_{s-r}b_{t+h_1-k}b_{s+h_1-r}+\sum_{\substack{r\le t\wedge t+h_1\wedge s+h_2\\k\le s}}b_{t-r}b_{s-k}b_{t+h_1-r}b_{s+h_1-r}\\
    &+\sum_{\substack{r\le  s\wedge t+h_1\wedge s+h_2\\k\le t}}b_{t-k}b_{s-r}b_{t+h_1-r}b_{s+h_1-r})z_{ir}^3z_{jk}+\sum_{r\le t\wedge t+h_1\wedge s+h_2}b_{t-r}b_{s-r}b_{t+h_1-r}b_{s+h_1-r}z_{ir}^3z_{jr}\}.
\end{align*}
Thus, by Condition (C2) (\romannumeral2) and the fact that $(T-h_2/2-\frac{3}{2}[T/2]+1/2)([T/2]-h_2)\ge T^2/9$ when $T$ is large enough, we have
\begin{align*}
    \mathbb{E}|\tilde{S}_{h_1,h_2,H}|=o(N^2T^{-2}).
\end{align*}
Furthermore, since $\tr(\bm\Sigma^2)\ge \tr^2(\bm\Sigma)/N\ge M_1^2N$ due to Condition (C3), we have
\begin{align*}
    \frac{\mathbb{E}\left|\tilde{S}_{0,0,H}+2\sum_{r=1}^M\tilde{S}_{0,r,H}+2\sum_{r=1}^M\tilde{S}_{r,0,H}+4\sum_{r=1}^M\sum_{s=1}^M\tilde{S}_{r,s,H}\right|}{\left(a_0^2+4\sum_{r=1}^Ma_0a_r+4\sum_{r=1}^M\sum_{s=1}^Ma_ra_s\right)\tr(\bm\Sigma^2)}=o(NT^{-2})=o(1).
\end{align*}

\noindent\textbf{Step 2.} Note that
\begin{align*}
    &(T-h_2/2-\frac{3}{2}[T/2]+1/2)([T/2]-h_2)\mathbb{E}(\tilde{S}_{h_1,h_2,L})\\
    =&\sum_{\substack{1\le t\le [T/2]-h_1\\t+[T/2]\le s\le T-h_2}}\{\sum_{1\le i,j\le N}[\bm\Sigma]_{ij}^2(\sum_{\substack{r\le t\wedge t+h_1\\k\le s\wedge s+h_2}}b_{t-r}b_{s-k}b_{t+h_1-r}b_{s+h_1-k}+\sum_{\substack{r\le t\wedge s+h_2\\k\le s\wedge t+h_1}}b_{t-r}b_{s-k}b_{t+h_1-k}b_{s+h_1-r})\\
    &+\sum_{1\le i,j\le N}[\bm\Sigma]_{ii}[\bm\Sigma]_{jj}\sum_{\substack{r\le t\wedge s\\k\le t+h_1\wedge s+h_2}}b_{t-r}b_{s-r}b_{t+h_1-k}b_{s+h_1-k}\\
    &+(\sum_{1\le i\ne j\le N}[\bm\Sigma]_{ii}[\bm\Sigma]_{jj}+2\sum_{1\le i\ne j\le N}[\bm\Sigma]_{ij}^2)\sum_{r\le t\wedge s\wedge t+h_1\wedge s+h_2}b_{t-r}b_{s-r}b_{t+h_1-r}b_{s+h_1-r}\}.
\end{align*}
Then, by Conditions (C2) (\romannumeral2) again, we have
\begin{align*}
    &(T-h_2/2-\frac{3}{2}[T/2]+1/2)([T/2]-h_2)\mathbb{E}(\tilde{S}_{h_1,h_2,L}-a_{h_1}a_{h_2}\tr(\bm\Sigma^2))\\
    =&\sum_{\substack{1\le t\le [T/2]-h_1\\t+[T/2]\le s\le T-h_2}}\{a_{s+h_2-t}a_{t+h_1-s}\tr(\bm\Sigma^2)+a_{t-s}a_{s-t+h_2-h_1}\tr^2(\bm\Sigma)\\
    &+(\sum_{1\le i\ne j\le N}[\bm\Sigma]_{ii}[\bm\Sigma]_{jj}+2\sum_{1\le i\ne j\le N}[\bm\Sigma]_{ij}^2)\sum_{r\le t\wedge s\wedge t+h_1\wedge s+h_2}b_{t-r}b_{s-r}b_{t+h_1-r}b_{s+h_1-r}\}\\
    =&o(N^2T^{-1}).
\end{align*}
Hence, $\mathbb{E}\{(\tilde{S}_{h_1,h_2,L}-a_{h_1}a_{h_2}\tr(\bm\Sigma^2))/\tr(\bm\Sigma^2)\}=o(NT^{-3})=o(1)$. Similarly, we can also verify that $\var\{(\tilde{S}_{h_1,h_2,L}-a_{h_1}a_{h_2}\tr(\bm\Sigma^2))/\tr(\bm\Sigma^2)\}=o(NT^{-2}+N^{-1})=o(1)$. Thus, $(\tilde{S}_{h_1,h_2,L}-a_{h_1}a_{h_2}\tr(\bm\Sigma^2))/\tr(\bm\Sigma^2)=o_p(1)$, which implies that 
\begin{align*}
    \frac{\tilde{S}_{0,0,L}+2\sum_{r=1}^M\tilde{S}_{0,r,L}+2\sum_{r=1}^M\tilde{S}_{r,0,L}+4\sum_{r=1}^M\sum_{s=1}^M\tilde{S}_{r,s,L}}{\left(a_0^2+4\sum_{r=1}^Ma_0a_r+4\sum_{r=1}^M\sum_{s=1}^Ma_ra_s\right)\tr(\bm\Sigma^2)}=1+o_p(1).
\end{align*}
In summary, we get $\tilde{\sigma}_T^2/\sigma_T^2\cp 1$.

According to the proof of Lemma \ref{error of e_t}, 
\begin{align*}
    |\tilde{S}_{h_1,h_2}-S_{h_1,h_2}|=&\left|\frac{\sum_{t=1}^{[T/2]-h_1}\sum_{s=t+[T/2]}^{T-h_2}\X_t^\top \X_s\X_{t+h_1}^\top \X_{s+h_2}}{(T-h_2/2-\frac{3}{2}[T/2]+1/2)([T/2]-h_2)}\left(1-\frac{\eta_t\eta_s\eta_{t+h_1}\eta_{s+h_2}}{e_te_se_{t+h_1}e_{s+h_2}}\right)\right|\\
    \le& CT^{-1}|\tilde{S}_{h_1,h_2}|,
\end{align*}
which implies that $\hat{\sigma}_T^2/\tilde{\sigma}_T^2=1+O_p(T^{-1})=1+o_p(1)$. In summary, we get $\hat{\sigma}_T^2/\sigma_T^2\cp 1$.

\subsubsection{Proof of Theorem 3}
\proof 
According to Lemmas \ref{consistence of mean estimator}-\ref{consistence of variance estimator}, we can easily obtain the result.

\subsection{Proof of Theorems 4-5}
Let $\D\equiv\diag\{\sigma_1,\cdots,\sigma_N\}$ be the diagonal matrix of the long-run covariance of $\bm\gamma_t$, $\O_M:=\Gamma_{0,M}+2\sum_{h=0}^M\Gamma_{h,M}$. Define $\bm R:=\D^{-1/2}\O_M\D^{-1/2}$. For a $d$-dimension vector $\bm Y=(Y_1,\cdots,Y_d)^\top$, define $\Vert\bm Y\Vert_{\infty}=\max_{1\le i\le d}|Y_i|$. For $X\in\mathcal{R}$, define $||X||_q=E[|X|^q]^{1/q}$.

\begin{lemma}\label{ld}
    Let $(Z_1,\cdots,Z_N)^\top$ be a zero mean multivariate normal random vector with covariance matrix $\Sigma=(\sigma_{ij})_{N\times N}$ and diagonal $\sigma_{ii}=1$ for $1\le i\le N$. Suppose that $\max_{1\le i<j\le N}|\sigma_{ij}|\le r<1$ and $\max_{1\leq j\leq N} \sum_{i=1}^N \sigma_{ij}^2\le c$ for some $r$ and $c$. Then for any $x\in \mathcal{R}$ as $N \to \infty$,
    \begin{align*}
        P\left(\max_{1\le i\le N}Z_i^2-2\log(N)+\log\{\log(N)\}\le x \right) \to \exp\left\{-\frac{1}{\sqrt{\pi}}\exp\left(-\frac{x}{2}\right)\right\}.
    \end{align*}
\end{lemma}
\proof See Lemma 6 in  \cite{TonyCai2014TwosampleTO}.

\begin{lemma}\label{md}
    Suppose $\left\{x_{i}\right\}_{i=1}^n$ is a p-dimensional M-dependent sequence. Let $n=(N+M)r,$ where $N \geq M$ and $N, M, r \rightarrow+\infty$ as $n \rightarrow+\infty$. Define the block sums
    \begin{align*}
        A_{i j}=\sum_{l=i N+(i-1) M-N+1}^{i N+(i-1) M} x_{l j},~~\text{and}~~B_{i j}=\sum_{l=i(N+M)-M+1}^{i(N+M)} x_{l j}.
    \end{align*}
    It is not hard to see that $\left\{A_{ij}\right\}_{i=1}^{r}$ and $\left\{B_{ij}\right\}_{i=1}^{r}$ with $1 \leq j \leq p$ are two sequences of i.i.d random variables. Let $V_{n j}=\sqrt{\sum_{i=1}^r(A_{i j}^{2}+B_{i j}^{2})}$. Assume that there exist $a_{j}, b_{j}>0$ such that
    \begin{align*}
        P\left(\sum_{i=1}^{n} x_{i j}>a_{j}\right) \leq 1 / 4, ~~ P\left(V_{n j}^{2}>b_{j}^{2}\right) \leq 1 / 4.
    \end{align*}
    Then we have
    \begin{align*}
        P\left(\left|\sum_{i=1}^{n} x_{i j}\right| \geq x\left(a_{j}+b_{j}+V_{n j}\right)\right) \leq 8 \exp \left(-x^{2} / 8\right)
    \end{align*}
    for any $1 \leq j \leq p .$ In particular, we can choose $b_{j}^{2}=4 \mathbb{E} V_{n j}^{2}$ and $a_{j}^{2}=2 b_{j}^{2}=8 \mathbb{E} V_{n j}^{2}$.
\end{lemma}
\proof See Lemma A.1 in \citet{ZC18}.

\begin{lemma}\label{error_1}
    Under Conditions (C1)-(C2) and (C7)-(C9), we have 
    \begin{align*}
        \mathbb{P}\left(T\left\| \D^{-1/2}\bar{\bm\gamma}_T\right\|_\infty^2-2\log(N)+\log\{\log(N)\}\right)\to \exp\left\{-\frac{1}{\sqrt{\pi}}\exp\left(-\frac{x}{2}\right)\right\}.
    \end{align*}
\end{lemma}
\proof Recalling the definition of $\bm \gamma_t:= \mathbb{E}(\bm \X_t|\z_{t-M},\cdots,\z_t)$ in the proof of Theorem 1, we can write it as $\bm\gamma_t=G(\mathcal{F}^t)$, where $\mathcal{F}^t=(\z_{t-M},\dots,\z_{t-1},\z_t)$ and $G=(g_1(\cdot),\dots,g_N(\cdot))^\top$ is an $\mathcal{R}^N$-valued measurable function. Define the functional dependence measure 
\begin{align*}
    \iota_{t,q,i}:=\Vert \gamma_{it}-g_i(\mathcal{F}^{t,\{0\}})\Vert_q,
\end{align*}
where $\mathcal{F}^{t,\{0\}}=(\z_{t-M},\dots,\z_{k-1},\z_k',\z_{k+1},\dots,\z_t)$ is a coupled version of $\mathcal{F}^t$ with $\z_k$ in $\mathcal{F}^t$ is replaced by $\z_k'$, and $\z_t,\z_t'$ are i.i.d. random elements. To account for the dependence in the process $\{\bm\gamma_{i.}\}_{i=1}^N$, we define the dependence adjusted norm
\begin{align*}
    \Vert\iota_{i.}\Vert_{q,a}:=\sup_{m\ge 0}(m+1)^a\sum_{t=m}^\infty \iota_{t,q,i}.
\end{align*}
To account for high dimensionality, we define the uniform and the overall dependence adjusted norm of $\{\bm\gamma_t\}_{t=1}^T$:
\begin{align*}
    \psi_{q,a}:=\max_{1\le i\le N}\Vert\iota_{i.}\Vert_{q,a},~~\text{and}~~\Upsilon_{q,a}:=\left(\sum_{i=1}^N\Vert\iota_{i.}\Vert_{q,a}^q\right)^{1/q}.
\end{align*}
Additionally, define the $\mathcal{L}^\infty$ functional dependence measure and its corresponding dependence adjusted norm of $\{\bm\gamma_t\}_{t=1}^T$:
\begin{align*}
    \omega_{t,q}:=\Vert \Vert\bm\gamma_t-G(\mathcal{F}^{t,\{0\}})\Vert_\infty \Vert_q~~\text{and}~~\Vert|\gamma_.|_\infty\Vert_{q,a}:=\sup_{m\ge 0}(1+m)^a\sum_{t=m}^\infty \omega_{t,q}.
\end{align*}
Following the same notations as \citet{ZW17}, define the following quantities 
\begin{align*}
    \Theta_{q,a}:=\Upsilon_{q,a}\wedge\{\Vert|\gamma_.|_\infty\Vert_{q,a}\log^{3/2}(N)\},~~&~~L_1:=\{\psi_{2,a}\psi_{2,0}\log^2(N)\}^{1/a},\\
    W_1:=(\psi_{3,0}^6+\psi_{4,0}^4)\log^7(NT),~~&~~W_2:=\psi_{2,a}^2\log^4(NT),\\
    N_1:=\{T/\log(N)\}^{q/2}/\Theta_{q,a}^q,~~&~~N_2:=T\log^{-2}(T)\psi_{2,a}^{-2}.
\end{align*}
Since $\{\bm\gamma\}_{t=1}^T$ is M-dependent and $\Vert\gamma_{it}\Vert_q$ is bounded for any positive integrate $q$ due to Condition (C7), we have
\begin{align*}
    \Theta_{q,a}=O\{M^{1+a}\log^2(N)\},~~&~~L_1=O[\{M^{2+a}\log^2(N)\}^{1/a}],\\
    W_1=O\{M^6\log^7(NT)\},~~&~~W_2=O\{M^{2a+2}\log^4(NT)\},\\
    N_1=O[\{T\log^{-5}(N)M^{-2a-2}\}^{q/2}],~~&~~
    N_2=O\{T\log^{-2}(N)M^{-2a-2}\}.
\end{align*}
Furthermore, by Condition (C9), we have $\Theta_{q,a}T^{1/q-1/2}\log^{3/2}(N)=o(1)$  and $L_1\max(W_1,W_2)=o(1)\min(N_1,N_2)$ for some $q\ge 4$ and $a>1/2-1/q$. So by Theorem 3.2 in \citet{ZW17}, we have
\begin{align*}\label{B.1}
    \sup_{x\ge 0}\left|\mathbb{P}\left(T\left\| \D^{-1/2}\bar{\bm\gamma}_T\right\|_\infty^2\ge x\right)-\mathbb{P}\left(\Vert\bm Z\Vert_\infty^2\ge x\right)\right|\to 0,\tag{B.1}
\end{align*}
where $\bm Z\sim N(\bm 0,\bm R)$.

By Condition (C8) and Lemma \ref{ld}, we have
\begin{align*}\label{B.2}
    \mathbb{P}(\Vert\bm Z\Vert_\infty^2-2\log(N)+\log\{\log(N)\}\le x)\to F(x).\tag{B.2}
\end{align*}
In summary, the desired result comes from \eqref{B.1} and \eqref{B.2}.

\begin{lemma}\label{error_2}
    Under Conditions (C1)-(C2) and (C7)-(C9), we have
    \begin{align*}
        T\left\|\D^{-1/2}(\bar{\X}_T-\bar{\bm\gamma}_T)\right\|_\infty^2=\max_{1\le i\le N}\frac{1}{T\sigma_i}\left\{\sum_{t=1}^T(X_{it}-\gamma_{it})\right\}^2=o_p(1).
    \end{align*}
\end{lemma}
\proof Let $[\bm\Sigma^{1/2}]_{(i)}$ be the $i$-th row of $\bm\Sigma^{1/2}$, then $X_{it}-\gamma_{it}=\sum_{k\le t-M+1}b_{t-k}[\bm\Sigma^{1/2}]_{(i)}\z_k$. Using the independence of $z_{it}$'s, we have
\begin{align*}
    \var\left\{\sum_{t=1}^T(X_{it}-\gamma_{it})\right\}=&\sum_{t_1=1}^T\sum_{t_2=1}^T\sum_{k_1\le t_1-M+1}\sum_{k_2\le t_2-M+1}b_{t_1-k_1}b_{t_2-k_2}\mathbb{E}([\bm\Sigma^{1/2}]_{(i)}\z_{k_1}[\bm\Sigma^{1/2}]_{(i)}\z_{k_2})\\
    =&\sum_{t_1=1}^T\sum_{t_2=1}^T\sum_{k\le (t_1\wedge t_2)-M+1}b_{t_1-k}b_{t_2-k}[\bm\Sigma]_{ii}\\
    \le &T^2\left(\sum_{k=M+1}^\infty |b_k|\right)^2[\bm\Sigma]_{ii}=o(T^2M^{-8})
\end{align*}
due to \eqref{A.10}. Thus, $T\left\|\D_0^{-1/2}(\bar{\X}_T-\bar{\bm\gamma}_T)\right\|_\infty^2=o_p(T^{-1}\cdot T^2M^{-8})=o_p(1)$.

\begin{lemma}\label{error_3}
    Under Conditions (C1)-(C2) and (C7)-(C9), we have
    \begin{align*}
        \max_{1\le i\le N}\frac{T(\hat{\alpha_i}-\tilde{\alpha}_i)^2}{\sigma_i}=\max_{1\le i\le N}T^{-1}\sigma_i^{-1}\left\{\sum_{t=1}^TX_{it}(\zeta_t/e_t-1)\right\}^2=O_p\{\sqrt{T^{-1}\log(N)}\}.
    \end{align*}
\end{lemma}
\proof Note that
\begin{align*}
    \max_{1\le i\le N}T^{-1}\sigma_i^{-1}\left\{\sum_{t=1}^TX_{it}(\zeta_t/e_t-1)\right\}^2\le&\max_{1\le i\le N}T^{-1}\sigma_i^{-1}\left\{\sum_{t=1}^T(X_{it}-\gamma_{it})(\zeta_t/e_t-1)\right\}^2\\
    &\quad+\max_{1\le i\le N}T^{-1}\sigma_i^{-1}\left\{\sum_{t=1}^T\gamma_{it}(\zeta_t/e_t-1)\right\}^2.
\end{align*}
For the first term on the right side of the above inequality, by Cauchy-Schwarz inequality, we have
\begin{align*}
    \left\{\sum_{t=1}^T(X_{it}-\gamma_{it})(\zeta_t/e_t-1)\right\}^2\le \sum_{t=1}^T(X_{it}-\gamma_{it})^2\cdot \sum_{s=1}^T(\zeta_s/e_s-1)^2.
\end{align*}
By \eqref{A.11}, we have $\sum_{s=1}^T(\zeta_s/e_s-1)^2=O_p(1)$. Next, following the proof of Lemma \ref{M-dependence_approximation}, we analyze $\sum_{t=1}^T(X_{it}-\gamma_{it})^2$.
\begin{align*}
    &\var\left\{\sum_{t=1}^T(X_{it}-\gamma_{it})^2\right\}\\
    =&\sum_{\substack{1\le t_1,t_2\le T\\(k_1,k_2,k_3,k_4)\in\mathcal{S} }}b_{t_1-k_1}b_{t_1-k_2}b_{t_2-k_3}b_{t_2-k_4}\mathbb{E}\{([\bm\Sigma^{1/2}]_{(i)}\z_{k_1}[\bm\Sigma^{1/2}]_{(i)}\z_{k_2}-\mathbb{I}_{(k_1=k_2)}[\bm\Sigma]_{ii})\\
    &\qquad\times ([\bm\Sigma^{1/2}]_{(i)}\z_{k_3}[\bm\Sigma^{1/2}]_{(i)}\z_{k_4}-\mathbb{I}_{(k_3=k_4)}[\bm\Sigma]_{ii})\}\\
    =&\sum_{\substack{1\le t_1,t_2\le T\\(k_1,k_2,k_3,k_4)\in\mathcal{S}\bigcap\mathcal{S}_1 }}b_{t_1-k_1}b_{t_1-k_1}b_{t_2-k_3}b_{t_2-k_3}\mathbb{E}[\{([\bm\Sigma^{1/2}]_{(i)}\z_{k_1})^2-[\bm\Sigma]_{ii}\}\{([\bm\Sigma^{1/2}]_{(i)}\z_{k_3})^2-[\bm\Sigma]_{ii}\}]\\
    &+\sum_{\substack{1\le t_1,t_2\le T\\(k_1,k_2,k_3,k_4)\in\mathcal{S}\bigcap\mathcal{S}_2 }}b_{t_1-k_1}b_{t_1-k_2}b_{t_2-k_1}b_{t_2-k_2}\mathbb{E}\{([\bm\Sigma^{1/2}]_{(i)}\z_{k_1}[\bm\Sigma^{1/2}]_{(i)}\z_{k_2})^2\}\\
    &+\sum_{\substack{1\le t_1,t_2\le T\\(k_1,k_2,k_3,k_4)\in\mathcal{S}\bigcap\mathcal{S}_3 }}b_{t_1-k_1}b_{t_1-k_2}b_{t_2-k_2}b_{t_2-k_1}\mathbb{E}\{([\bm\Sigma^{1/2}]_{(i)}\z_{k_1}[\bm\Sigma^{1/2}]_{(i)}\z_{k_2})^2\}\\
    &+\sum_{\substack{1\le t_1,t_2\le T\\(k_1,k_2,k_3,k_4)\in\mathcal{S}\bigcap\mathcal{S}_4 }}b_{t_1-k_1}b_{t_1-k_1}b_{t_2-k_1}b_{t_2-k_1}\mathbb{E}[\{([\bm\Sigma^{1/2}]_{(i)}\z_{k_1})^2-[\bm\Sigma]_{ii}\}^2]\\
    =&\left(\sum_{\substack{1\le t_1,t_2\le T\\(k_1,k_2,k_3,k_4)\in\mathcal{S}\bigcap\mathcal{S}_2 }}b_{t_1-k_1}b_{t_1-k_2}b_{t_2-k_1}b_{t_2-k_2}+\sum_{\substack{1\le t_1,t_2\le T\\(k_1,k_2,k_3,k_4)\in\mathcal{S}\bigcap\mathcal{S}_3 }}b_{t_1-k_1}b_{t_1-k_2}b_{t_2-k_2}b_{t_2-k_1}\right)[\bm\Sigma]_{ii}^2\\
    &+\sum_{\substack{1\le t_1,t_2\le T\\(k_1,k_2,k_3,k_4)\in\mathcal{S}\bigcap\mathcal{S}_4 }}b_{t_1-k_1}b_{t_1-k_1}b_{t_2-k_1}b_{t_2-k_1}\left(\mu_4\sum_{j=1}^N[\bm\Sigma^{1/2}]_{ij}^4+5[\bm\Sigma]_{ii}^2\right)\\
    \lesssim&\sum_{t_1,t_2\ge 0}|b_{t_1}b_{t_2}|\sum_{s_1,s_2\ge M}|b_{s_1}b_{s_2}|+\sum_{t_1\ge 0,s_2\ge M}|b_{t_1}b_{s_2}|\sum_{t_2\ge 0,s_1\ge M}|b_{t_2}b_{s_1}|+\frac{N}{T}\sum_{t_1,t_2\ge 0}\sum_{s_1,s_2\ge M}|b_{t_1}b_{t_2}b_{s_1}b_{s_2}|\\
    =&o(M^{-8}),
\end{align*}
where $\mathcal{S}=\{(k_1,k_2,k_3,k_4):k_1,k_2\le t_1-M+1,k_3,k_4\le t_2-M+1\}$, $\mathcal{S}_1=\{(k_1,k_2,k_3,k_4):k_1=k_2\ne k_3=k_4\}$, $\mathcal{S}_2=\{(k_1,k_2,k_3,k_4):k_1=k_3\ne k_2=k_4\}$, $\mathcal{S}_3=\{(k_1,k_2,k_3,k_4):k_1=k_4\ne k_2=k_3\}$ and $\mathcal{S}_4=\{(k_1,k_2,k_3,k_4):k_1=k_2= k_3=k_4\}$. Therefore, $\sum_{t=1}^T(X_{it}-\gamma_{it})^2=o_p(M^{-4})$. Furthermore, we get $\max_{1\le i\le N}T^{-1}\sigma_i^{-1}\left\{\sum_{t=1}^T(X_{it}-\gamma_{it})(\zeta_t/e_t-1)\right\}^2=o_p(T^{-1}M^{-4})$. 

For the second term, by Cauchy-Schwarz inequality again, we have
\begin{align*}
    \left\{\sum_{t=1}^T\gamma_{it}(\zeta_t/e_t-1)\right\}^2\le\sum_{t=1}^T\gamma_{it}^2\cdot \sum_{s=1}^T(\zeta_t/e_t-1)^2.
\end{align*}
Let $T=(M^*+M)r$ where $M^*\ge M$ and $M^*,M,r\to\infty$ as $T\to\infty$. Define the block sums
\begin{align*}
    A_{it}=\sum_{l=(t-1)(M^*+M)+1}^{tM^*+(t-1)M}\gamma_{il}^2,~~\text{and}~~B_{it}=\sum_{l=tM^*+(t-1)M+1}^{t(M^*+M)}\gamma_{il}^2.
\end{align*}
Since $\{\bm\gamma_t\}_{t=1}^T$ is M-dependent, $\{A_{il}\}_{l=1}^r$ and $\{B_{il}\}_{l=1}^r$ with $1\le i\le N$ are two sequence of i.i.d. random variables. Let $V_{iT}=\{\sum_{l=1}^r(A_{il}^2+B_{il}^2)\}^{1/2}$. Then, by Lemma \ref{md}, we have
\begin{align*}
    \mathbb{P}\left(\sum_{t=1}^T\gamma_{it}^2\ge x\{(2+2\sqrt{2})\sqrt{\mathbb{E}(V_{iT}^2)}+V_{iT}\}\right)\le 8\exp(-x^2/8).
\end{align*}
Note that $(2+2\sqrt{2})\sqrt{\mathbb{E}(V_{iT}^2)}+V_{iT}\le CT^{1/2}$ for some positive constant $C$, then we have
\begin{align*}
    \mathbb{P}\left(\max_{1\le i\le N}\sum_{t=1}^T\gamma_{it}^2\ge Cx\right)\le \sum_{i=1}^N\mathbb{P}\left(\sum_{t=1}^T\gamma_{it}^2\ge Cx\right)\le 8N\exp(-\frac{x^2}{8T})\to 0
\end{align*}
by setting $x=10\sqrt{T\log(N)}$. Hence, $\max_{1\le i\le N}\sum_{t=1}^T\gamma_{it}^2=O_p\{\sqrt{T\log(N)}\}$ and 
\begin{align*}
    \max_{1\le i\le N}T^{-1}\sigma_i^{-1}\left\{\sum_{t=1}^T\gamma_{it}(\zeta_t/e_t-1)\right\}^2=O_p\{\sqrt{T^{-1}\log(N)}\}.
\end{align*}
In summary, we have
\begin{align*}
    \max_{1\le i\le N}\frac{T(\hat{\alpha_i}-\tilde{\alpha}_i)^2}{\sigma_i}=o_p(T^{-1}M^{-4})+O_p\{\sqrt{T^{-1}\log(N)}\}=O_p\{\sqrt{T^{-1}\log(N)}\}.
\end{align*}

\begin{lemma}\label{error_4}
    Under Conditions (C1)-(C2) and (C7)-(C9), we have
    \begin{align*}
        \max_{1\le i\le N}|\hat{\sigma}_i-\sigma_i|=O_p(T^{-\delta}) 
    \end{align*}
    for sufficiently some $\delta>0$.
\end{lemma}
\proof According to \eqref{A.13}, we can rewrite $\hat{\phi}_{i,h}$ as
\begin{align*}
    \hat{\phi}_{i,h}=&\frac{1}{T-h}\sum_{t=h+1}^T\{\varepsilon_{it}\varepsilon_{i,t-h}+(\varepsilon_{it}+\varepsilon_{i,t-h})(\alpha_i-\hat{\alpha}_i)+\varepsilon_{it}(\bm\beta_i-\hat{\bm\beta}_i)^\top \f_{t-h}+\varepsilon_{i,t-h}(\bm\beta_i-\hat{\bm\beta}_i)^\top \f_t\\
    & +(\alpha_i-\hat{\alpha}_i)^2+(\alpha_i-\hat{\alpha}_i)(\bm\beta_i-\hat{\bm\beta}_i)^\top(\f_t+\f_{t-h})+(\bm\beta_i-\hat{\bm\beta}_i)^\top \f_t(\bm\beta_i-\hat{\bm\beta}_i)^\top \f_{t-h}\}\eta_t\eta_{t-h}.
\end{align*}
Define $\tilde{\phi}_{i,h}:=\frac{1}{T-h}\sum_{t=h+1}^T\varepsilon_{it}\varepsilon_{i,t-h}\eta_t\eta_{t-h}$, $\phi_{i,h}^{NG}:=\frac{1}{T-h}\sum_{t=h+1}^T\gamma_{it}\gamma_{i,t-h}$ and $\phi_{i,h}:=\mathbb{E}(\gamma_{i0}\gamma_{ih})$, then $\sigma_i=\sum_{h\in\mathcal{M}}\phi_{i,h}$ and $\hat{\sigma}_i-\sigma_i=\sum_{h\in\mathcal{M}}(\hat{\phi}_{i,h}-\phi_{i,h})$. By \eqref{A.14}, we have
\begin{align*}
    \sqrt{T}\max_{1\le i\le N}\left\|\hat{\phi}_{i,h}-\tilde{\phi}_{i,h}\right\|_{q/2}=O\{\log(N)T^{-1/2}\}=o(1).
\end{align*}
By Lemma E.5 in \citet{J15}, we have for any positive integrate $q>4$,
\begin{align*}
    \sqrt{T}\max_{1 \leq i \leq N}\left\|\phi_{i,h}^{NG}-\phi_{i,h}\right\|_{q/2}=O(1).
\end{align*}
Following the proof of Lemma \ref{M-dependence_approximation}, we can prove $\frac{1}{T-h}\sum_{t=h+1}^TX_{it}X_{i,t-h}-\phi_{i,h}^{NG}=o_p(M^{-4})$, which together with \eqref{A.11}, yields that 
\begin{align*}
    \sqrt{T}\max_{1 \leq i \leq N}\left\|\phi_{i,h}^{NG}-\tilde{\phi}_{i,h}\right\|_{q/2}=o(T^{1/2}M^{-4})=o(1).
\end{align*}
In summary, we have 
\begin{align*}
    \sqrt{T}\max_{1 \leq i \leq N}\left\|\hat{\phi}_{i,h}-\phi_{i,h}\right\|_{q/2}=O(1).
\end{align*}
It follows that for large enough $N, T,$ we have
\begin{align*}
    \mathbb{P}\left(\max_{1\le i\le N}|\hat{\sigma}_i-\sigma_i|\ge T^{-\delta}\right)
    \le&\sum_{i=1}^N\mathbb{P}\left(\left|\sum_{h\in\mathcal{M}}(\hat{\phi}_{i,h}-\phi_{i,h})\right|\ge T^{-\delta}\right)\\
    \le&\sum_{i=1}^N\sum_{h=1}^M\mathbb{P}\left(\left|\hat{\phi}_{i,h}-\phi_{i,h}\right|\ge \frac{T^{-\delta}}{2M}\right)\\
    \le&\sum_{i=1}^N\sum_{h=1}^M\left(\frac{T^{-\delta}}{2M}\right)^{-q/2}\left\|\hat{\phi}_{i,h}-\phi_{i,h}\right\|_{q/2}^{q/2},
\end{align*}
which is bounded by $NT^{\delta q/2}M^{q/2+1}T^{-q/4}\lesssim T^{-C}$, $C>0$ for sufficiently small $\delta>0$ and large $q$.

\subsubsection{Proof of Theorem 4}
\proof Note that
\begin{align*}\label{B.3}
    &\left|\max_{1\le i\le N}\frac{T(\hat{\alpha}_i-\alpha_i)^2}{\hat{\sigma}_i}-\max_{1\le i\le N}\frac{T(\hat{\alpha}_i-\alpha_i)^2}{\sigma_i}\right|\\
    \le &\max_{1\le i\le N}\frac{T(\hat{\alpha}_i-\alpha_i)^2}{\sigma_i}\cdot\max_{1\le i\le N}\left|\frac{\sigma_i}{\hat{\sigma}_i}-1\right|\\
    \le&2\left\{\max_{1\le i\le N}\frac{T(\hat{\alpha}_i-\tilde{\alpha}_i)^2}{\sigma_i}+\max_{1\le i\le N}\frac{T(\tilde{\alpha}_i-\alpha_i)^2}{\sigma_i}\right\}\cdot\max_{1\le i\le N}\left|\frac{\sigma_i}{\hat{\sigma}_i}-1\right|,\tag{B.3}
\end{align*}
\begin{align*}\label{B.4}
    &\left|\max_{1\le i\le N}\frac{T(\hat{\alpha}_i-\alpha_i)^2}{\sigma_i}-\max_{1\le i\le N}\frac{T(\tilde{\alpha}_i-\alpha_i)^2}{\sigma_i}\right|\\
    \le&\max_{1\le i\le N}\frac{T(\hat{\alpha}_i-\tilde{\alpha}_i)^2}{\sigma_i}+2\left\{\max_{1\le i\le N}\frac{T(\tilde{\alpha}_i-\alpha_i)^2}{\sigma_i}\cdot\max_{1\le i\le N}\frac{T(\hat{\alpha}_i-\tilde{\alpha}_i)^2}{\sigma_i}\right\}^{1/2},\tag{B.4}
\end{align*}
and
\begin{align*}\label{B.5}
    \max_{1\le i\le N}\frac{T(\tilde{\alpha}_i-\alpha_i)^2}{\sigma_i}=T\left\|\D^{-1/2}\bar{\X}_T\right\|_\infty^2\le T\left\|\D^{-1/2}\bar{\bm\gamma}_T\right\|_\infty^2+T\left\|\D^{-1/2}(\bar{\X}_T-\bar{\bm\gamma}_T)\right\|_\infty^2.\tag{B.5}
\end{align*}
By Lemmas \ref{error_1}-\ref{error_4}, we have
\begin{align*}
    P\left(\max_{1\le i \le N}\frac{T(\hat{\alpha}_i-\alpha_i)^2}{\hat{\sigma}_i}-2\log(N)+\log\{\log(N)\}\le x \right) \to \exp\left\{-\frac{1}{\sqrt{\pi}}\exp\left(-\frac{x}{2}\right)\right\}.
\end{align*}
Here we complete the proof.

\subsection{Proof of Theorem 5}
\proof According to the proof of Theorem 4, we have
\begin{align*}
    P\left(\max_{1\le i \le N}\frac{T(\hat{\alpha}_i-\alpha_i)^2}{\hat{\sigma}_i}-2\log(N)+\log\{\log(N)\}\le x \right) \to \exp\left\{-\frac{1}{\sqrt{\pi}}\exp\left(-\frac{x}{2}\right)\right\}.
\end{align*}
Thus,
\begin{align*}
    P\left(\max_{1\le i \le N}\frac{T(\hat{\alpha}_i-\alpha_i)^2}{\hat{\sigma}_i}\le 2\log(N)-\frac{1}{2}\log\{\log(N)\} \right) \to 1,
\end{align*}
when we set $x=\frac{1}{2}\log \{\log (N)\}$.  By the triangle inequality, we have
\begin{align*}
    \max_{1\le i \le N}\frac{T\hat{\alpha}_i^2}{\hat{\sigma}_i} \ge&  \max_{1\le i \le N}\frac{T{\alpha}_i^2}{2\hat{\sigma}_i} -\max_{1\le i \le N}\frac{T(\hat{\alpha}_i-\alpha_i)^2}{\hat{\sigma}_i}\\
    \ge &\max_{1\le i \le N}\frac{T{\alpha}_i^2}{2{\sigma}_i} -\max_{1\le i \le N}\frac{T(\hat{\alpha}_i-\alpha_i)^2}{\hat{\sigma}_i}-\max_{1\le i\le N}\left(1-\frac{\sigma_i}{\hat{\sigma}_i}\right)\frac{T\alpha_i^2}{2\sigma_i}\\
    \ge&\max_{1\le i \le N}\frac{T{\alpha}_i^2}{2{\sigma}_i}-2\log(N)+\frac{1}{2}\log\{\log(N)\}-T^{-\delta}\max_{1\le i\le N}\frac{T\alpha_i^2}{2\sigma_i}
\end{align*}
with probability tending to one by Lemma \ref{error_4}. If $\max_{1\le i \le N}\frac{T{\alpha}_i^2}{2{\sigma}_i}=O\{\log(N)\}$,
\begin{align*}
    \max_{1\le i \le N}\frac{T\hat{\alpha}_i^2}{\hat{\sigma}_i}\ge& 4\log(N)-2\log(N)+\frac{1}{2}\log\{\log(N)\}-O(T^{-\delta}\log(N))\\
    \ge& 2\log(N)-\log\{\log(N)\}+q_{\gamma},
\end{align*}
which implies $P(\Phi_{\gamma}=1)\to 1$. If $\log^{-1}(N)\max_{1\le i \le N}\frac{T{\alpha}_i^2}{2{\sigma}_i} \to \infty$,
\begin{align*}
    \max_{1\le i \le N}\frac{T\hat{\alpha}_i^2}{\hat{\sigma}_i} \ge C\max_{1\le i \le N}\frac{T\hat{\alpha}_i^2}{\sigma_i} \ge 2\log(N)-\log\{\log(N)\}+q_{\gamma},
\end{align*}
where the first inequality are obtained by Lemma \ref{error_4}. Thus, $P(\Phi_{\gamma}=1)$ converge to one. Following the proof of Lemma \ref{error of variance}, we can prove $\sigma_i/\sigma_i^0=1+o(1)$. Then we complete the proof.

\subsection{Proof of Theorems 6-7}
\begin{lemma}\label{as}
    Let $\{(U,U_N,\widetilde{U}_N)\in\mathbb{R}^3;N\geq 1\}$ and $\{(V,V_N,\widetilde{V}_N)\in\mathbb{R}^3;N\geq 1\}$ be two sequences of random variables with $U_N\stackrel{d}{\to} U$ and $V_N\stackrel{d}{\to} V$ as $N\to\infty$. Assume $U$ and $V$ are continuous random variables. We assume that 
    \begin{equation*}
    	\widetilde{U}_N=U_N+o_p(1)~~\text{and}~~ \widetilde{V}_N=V_N+o_p(1).
    \end{equation*}
    If $U_N$ and $V_N$ are asymptotically independent, then $\widetilde{U}_N$ and $\widetilde{V}_N$ are also asymptotically independent.
\end{lemma}
\proof See Lemma 7.10 in \citet{Feng2022AsymptoticIO}.

\begin{lemma}\label{in}
    Let $\bm{X}\sim N(\bm{\mu},\bm{\Sigma})$ with  invertible $\bm{\Sigma}$, and partition $\bm{X}$, $\bm{\mu}$ and $\bm{\Sigma}$ as 
    \begin{equation*}
    	\bm{X}=\left(\begin{array}{l}
    	\bm{X}_1\\\bm{X}_2
    	\end{array} \right), \quad
    	\bm{\mu}=\left(\begin{array}{l}
    	\bm{\mu}_1\\\bm{\mu}_2
    	\end{array} \right),\quad
    	\bm{\Sigma}=\left(\begin{array}{ll}
    	\bm{\Sigma}_{11}&\bm{\Sigma}_{12}\\
    	\bm{\Sigma}_{21}&\bm{\Sigma}_{22}
    	\end{array} \right).
    \end{equation*}
    Then $\bm{X}_2-\bm{\Sigma}_{21}\bm{\Sigma}_{11}^{-1}\bm{X}_1\sim N(\bm{\mu}_2-\bm{\Sigma}_{21}\bm{\Sigma}_{11}^{-1}\bm{\mu}_1,\bm{\Sigma}_{22\cdot 1})$ and is independent of $\bm{X}_1$, where $\bm{\Sigma}_{22\cdot 1}=\bm{\Sigma}_{22}-\bm{\Sigma}_{21}\bm{\Sigma}_{11}^{-1}\bm{\Sigma}_{12}$.
\end{lemma}
\proof See Theorem 1.2.11 in \citet{Muirhead1982AspectsOM}.

\subsubsection{Proof of Theorem 6}
\proof According to \eqref{A.12} and Lemmas \ref{clt}-\ref{consistence of variance estimator}, we have
\begin{align*}
    \frac{T_{\SUM}-\hat{\mu}_T}{\hat{\sigma}_T}=\frac{\bar{\bm\gamma}_T^\top\bar{\bm\gamma}_T-T^{-1}\tr(\O_{T,M})}{\sqrt{2T^{-2}\tr(\O_{T,M}^2)}}+o_p(1).
\end{align*}
Recalling the definitions in \eqref{A.1} and \eqref{A.3}, according to the proof of \eqref{A.7}-\eqref{A.9}, we have
\begin{align*}\label{C.1}
    \frac{\bar{\bm\gamma}_T^\top\bar{\bm\gamma}_T-T^{-1}\tr(\O_{T,M})}{\sqrt{2T^{-2}\tr(\O_{T,M}^2)}}=W(\bm\xi_1,\dots,\bm\xi_{q_T})+o_p(1).\tag{C.1}
\end{align*}
Here, $\bm\xi_t,t=1,\dots,q_t$ are independent and
\begin{align*}
    W(\bm\xi_1,\dots,\bm\xi_{q_T})=\frac{1}{T^2}\sum_{1\le t<s\le q_T}\sum_{k=(t-1)w_T+1}^{tw_T-M}\sum_{l=(s-1)w_T+1}^{sw_T-M}\bm\gamma_k^\top\bm\gamma_l/\sqrt{2T^{-2}\tr(\O_{T,M}^2)}.
\end{align*}
According to \eqref{B.3}-\eqref{B.5} and Lemmas \ref{error_1}-\ref{error_4}, we have
\begin{align*}\label{C.2}
    T_{\MAX}=T\left\|\D^{-1/2}\bar{\bm\gamma}_T\right\|_\infty^2+o_p(1).\tag{C.2}
\end{align*}
Hence, by Lemma \ref{as}, it suffices to show that $W(\bm\xi_1,\dots,\bm\xi_{q_T})$ and $T\left\|\D^{-1/2}\bar{\bm\gamma}_T\right\|_\infty^2$ are asymptotically independent.

\noindent\textbf{Case \uppercase\expandafter{\romannumeral1}.} Investigate the asymptotic independence when $\bm\gamma_t$'s are M-dependent Gaussian.

For any fixed $x,y\in\mathbb{R}$, define 
\begin{align*}
    &A_N=A_N(x):=\left\{W(\bm\xi_1,\dots,\bm\xi_{q_T})\le x\right\},~~\text{and}\\
    &B_i=B_i(y):=\left\{T^{-1}\sigma_i^{-1}\left(\sum_{t=1}^T\gamma_{it}\right)^2>2\log(N)-\log\{\log(N)\}+y\right\}
\end{align*}
for $i=1,...,N$. Then $\mathbb{P}(A_N)\to\Phi(x)$ and $\mathbb{P}(\cup_{i=1}^NB_i)\to1-F(y)$. Our goal is to prove that 
\begin{equation*}
    \mathbb{P}\left(\bigcup\limits_{i=1}^NA_NB_i \right)\to \Phi(x)\{1-F(y)\}. 
\end{equation*}
    	
For each $d\ge 1$, define
\begin{align*}
    &\zeta(N,d):=\sum_{1\le i_1<...<i_d\le N}|\mathbb{P}(A_NB_{i_1}...B_{i_d})-\mathbb{P}(A_N)\mathbb{P}(B_{i_1}...B_{i_d}) |, \\
    &H(N,d):=\sum_{1\le i_1<...<i_d\le N}|\mathbb{P}(B_{i_1}...B_{i_d})|. 
\end{align*}
By the inclusion-exclusion principle, we observe that for any integer $k\ge1$,
\begin{align*}
    &\mathbb{P}\left(\bigcup\limits_{i=1}^NA_NB_i \right)\\
    \le& \sum_{1\le i_1\le N}\mathbb{P}(A_NB_{i_1})-\sum_{1\le i_1<i_2\le N}\mathbb{P}(A_NB_{i_1}B_{i_2})+...+\sum_{1\le i_1<...<i_{2k+1}\le N}\mathbb{P}(A_NB_{i_1}...B_{i_{2k+1}})\\
    \le& \mathbb{P}(A_N)\left\lbrace \sum_{1\le i_1\le N}\mathbb{P}(B_{i_1})-\sum_{1\le i_1<i_2\le N}\mathbb{P}(B_{i_1}B_{i_2})+...-\sum_{1\le i_1<...<i_{2k}\le N}\mathbb{P}(B_{i_1}...B_{i_{2k}})\right\rbrace  \\
    &\qquad+\sum_{d=1}^{2k}\zeta(N,d)+H(N,2k+1)\\
    \le& \mathbb{P}(A_N)\mathbb{P}\left(\bigcup\limits_{i=1}^NB_i \right)+\sum_{d=1}^{2k}\zeta(N,d)+H(N,2k+1).
\end{align*} 
According to the proof of Lemma \ref{error_1} and the proof of Lemma 6 in \citet{CLX14}, we have for each $d$,
\begin{equation*}
    \lim\limits_{N\to\infty}H(N,d)=\pi^{-d/2}\frac{1}{d!}\exp\left(-\frac{dx}{2} \right). 
\end{equation*}
We claim that for each $d$,
\begin{equation*}\label{C.3}
    \lim\limits_{N\to\infty}\zeta(N,d)\to 0.\tag{C.3}
\end{equation*}
Then, by letting $k\to \infty$, we have 
\begin{equation*}
    \limsup\limits_{N\to\infty}\mathbb{P}\left(\bigcup\limits_{i=1}^NA_NB_i \right)\le \Phi(x)\{1-F(y)\}.
\end{equation*}
Likewise, we have 
\begin{equation*}
    \liminf\limits_{N\to\infty}\mathbb{P}\left(\bigcup\limits_{i=1}^NA_NB_i \right)\ge \Phi(x)\{1-F(y)\}.
\end{equation*}
Hence, the desired result follows.

It remains to prove that the claim \eqref{C.3} indeed holds. For each $t$, let $\bm{\gamma}_{(1),t}=(\gamma_{i_1,t},...,\gamma_{i_d,t})'$, $\bm{\gamma}_{(2),t}=(\gamma_{i_{d+1},t},...,\gamma_{i_N,t})'$, and for $a,b\in\{1,2\}$, let $\bm{\Sigma}_{ab}=\cov(\bm{\gamma}_{(a),t},\bm{\gamma}_{(b),t})$. By Lemma \ref{in}, $\bm{\gamma}_{(2),t}$ can be decomposed as $\bm{\gamma}_{(2),t}=\bm{U}_t+\bm{V}_t$, where $\bm{U}_t=\bm{\gamma}_{(2),t}-\bm{\Sigma}_{21}\bm{\Sigma}_{11}^{-1}\bm\gamma_{(1),t}$ and $\bm{V}_t=\bm{\Sigma}_{21}\bm{\Sigma}_{11}^{-1}\bm{\gamma}_{(1),t}$ satisfying 
\begin{equation*}
    \bm{U}_t\sim N(\bm{0},\bm{\Sigma}_{22}-\bm{\Sigma}_{21}\bm{\Sigma}_{11}^{-1}\bm{\Sigma}_{12}),\quad\bm{V}_t\sim N(\bm{0},\bm{\Sigma}_{21}\bm{\Sigma}_{11}^{-1}\bm{\Sigma}_{12})~~\text{and}~~
    \bm{U}_t~~\text{and}~~\bm{\gamma}_{(1),t}~~\text{are independent.}
\end{equation*}
Thus, we have
\begin{align*}
    &\frac{1}{T^2}\sum_{1\le t<s\le q_T}\sum_{k=(t-1)w_T+1}^{tw_T-M}\sum_{l=(s-1)w_T+1}^{sw_T-M}\bm\gamma_k^\top\bm\gamma_l\\
    =&\frac{1}{T^2}\sum_{1\le t<s\le q_T}\sum_{k=(t-1)w_T+1}^{tw_T-M}\sum_{l=(s-1)w_T+1}^{sw_T-M}\left\{\bm U_k^\top\bm U_l+\bm\gamma_{(1),k}^\top\bm\gamma_{(1),l}+2\bm V_k^\top\bm U_l+\bm V_k^\top\bm V_l\right\}\\
    =:&\varphi^*_1+\Theta_1+\Theta_2+\Theta_3=:\varphi^*_1+\varphi^*_2.
\end{align*}
    	
We claim that for any $\epsilon>0$, $\exists$ a sequence of constants $c:=c_N>0$ with $c_N\to \infty$ s.t.  
\begin{align*}\label{C.4}
    \mathbb{P}\left(|\Theta_k|\ge\epsilon\sqrt{2T^{-2}\tr(\O_{T,M}^2)}\right)\le N^{-c}, k=1,2,3,\tag{C.4}
\end{align*}
for sufficiently large $N$. Consequently, $\mathbb{P}(|\varphi^*_2/\sqrt{2T^{-2}\tr(\O_{T,M}^2)}|\ge\epsilon)\le N^{-c}$. Furthermore,
\begin{align*}
    &\mathbb{P}\left(A_N(x)B_{i_1}...B_{i_d}\right)\\
    \le&\mathbb{P}\left(A_N(x)B_{i_1}...B_{i_d}, |\varphi^*_2/\sqrt{2T^{-2}\tr(\O_{T,M}^2)}|<\epsilon\right)+N^{-c}\\
    \le&\mathbb{P}\left(|\varphi^*_1/\sqrt{2T^{-2}\tr(\O_{T,M}^2)}|<\epsilon+x, B_{i_1}...B_{i_d} \right)+N^{-c}\\
    =&\mathbb{P}\left(|\varphi^*_1/\sqrt{2T^{-2}\tr(\O_{T,M}^2)}|<\epsilon+x \right)\mathbb{P}\left( B_{i_1}...B_{i_d} \right)+N^{-c}\\
    \le&\left\lbrace \mathbb{P}\left(|\varphi^*_1/\sqrt{2T^{-2}\tr(\O_{T,M}^2)}|<\epsilon+x,|\varphi^*_2/\sqrt{2T^{-2}\tr(\O_{T,M}^2)}|<\epsilon  \right)+N^{-c}\right\rbrace \mathbb{P}\left( B_{i_1}...B_{i_d} \right)+N^{-c}\\
    \le& \mathbb{P}\left(A_N(x+2\epsilon)\right)\mathbb{P}\left( B_{i_1}...B_{i_d} \right)+2N^{-c}.
\end{align*}
Likewise,
\begin{align*}
    \mathbb{P}\left(A_N(x)B_{i_1}...B_{i_d}\right)
    \ge \mathbb{P}\left(A_N(x-2\epsilon)\right)\mathbb{P}\left( B_{i_1}...B_{i_d} \right)-2N^{-c}.
\end{align*}
Hence,
\begin{align*}
    \left| \mathbb{P}\left(A_N(x)B_{i_1}...B_{i_d}\right)
    -\mathbb{P}\left(A_N(x)\right)\mathbb{P}\left( B_{i_1}...B_{i_d} \right)\right|\le \Delta_{N,\epsilon}\cdot \mathbb{P}\left( B_{i_1}...B_{i_d}\right)+2N^{-c},
\end{align*}
where 
\begin{align*}
    \Delta_{N,\epsilon}
    &=\left|\mathbb{P}\left(A_N(x)\right)-\mathbb{P}\left(A_N(x+2\epsilon)\right) \right|+\left|\mathbb{P}\left(A_N(x)\right)-\mathbb{P}\left(A_N(x-2\epsilon)\right) \right|\\
    &=\mathbb{P}\left(A_N(x+2\epsilon)\right)-\mathbb{P}\left(A_N(x-2\epsilon)\right)  
\end{align*}
since $\mathbb{P}(A_N(x))$ is increasing in $x$. By running over all possible combinations of $1\le i_1<...<i_d\le N$, we have
\begin{align*}
    \zeta(N,d)\le \Delta_{N,\epsilon}\cdot H(N,d)+2\binom{N}{d}\cdot N^{-c}.
\end{align*}
Since $\mathbb{P}(A_N(x))\to\Phi(x)$, we have $\lim_{\epsilon\downarrow 0}\limsup_{N\to\infty}\Delta_{N,\epsilon}=\lim_{\epsilon\downarrow 0}\{\Phi(x+2\epsilon)-\Phi(x-2\epsilon)\}=0$. Since for each $d\ge 1$, $H(N,d)\to \pi^{-1/2}\exp(-dx/2)/d!$ as $N\to\infty$, we have $\limsup_{N\to\infty}H(N,d)<\infty$. Due to the fact that $\binom{N}{d}\le N^d$ for fixed $d\ge 1$, first sending $N\to\infty$ and then sending $\epsilon\downarrow0$, we get \eqref{C.3}.
    	
It remains to prove that the claim \eqref{C.4} indeed holds. According to Condition (C4) and the proof of Lemma \ref{error of variance}, there exist constants $0<c_M<C_M<\infty$ such that
\begin{align*}\label{C.5}
    c_M^2T^{-2}\tr(\Gamma_{0,M}^2)\le 2T^{-2}\tr(\O_{T,M}^2)\le C_M^2T^{-2}\tr(\Gamma_{0,M}^2).\tag{C.5}
\end{align*}
Define $\sigma_d^2=c^2T^{-2}\tr(\bm{\Sigma}_{11}^2)$ with $c_M\le c\le C_M$, then
\begin{align*}
    &\mathbb{P}\left(|\Theta_1|\ge\epsilon\sqrt{T^{-1}\tr(\O_{T,M}^2)}\right)\\
    =&\mathbb{P}\left( \left| \frac{1}{T^2}\sum_{1\le t<s\le q_T}\sum_{k=(t-1)w_T+1}^{tw_T-M}\sum_{l=(s-1)w_T+1}^{sw_T-M}\bm\gamma_{(1),k}^\top\bm\gamma_{(1),l}\right| \ge\epsilon\sqrt{T^{-1}\tr(\O_{T,M}^2)}\right)\\
    =&\mathbb{P}\left( \left|\frac{1}{\sigma_d T^2}\sum_{1\le t<s\le q_T}\sum_{k=(t-1)w_T+1}^{tw_T-M}\sum_{l=(s-1)w_T+1}^{sw_T-M}\bm\gamma_{(1),k}^\top\bm\gamma_{(1),l}\right| \ge\epsilon'\sqrt{\frac{\tr(\Gamma_{0,M}^2)}{\tr(\bm{\Sigma}_{11}^2)}}\right)\\
    \le&\exp\left\{ -\epsilon' \sqrt{\frac{\tr(\Gamma_{0,M}^2)}{\tr(\bm{\Sigma}_{11}^2)}} \right\}  \cdot \mathbb{E}\left(\exp\left|\frac{1}{\sigma_d T^2}\sum_{1\le t<s\le q_T}\sum_{k=(t-1)w_T+1}^{tw_T-M}\sum_{l=(s-1)w_T+1}^{sw_T-M}\bm\gamma_{(1),k}^\top\bm\gamma_{(1),l}\right| \right)\\
    \le&\exp\left\{ -\epsilon' \sqrt{\frac{\tr(\Gamma_{0,M}^2)}{\tr(\bm{\Sigma}_{11}^2)}} \right\}  \cdot \log(T),
\end{align*}
where the last inequality follows since 
\begin{align*}
    \sigma_d^{-1}T^{-2}\sum_{1\le t<s\le q_T}\sum_{k=(t-1)w_T+1}^{tw_T-M}\sum_{l=(s-1)w_T+1}^{sw_T-M}\bm\gamma_{(1),k}^\top\bm\gamma_{(1),l}/\log\{\log(T)\}\to 0,~~\text{a.s.}
\end{align*}
by the law of the iterated logarithm of zero-mean square integrable martingale (see Theorem 4.8 in \citet{Hall1980MartingaleLT}). Similarly,
\begin{align*}
    \mathbb{P}\left(|\Theta_2|\ge\epsilon\sqrt{T^{-1}\tr(\O_{T,M}^2)}\right)\le\exp\left\{ -\frac{\epsilon'}{2}\sqrt{\frac{\tr(\Gamma_{0,M}^2)}{\tr(\bm{\Sigma}_{22\cdot 1}\bm{\Sigma}_{21}\bm{\Sigma}_{11}^{-1}\bm{\Sigma}_{12})}}\right\} \cdot \log(T),
\end{align*}
where $\bm{\Sigma}_{22\cdot 1}=\bm{\Sigma}_{22}-\bm{\Sigma}_{21}\bm{\Sigma}_{11}^{-1}\bm{\Sigma}_{12}$, and
\begin{align*}
    \mathbb{P}\left(|\Theta_3|\ge\epsilon\sqrt{T^{-1}\tr(\O_{T,M}^2)}\right)\le\exp\left\lbrace -\frac{\epsilon'}{2}\sqrt{\frac{\tr(\Gamma_{0,M}^2)}{\tr[(\bm{\Sigma}_{21}\bm{\Sigma}_{11}^{-1}\bm{\Sigma}_{12})^2]}}\right\rbrace  \cdot \log(T).
\end{align*}
Here, we complete the proof of asymptotic independence under Gaussian case.
\\~\\
\noindent\textbf{Case \uppercase\expandafter{\romannumeral2}.} Investigate the asymptotic independence when $\bm\gamma_t$'s are M-dependent non-Gaussian. For $\bm{X}=(x_1,...,x_q)'\in\mathbb{R}^q$, we consider a smooth approximation of the maximum function $\bm{X}\to \max_{1\le i\le q}x_i$, namely,
\begin{align*}
    F_{\beta}(\bm{X})=\beta^{-1}\log\left\lbrace\sum_{i=1}^q\exp(\beta x_i) \right\rbrace , 
\end{align*}
where $\beta>0$ is the smoothing parameter that controls the level of approximation. An elementary calculation shows that $\forall \bm{X} \in\mathbb{R}^q$,
\begin{align*}
    0\le F_{\beta}(\bm{X})-\max_{1\le i\le q}x_i\le \beta^{-1}\log(q),
\end{align*}
see \citet{Chernozhukov2013InferenceOC}. W.L.O.G. assume that $\sigma_i=1$ for $i=1,\dots,N$. Define
\begin{align*}
    V(\bm{\gamma}_1,\dots,\bm{\gamma}_T):=F_{\beta}(\sqrt{T}\D^{-1/2}\bar{\bm\gamma}_T)=\beta^{-1}\log\left\lbrace\sum_{i=1}^N\exp\left(\beta T^{-1/2}\sum_{t=1}^T\gamma_{it} \right)  \right\rbrace .
\end{align*}
According to \eqref{C.1}-\eqref{C.2} and Lemma \ref{as}, setting $\beta=T^{1/8}\log(N)$, it suffices to show that
\begin{align*}
    \mathbb{P}\left(W(\bm\xi_1,\dots,\bm\xi_{q_T})\le x,V(\bm{\gamma}_1,\dots,\bm{\gamma}_T)\le \sqrt{2\log(N)-\log\{\log(N)\}+y} \right)\to\Phi(x)F(y). 
\end{align*}
Let $\{\bm\delta_t'\}_{t=1}^T$ be a Gaussian sequence which is independent of $\{\bm\gamma_t\}_{t=1}^T$ and preserves the auto-covariance structure of $\{\bm\gamma_t\}_{t=1}^T$. Define $\bm\zeta_t':=\frac{1}{w_T-M}\sum_{k=(t-1)w_T+1}^{tw_T-M}\bm\delta_k'$. According to the results of Case \uppercase\expandafter{\romannumeral1}, it remains to show that $(W(\bm\xi_1,\dots,\bm\xi_{q_T}), V(\bm{\gamma}_1,\dots,\bm{\gamma}_T))$ has the same limiting distribution as $(W(\bm\zeta_1',\dots,\bm\zeta_{q_T}'),V(\bm\delta_1',\dots,\bm\delta_T'))$. Similar to the Step 1 in the proof of Lemma \ref{Gaussian_approximation}, it suffices to show that 
\begin{align*}
    \mathbb{E}\{f(W(\bm\xi_1,\dots,\bm\xi_{q_T}), V(\bm{\gamma}_1,\dots,\bm{\gamma}_T))\}-\mathbb{E}\{f(W(\bm\zeta_1',\dots,\bm\zeta_{q_T}'),V(\bm\delta_1',\dots,\bm\delta_T'))\}\to 0,
\end{align*}
for every $f\in \mathcal{C}_b^3(\mathbb{R})$ as $(N,T)\to \infty$. Here, $\mathcal{C}_b^3(\mathbb{R})$ is the class of bounded functions with bounded and continuous derivatives up to order 3. 

Recall the definitions of $W_k$ and $W_{k,0}$ in \eqref{A.4} and define 
\begin{align*}
    &V_k:=V(\bm{\gamma}_1,\dots,\bm{\gamma}_{k-1},\bm{\delta}_k',\dots,\bm{\delta}_T')~~\text{for}~~k=1,\dots,T+1,\\
    &V_{k,0}=\beta^{-1}\log\left[\sum_{i=1}^N\exp\left\lbrace\beta T^{-1/2}\left(\sum_{t=1}^{k-1}\gamma_{it} +\sum_{t=k+1}^T\delta_{it}' \right)  \right\rbrace  \right], 
\end{align*}
which only rely on $\mathcal{F}_k:=\sigma\{\bm{\gamma}_1,...,\bm{\gamma}_{k-1},\bm{\delta}_{k+1}',...,\bm{\delta}_T'\}$. 
Then 
\begin{align*}
    &|\mathbb{E}\{f(W(\bm\xi_1,\dots,\bm\xi_{q_T}), V(\bm{\gamma}_1,\dots,\bm{\gamma}_T))\}-\mathbb{E}\{f(W(\bm\zeta_1',\dots,\bm\zeta_{q_T}'),V(\bm\delta_1',\dots,\bm\delta_T'))\}|\\
    \le&\sum_{k=1}^T|\mathbb{E}\{f(W_k,V_k)\}-\mathbb{E}\{f(W_{k+1},V_{k+1})\}|.
\end{align*}
By Taylor's expansion, we have
\begin{align*}
    &f(W_k,V_k)-f(W_{k,0},V_{k,0})\\
    =&f_1(W_{k,0},V_{k,0})(W_k-W_{k,0})+f_2(W_{k,0},V_{k,0})(V_k-V_{k,0})\\
    &+\frac{1}{2}f_{11}(W_{k,0},V_{k,0})(W_k-W_{k,0})^2+\frac{1}{2}f_{22}(W_{k,0},V_{k,0})(V_k-V_{k,0})^2\\
    &+\frac{1}{2}f_{12}(W_{k,0},V_{k,0})(W_k-W_{k,0})(V_k-V_{k,0})\\
    &+O(|W_k-W_{k,0}|^3)+O(|V_k-V_{k,0}|^3),
\end{align*}
and
\begin{align*}
    &f(W_{k+1},V_{k+1})-f(W_{k,0},V_{k,0})\\
    =&f_1(W_{k,0},V_{k,0})(W_{k+1}-W_{k,0})+f_2(W_{k,0},V_{k,0})(V_{k+1}-V_{k,0})\\
    &+\frac{1}{2}f_{11}(W_{k,0},V_{k,0})(W_{k+1}-W_{k,0})^2+\frac{1}{2}f_{22}(W_{k,0},V_{k,0})(V_{k+1}-V_{k,0})^2\\
    &+\frac{1}{2}f_{12}(W_{k,0},V_{k,0})(W_{k+1}-W_{k,0})(V_{k+1}-V_{k,0})\\
    &+O(|W_{k+1}-W_{k,0}|^3)+O(|V_{k+1}-V_{k,0}|^3),
\end{align*}
where $f=f(x,y)$, $f_1=\partial f/\partial x$, $f_2=\partial f/\partial y$, $f_{11}=\partial^2f/\partial x^2 $, $f_{22}=\partial^2f/\partial y^2 $ and $f_{12}=\partial^2f/\partial x\partial y $. By \eqref{A.5}, we have
\begin{align*}
    &\mathbb{E}\{f_1(W_k,V_{k,0})(W_{k+1}-W_{k,0})\}=\mathbb{E}\{f_1(W_{k,0},V_{k,0})(W_{k+1}-W_{k,0})\},~~\text{and}\\
    &\mathbb{E}\{f_{11}(W_{k,0},V_{k,0})(W_k-W_{k,0})^2\}=\mathbb{E}\{f_{11}(W_{k,0},V_{k,0})(W_{k+1}-W_{k,0})^2\}.
\end{align*}
    	
Next consider $V_k-V_{k,0}$. Let $v_{k,0,i}=T^{-1/2}\sum_{t=1}^{k-1}\gamma_{it}+T^{-1/2}\sum_{t=k+1}^T\delta_{it}'$, $v_{k,i}=v_{k,0,i}+T^{-1/2}\delta_{ik}'$, $v_{k+1,i}=v_{k,0,i}+T^{-1/2}\gamma_{ik}$, $\bm{v}_{k,0}=(v_{k,0,1},...,v_{k,0,N})'$ and $\bm{v}_k=(v_{k,1},...,v_{k,N})'$. By Taylor's expansion, we have
\begin{align*}\label{C.7}
    &V_k-V_{k,0}\\
    =&\sum_{i=1}^N\partial_iF_\beta(\bm{v}_{k,0})(v_{k,i}-v_{k,0,i})+\frac{1}{2}\sum_{i=1}^N\sum_{j=1}^N\partial_i\partial_jF_\beta(\bm{v}_{k,0})(v_{k,i}-v_{k,0,i})(v_{k,j}-v_{k,0,j})\\
    &+\frac{1}{6}\sum_{i=1}^N\sum_{j=1}^N\sum_{l=1}^N\partial_i\partial_j\partial_lF_\beta(\bm{v}_{k,0}+\delta(v_k-v_{k,0}))(v_{k,i}-v_{k,0,i})(v_{k,j}-v_{k,0,j})(v_{k,l}-v_{k,0,l}),\tag{C.7}
\end{align*}
for some $\delta\in(0,1)$. Again, due to $\mathbb{E}(\bm{\gamma}_t)=\mathbb{E}(\bm{\delta}_t')=0$ and $\mathbb{E}(\bm{\gamma}_t\bm{\gamma}_t^\top)=\mathbb{E}(\bm{\delta}_t\bm{\delta}_t^{\prime\top})$, it can be verified that 
\begin{align*}
    \mathbb{E}(v_{k,i}-v_{k,0,i}|\mathcal{F}_k)=\mathbb{E}(v_{k+1,i}-v_{k,0,i}|\mathcal{F}_k)~~\text{and}~~ 
    \mathbb{E}\{(v_{k,i}-v_{k,0,i})^2|\mathcal{F}_k\}=\mathbb{E}\{(v_{k+1,i}-v_{k,0,i})^2|\mathcal{F}_k\}.
\end{align*}
By Lemma A.2 in \citet{Chernozhukov2013InferenceOC}, we have
\begin{align*}
    \left|\sum_{i=1}^N\sum_{j=1}^N\sum_{l=1}^N\partial_i\partial_j\partial_lF_\beta(\bm{v}_{k,0}+\delta(\bm{v}_k-\bm{v}_{k,0})) \right|\le C\beta^2 
\end{align*}
for some positive constant $C$. By Condition (C2) and the definition of $\bm\gamma_t$, we have $\mathbb{P}(\max_{i,t}|\gamma_{it}|>C\log(NT))\to 0$, and since $\bm\delta_t'$ are Gaussian, $\mathbb{P}(\max_{i,t}|\delta_{it}'|>C\log(NT))\to 0$. Hence,
\begin{align*}
    &\left|\sum_{i=1}^N\sum_{j=1}^N\sum_{l=1}^N\partial_i\partial_j\partial_lF_\beta(\bm{v}_{k,0}+\delta(\bm{v}_k-\bm{v}_{k,0}))(v_{k,i}-v_{k,0,i})(v_{k,j}-v_{k,0,j})(v_{k,l}-v_{k,0,l}) \right|\\
    \le& C\beta^2T^{-3/2}\log^3(NT)
\end{align*}
holds with probability approaching one. Consequently, we have with probability approaching one,
\begin{equation*}
    \left|\mathbb{E}\{f_2(W_{k,0},V_{k,0})(V_k-V_{k,0})\}-\mathbb{E}\{f_2(W_{k,0},V_{k,0})(V_{k+1}-V_{k,0})\} \right|\le C\beta^2T^{-3/2}\log^3(NT). 
\end{equation*}
Similarly, it can be verified that
\begin{equation*}
    \left|\mathbb{E}\{f_{22}(W_{k,0},V_{k,0})(V_k-V_{k,0})^2\}-\mathbb{E}\{f_2(W_{k,0},V_{k,0})(V_{k+1}-V_{k,0})^2\} \right|\le C\beta^2T^{-3/2}\log^3(NT),
\end{equation*}
and 
\begin{align*}
    &\left|\mathbb{E}\{f_{12}(W_{k,0},V_{k,0})(W_k-W_{k,0})(V_k-V_{k,0})\}-\mathbb{E}\{f_{12}(W_{k,0},V_{k,0})(W_{k+1}-W_{k,0})(V_{k+1}-V_{k,0})\} \right|\\
    \le& C\beta^2T^{-3/2}\log^3(NT). 
\end{align*}
Again, Lemma A.2 in \citet{Chernozhukov2013InferenceOC}, together with  \eqref{C.7}, implies that  $\mathbb{E}(|V_k-V_{k,0}|^3)=O\{T^{-3/2}\log^3(NT)\}$. By \eqref{A.6}, we have $\mathbb{E}(|W_k-W_{k,0}|^4)=O(T^{-2})$, thus 
\begin{equation*}
    \sum_{k=1}^T\mathbb{E}(|W_k-W_{k,0}|^3)\le \sum_{k=1}^T\{\mathbb{E}(|W_k-W_{k,0}|^4)\}^{3/4}=O(T^{-1/2}).
\end{equation*}
Combining all facts together, we conclude that
\begin{equation*}
    \sum_{k=1}^T|\mathbb{E}\{f(W_k,V_k)\}-\mathbb{E}\{f(W_{k+1},V_{k+1})\}|=O\{\beta^2T^{-3/2}\log^3(NT)\}+O(T^{-1/2})\to 0,
\end{equation*}
as $(N,T)\to \infty$, due to Condition (C9). Then the proof is complete.

\subsubsection{Proof of Theorem 7}
\proof It suffices to show the conclusion holds for Gaussian $\bm\gamma_t$'s. Let $\mathcal{A}:=\{1\le i\le N:\alpha_i\ne 0\}$. Define $\bm\gamma_{\mathcal{A},t}:=(\gamma_{it},i\in\mathcal{A})^\top$ and $\bm\gamma_{\mathcal{A}^c,t}:=(\gamma_{it},i\in\mathcal{A}^c)^\top$. By the proof of Theorems 1-3, we have under $H_1$,
\begin{align*}
    \frac{T_{\SUM}-\hat{\mu}_T}{\hat{\sigma}_T}=&W(\bm\xi_1,\dots,\bm\xi_{q_T})+\frac{\bma^\top\bma}{\sqrt{2T^{-2}\tr(\O_{T,M}^2)}}+o_p(1)\\
    =&W_{\mathcal{A}}+W_{\mathcal{A}^c}+\frac{\bma^\top\bma}{\sqrt{2T^{-2}\tr(\O_{T,M}^2)}}+o_p(1),
\end{align*}
with
\begin{align*}
    &W(\bm\xi_1,\dots,\bm\xi_{q_T})=\frac{1}{T^2}\sum_{1\le t<s\le q_T}\sum_{k=(t-1)w_T+1}^{tw_T-M}\sum_{l=(s-1)w_T+1}^{sw_T-M}\bm\gamma_k^\top\bm\gamma_l/\sqrt{2T^{-2}\tr(\O_{T,M}^2)},\\
    &W_{\mathcal{A}}:=\frac{1}{T^2}\sum_{1\le t<s\le q_T}\sum_{k=(t-1)w_T+1}^{tw_T-M}\sum_{l=(s-1)w_T+1}^{sw_T-M}\sum_{i\in\mathcal{A}}\gamma_{ik}\gamma_{il}/\sqrt{2T^{-2}\tr(\O_{T,M}^2)},~~\text{and}\\
    &W_{\mathcal{A}^c}:=\frac{1}{T^2}\sum_{1\le t<s\le q_T}\sum_{k=(t-1)w_T+1}^{tw_T-M}\sum_{l=(s-1)w_T+1}^{sw_T-M}\sum_{i\in\mathcal{A}^c}\gamma_{ik}\gamma_{il}/\sqrt{2T^{-2}\tr(\O_{T,M}^2)}.
\end{align*}
By the proof of Theorems 4-5, we have under $H_1$,
\begin{align*}
    T_{\MAX}=&T\left\|\D^{-1/2}\bar{\bm\gamma}_T\right\|_\infty^2+\max_{1\le i\le N}\frac{T\alpha_i^2}{\sigma_i}+o_p(1)\\
    =&\max_{i\in\mathcal{A}}\frac{1}{T\sigma_i}\left(\sum_{t=1}^T\gamma_{it}\right)^2+\max_{i\in\mathcal{A}^c}\frac{1}{T\sigma_i}\left(\sum_{t=1}^T\gamma_{it}\right)^2+\max_{1\le i\le N}\frac{T\alpha_i^2}{\sigma_i}+o_p(1).
\end{align*}
According to the proof of Theorem 6, we have known that $W_{\mathcal{A}^c}$ and $\max_{i\in\mathcal{A}^c}T^{-1}\sigma_i^{-1}(\sum_{t=1}^T\gamma_{it})^2$ are asymptotically independent. Hence, it suffices to show that $W_{\mathcal{A}^c}$ is asymptotically independent of $T^{-1}\sigma_i^{-1}(\sum_{t=1}^T\gamma_{it})^2$ for $i\in\mathcal{A}$.

Define $\bm{\Sigma}_{\mathcal{A},\mathcal{A}^c}:=\cov(\bm{\gamma}_{\mathcal{A},t}, \bm{\gamma}_{\mathcal{A}^c,t})$. By Lemma \ref{in}, $\bm{\gamma}_{\mathcal{A}^c,t}$ can be decomposed as $\bm{\gamma}_{\mathcal{A}^c,t}=\bm{u}_t+\bm{v}_t$, where $\bm{u}_t=\bm{\gamma}_{\mathcal{A}^c,t}-\bm{\Sigma}_{\mathcal{A}^c,\mathcal{A}}\bm{\Sigma}_{\mathcal{A},\mathcal{A}}^{-1}\bm{\gamma}_{\mathcal{A},t}$ and $\bm{v}_t=\bm{\Sigma}_{\mathcal{A}^c,\mathcal{A}}\bm{\Sigma}_{\mathcal{A},\mathcal{A}}^{-1}\bm{\gamma}_{\mathcal{A},t}$ satisfying that $\bm{u}_t\sim N(\bm{0},\bm{\Sigma}_{\mathcal{A}^c,\mathcal{A}^c}-\bm{\Sigma}_{\mathcal{A}^c,\mathcal{A}}\bm{\Sigma}_{\mathcal{A},\mathcal{A}}^{-1}\bm{\Sigma}_{\mathcal{A},\mathcal{A}^c})$, $\bm{v}_t\sim N(\bm{0},\bm{\Sigma}_{\mathcal{A}^c,\mathcal{A}}\bm{\Sigma}_{\mathcal{A},\mathcal{A}}^{-1}\bm{\Sigma}_{\mathcal{A},\mathcal{A}^c})$ and 
\begin{equation*}\label{C.6}
    \bm{u}_t~~\text{and}~~\bm{\gamma}_{\mathcal{A},t}~~\text{are independent}.\tag{C.6}
\end{equation*}
Then, we have
\begin{align*}
    W_{\mathcal{A}^c}=\frac{1}{T^2}\sum_{1\le t<s\le q_T}\sum_{k=(t-1)w_T+1}^{tw_T-M}\sum_{l=(s-1)w_T+1}^{sw_T-M}(\bm{u}_k'\bm{u}_l+2\bm{u}_k'\bm{v}_l +\bm{v}_k'\bm{v}_l)/\sqrt{2T^{-2}\tr(\O_{T,M}^2)}.
\end{align*}
By using the arguments similar to those in the proof of \eqref{C.4}, we have 
\begin{align*}
    &\mathbb{P}\left(\frac{1}{T^2}\sum_{1\le t<s\le q_T}\sum_{k=(t-1)w_T+1}^{tw_T-M}\sum_{l=(s-1)w_T+1}^{sw_T-M}2\bm{u}_k'\bm{v}_l\ge \epsilon \sqrt{2T^{-2}\tr(\O_{T,M}^2)} \right)\le\log(T)\exp\left(-c_{\epsilon}\sqrt{\frac{N}{|\mathcal{A}|}}\right)\to 0,\\
    &\mathbb{P}\left(\frac{1}{T^2}\sum_{1\le t<s\le q_T}\sum_{k=(t-1)w_T+1}^{tw_T-M}\sum_{l=(s-1)w_T+1}^{sw_T-M}\bm{v}_k'\bm{v}_l\ge \epsilon \sqrt{2T^{-2}\tr(\O_{T,M}^2)} \right)\le\log(T)\exp\left(-c_{\epsilon}\sqrt{\frac{N}{|\mathcal{A}|}}\right)\to 0,
\end{align*}
due to $|\mathcal{A}|=o[N/\log^2\{\log(N)\}]$ and Condition (C9). Consequently, we conclude that 
\begin{align*}
    W_{\mathcal{A}^c}=\frac{1}{T^2}\sum_{1\le t<s\le q_T}\sum_{k=(t-1)w_T+1}^{tw_T-M}\sum_{l=(s-1)w_T+1}^{sw_T-M}\bm{u}_k'\bm{u}_l/\sqrt{2T^{-2}\tr(\O_{T,M}^2)}+o_p(1),
\end{align*}
which, together with Lemma \ref{as} and \eqref{C.6}, implies that $W_{\mathcal{A}^c}$ is asymptotically independent of $T^{-1}\sigma_i^{-1}(\sum_{t=1}^T\gamma_{it})^2,\forall i\in\mathcal{A}$.

\bibliographystyle{apa}
\bibliography{Refers}
\end{document}